\documentclass[oneside,graybox,envcountchap,sectrefs]{svmult}

\usepackage{mathptmx}
\usepackage{helvet}
\usepackage{courier}
\usepackage{braket} 
\usepackage{algorithmicx}
\usepackage{algpseudocode} 
\usepackage{amsfonts}
\usepackage{amsmath}
\usepackage{simplewick}
\usepackage{type1cm}         
\usepackage{exercise}
\usepackage{makeidx}         
\usepackage{graphicx}        
\usepackage{multicol}        
\usepackage[bottom]{footmisc}
\usepackage[usenames,dvipsnames,x11names]{xcolor}
 \usepackage{listings}
 \usepackage{epic}
 \usepackage{eepic}
 \usepackage{a4wide}
 \usepackage{color}
 \usepackage{amsmath}
 \usepackage{amssymb}
 \usepackage[T1]{fontenc}
 \usepackage{cite} 
 \usepackage{shadow}
 \usepackage{hyperref}
 \usepackage{bezier}
 \usepackage{pstricks}
\usepackage{textcomp,type1ec,pdfpages}
\usepackage{bera}
\usepackage{slashed}
\usepackage{environ}
\usepackage{esvect}
\usepackage{xspace}
\usepackage{bm}

\usepackage{chapterbib}
\definecolor{dkgreen}{rgb}{0,0.6,0}
\definecolor{gray}{rgb}{0.5,0.5,0.5}
\definecolor{mauve}{rgb}{0.58,0,0.82}

\definecolor{gray}{gray}{0.5}
\definecolor{green}{rgb}{0,0.5,0}

\lstnewenvironment{Python}[1]{
\lstset{
language=python,
basicstyle=\footnotesize\setstretch{1},
stringstyle=\color{red},
showstringspaces=false,
alsoletter={1234567890},
otherkeywords={\ , \}, \{},
keywordstyle=\color{blue},
emph={access,and,break,class,continue,def,del,elif ,else,%
except,exec,finally,for,from,global,if,import,in,is,%
lambda,not,or,pass,print,raise,return,try,while},
emphstyle=\color{black}\bfseries,
emph={[2]True, False, None, self},
emphstyle=[2]\color{red},
emph={[3]from, import, as},
emphstyle=[3]\color{blue},
upquote=true,
morecomment=[s]{"""}{"""},
commentstyle=\color{dkgreen}\slshape, 
emph={[4]1, 2, 3, 4, 5, 6, 7, 8, 9, 0},
emphstyle=[4]\color{blue},
framexleftmargin=1mm, framextopmargin=1mm, rulesepcolor=\color{blue},
breakatwhitespace=true,
tabsize=2
}}{}

\lstnewenvironment{C++}[1]{
\lstset{
language=c++,
basicstyle=\footnotesize\setstretch{1},
stringstyle=\color{red},
showstringspaces=false,
alsoletter={1234567890},
otherkeywords={\ , \}, \{},
keywordstyle=\color{blue},
emph={access,and,break,class,continue,def,del,elif ,else,%
except,exec,finally,for,from,global,if,import,in,is,%
lambda,not,or,pass,print,raise,return,try,while},
emphstyle=\color{black}\bfseries,
emph={[2]True, False, None, self},
emphstyle=[2]\color{red},
emph={[3]from, import, as},
emphstyle=[3]\color{blue},
upquote=true,
morecomment=[s]{"""}{"""},
commentstyle=\color{dkgreen}\slshape, 
emph={[4]1, 2, 3, 4, 5, 6, 7, 8, 9, 0},
emphstyle=[4]\color{blue},
framexleftmargin=1mm, framextopmargin=1mm, rulesepcolor=\color{blue},
breakatwhitespace=true,
tabsize=2
}}{}
\usepackage{calrsfs}
\DeclareMathAlphabet{\pazocal}{OMS}{zplm}{m}{n}
\newcommand {\bea}{\begin{eqnarray}}
\newcommand {\eea}{\end{eqnarray}}
\newcommand {\be}{\begin{equation}}
\newcommand {\ee}{\end{equation}}
\newcommand{\beq}{\begin{eqnarray}}
\newcommand{\eeq}{\end{eqnarray}}
\newcommand {\bc}{\begin{center}}
\newcommand {\ec}{\end{center}}

\def\lsim{\mathrel{\rlap{\lower4pt\hbox{\hskip1pt$\sim$}}
    \raise1pt\hbox{$<$}}}               
\def\gsim{\mathrel{\rlap{\lower4pt\hbox{\hskip1pt$\sim$}}
    \raise1pt\hbox{$>$}}}  
    
\graphicspath{{./Chapter11-figures/}}

\makeindex             

\begin{document}
\frontmatter


\tableofcontents
\mainmatter

%
%

\title{Self-consistent Green's function approaches}
\author{Carlo~Barbieri and Arianna Carbone}
\institute{
Carlo Barbieri  \at 1 Department of Physics, University of Surrey, Guildford GU2 7XH, UK \email{C.Barbieri@surrey.ac.uk},
\and Arianna Carbone   \at 2 Institut f\"ur Kernphysik, Technische Universit\"at Darmstadt, 64289 Darmstadt, Germany, and 
\\ ExtreMe Matter Institute EMMI, GSI Helmholtzzentrum f\"ur Schwerionenforschung GmbH, 64291 Darmstadt, Germany, \email{arianna@theorie.ikp.physik.tu-darmstadt.de}
}
\maketitle
\abstract{
We present the fundamental techniques and working equations of many-body Green's function theory for calculating ground state properties and the spectral strength.  Green's function methods closely relate to other polynomial scaling approaches discussed in chapters~8 and ~10. However, here we aim directly at a global view of the many-fermion structure.
  We  derive the working equations for calculating many-body propagators, using both the Algebraic Diagrammatic Construction technique and the self-consistent formalism at finite temperature.  Their implementation is discussed, as well as the inclusion of three-nucleon interactions.
The self-consistency feature is essential to guarantee thermodynamic consistency.
The  pairing and neutron matter models introduced in previous chapters are solved and compared with the other methods in this book.}

\allowdisplaybreaks[1]

\section{Introduction}

{\em Ab initio} methods that present polynomial scaling with the number of particles have proven highly useful in reaching
finite systems of rather large sizes up to medium mass nuclei~\cite{ch11_Soma2014s2n,ch11_Barbieri2014qmbt17,ch11_CCM_2014rev} 
and even infinite matter~\cite{ch11_Frick2003,ch11_Baardsen2013ccNM,ch11_Carbone2014}. Most approaches of this type that are discussed in previous chapters aim at the direct evaluation of the ground state energy of the system, where several other quantities of interest can be addressed in a second stage through the equation of motion and particle removal or attachment techniques.
Here, we will follow a different route and focus on gaining from the start a global  view of the spectral structure of a system of fermions. Our approach will be that of calculating directly the self-energy (also know as {\em mass operator}), which describes the 
complete response of a particle embedded in the true ground state of the system. This not only provides an effective in medium interaction for the nucleon, but it is also the optical potential for elastic scattering and it yields the spectral information relative to the attachment and removal of a particle.   Once the  one body Green's function has been obtained, the total energy of the system is calculated, as the final step, by means of appropriate sum rules~\cite{ch11_Koltun1974KSR,ch11_dickhoff2004}.

Two main approaches have become standard choices for calculations of Green's functions in nuclear many-body theory.
The Algebraic Diagrammatic Construction (ADC) method, that was originally devised for quantum chemistry applications~\cite{ch11_Schirmer1982ADC2,ch11_Schirmer1983ADCn}, has proven to be optimal for discrete bases, as it is normally necessary to exploit for finite nuclei. However, this can also be applied to fermion gases in a box with periodic boundary conditions, which  simplifies the analysis even more thanks to translational invariance. We will focus on the case of infinite nucleonic matter and provide an example of a working numerical code.
ADC($n$) methods are part of a larger class of approaches based on intermediate-state representations (ISRs) to which also the equation-of-motion coupled cluster belongs~\cite{ch11_Mertins1996ISR1,ch11_Mertins1996ISR2}.
The other method consists in solving directly the nucleon-nucleon ladder scattering matrix for dressed particles in the medium, which can be done effectively in a finite temperature formalism~\cite{ch11_Frick2004PhD,ch11_Rios2007PhD}. Hence, this makes possible to study thermodynamic properties of the infinite and liquid matter. For these studies to be reliable, it is mandatory to ensure the satisfaction of fundamental conservation laws and to maintain thermodynamic consistency in the infinite size limit. We show here how to achieve this by preforming fully {\em self-consistent} calculations of the Green's function. In this context, `self-consistency' means that the input information about the ground state and excitations of the systems no longer depend on any reference state but instead it is taken directly from the computed correlated wave function (or propagator, in our case). To achieve this, the computed spectral function is fed back into the working equations and calculations are repeated until a consistency between input and output is obtained.  This approach 
is referred to as self-consistent Green's function (SCGF) method and it is always implemented, partially or in full, for nuclear structure applications.

Very recent advances in computational applications concern the extension of SCGF theory to the Gorkov-Nambu formalism for the breaking on particle number symmetry~\cite{ch11_VdSluys1993,ch11_Soma2011GkvI,ch11_Idini2012}. This allows to  treat pairing systematically in systems with degenerate reference states and, therefore, to calculate open-shell nuclei directly. As a result, these developments have  opened the possibility of  studying large set of semi-magic nuclei that were previously beyond the reach of {\emph ab initio} theory. We will not discuss the Gorkov-SCGF method here, but we will focus on the fundamental features of the standard approaches instead. The interested reader is referred to recent literature on the topic~\cite{ch11_Soma2011GkvI,ch11_Soma2013rc,ch11_Soma2014Lanc,ch11_Cipollone2015OxChain}.

In the process of discussing the relevant working equations  of SCGF theory, we will also deal with applications to the same pairing model and the neutron matter with the Minnesota potential already discussed in chapters~8 and~9.  Together with presenting the most important steps for their numerical implementations, this book provides two examples of working codes in FORTRAN and C++ that can solve these models.  Results for the self-energy and spectral functions  should serve to gain a deeper understanding of the many-body physics that is embedded in the SCGF method.
 In discussing this, we will also  benchmark the SCGF results with those obtained in other chapters of this book: coupled cluster (chapter~8),  Monte Carlo (chapter~9) and in-medium similarity renormalization group (chapter 10). 

%

\section{Many-body Green's function theory}
\label{sec:scgf_defs}


This chapter will focus on many-body {\em  Green's functions}, which are also referred to as {\em propagators}. These are defined  in the 
second quantization formalism by assuming the knowledge of the true ground state $\vert\Psi^A_0\rangle$ of a target system of $A$ nucleons, which
is taken to be a vacuum of excitations.  The one-body Green's function (or propagator) is then defined as~\cite{ch11_FetterWalecka,ch11_Dickhoff2008}:
\begin{equation}
 i\hbar \; g_{\alpha\beta}(t - t') =  \langle\Psi^A_0\vert  \pazocal{T} [ a_\alpha(t)   a^\dagger_\beta(t') ]  \vert\Psi^A_0\rangle \; ,
 \label{eq:g1Time}
\end{equation}
where $\pazocal{T}$ is the time ordering operator, $a^\dagger_\alpha(t)$~($a_\alpha(t)$) are the creation (annihilation) operators in Heisenberg picture,
and  greek indices $\alpha$,~$\beta$,~... label a complete single particle basis that defines our  model space. These can be the continuum momentum or coordinate spaces or any discrete set of single particle states. Note that $ g(t - t')$ depends only on the time difference $t-t'$ due to time translation invariance.
For $t>t'$, Eq.~\eqref{eq:g1Time} gives the probability amplitude to add a particle to $\vert\Psi^A_0\rangle$ in state $\beta$  at time $t'$ and then to let it propagate to reach  state $\alpha$ at a later time $t$. Vice versa, for $t<t'$ a particle is removed from state $\alpha$ at $t$ and added to $\beta$ at $t'$.

In spite of being the simplest type of propagator, the one-body Green's function does contain a wealth of information regarding single particle behavior inside the many-body system, one-body observables, the total binding energy, and even elastic nucleon-nucleus scattering.
The propagation of a particle or a hole excitation corresponds to the time evolution of an intermediate many-body system with $A+1$ or $A-1$ particles.
One can better  understand the physics information included in Eq.~\eqref{eq:g1Time} from considering the eigenstates $\vert\Psi^{A+1}_n\rangle$, $\vert\Psi^{A-1}_k\rangle$ and eigenvalues $E^{A+1}_n$, $E^{A-1}_k$ of these intermediate systems. By expanding on these eigenstates and Fourier transforming from time to frequency, one arrives at the spectral representation of the one-body Green's function:
\begin{align}
 g_{\alpha\beta}(\omega) ={}& \int {\rm d}\tau \; e^{i\omega\tau} g_{\alpha\beta}(\tau)
 \nonumber \\
 ={}&
 \sum_n  \frac{ 
          \langle\Psi^A_0\vert  	a_\alpha   \vert\Psi^{A+1}_n\rangle
          \langle\Psi^{A+1}_n\vert  a^\dagger_\beta  \vert\Psi^A_0\rangle
              }{\hbar\omega - (E^{A+1}_n - E^A_0) + {\rm i} \eta }
 ~+~ \sum_k \frac{
 		  \langle\Psi^A_0\vert  	a^\dagger_\beta   \vert\Psi^{A-1}_k\rangle
          \langle\Psi^{A-1}_k\vert  a_\alpha  \vert\Psi^A_0\rangle
              }{\hbar\omega - (E^A_0 - E^{A-1}_k) - {\rm i} \eta } \; ,
 \nonumber \\
  \equiv{}& \sum_n \frac{(\pazocal{X}^{n}_\alpha)^*  \pazocal{X}^{n}_\beta}{\hbar\omega  - \varepsilon^+_{n} + i\eta} 
        ~+~ \sum_k \frac{\pazocal{Y}^{k}_\alpha  (\pazocal{Y}^{k}_\beta)^*}{\hbar\omega  - \varepsilon^-_{k} - i\eta}  \; ,
\label{eq:g1Leh}
\end{align}
where the operators $a^\dagger_\alpha$ and~$a_\alpha$ are now in Sch\"ordinger picture. 
Eq.~\eqref{eq:g1Leh} was derived by a number of authors in the 1950s but is usually referred to as the `Lehmann' representation in many-body physcis~\cite{ch11_Umezawa1951spRep,ch11_Kallen1952SpRep,ch11_Lehmann1954}.
For the rest of this chapter (with the only exception of Appendix 1) we will work in dimensionless $\hbar=c=1$ units to avoid carrying over unnecessary  $\hbar$ terms.
From Eq.~\eqref{eq:g1Leh}, we see that the poles of the Green's function,
\hbox{$\varepsilon_n^{+}\equiv(E^{A+1}_n - E^A_0)$} and  \hbox{$\varepsilon_k^{-}\equiv(E^A_0 - E^{A-1}_k)$},
are one-nucleon addition and removal energies, respectively.  Note that these are generically referred to in the literature as ``separation'' or ``quasiparticle'' energies although the first naming should normally refer to transitions involving only ($A\pm1$)-nucleon ground states.  We will use the second convention in the following, unless the two naming are strictly equivalent.
In the last line of Eq.~\eqref{eq:g1Leh} we have also introduced short notations for the spectroscopic amplitudes associated with the addition
($\pazocal{X}^{n}_\alpha \equiv \langle\Psi^{A+1}_n\vert  a^\dagger_\alpha  \vert\Psi^A_0\rangle$) and the removal
($\pazocal{Y}^{k}_\alpha \equiv \langle\Psi^{A-1}_k\vert  a_\alpha  \vert\Psi^A_0\rangle$)  of a particle to and from the initial ground state~$\vert\Psi^A_0\rangle$.
We will use the latin letter  $n$ to label one-particle excitations and to distinguish them from one-hole states that are indicated by $k$ instead.
 This compact form will simplify deriving the working formalism in the following sections.

\begin{figure}[t]
\begin{center}
\includegraphics[width=0.42\textwidth]{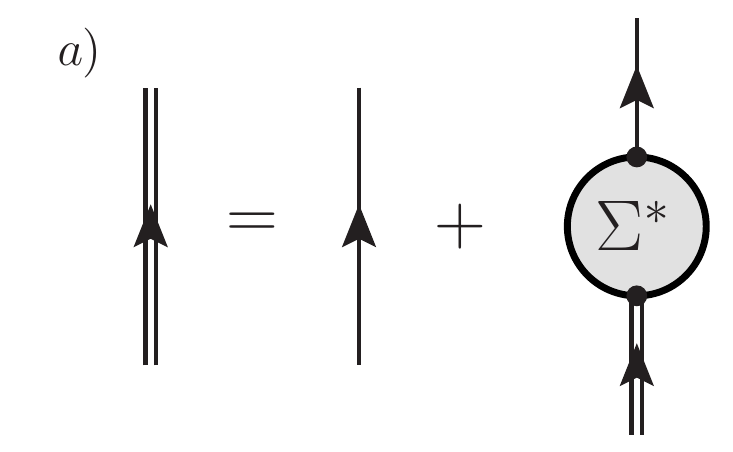}   \hspace{0.08\textwidth} 
\includegraphics[width=0.42\textwidth]{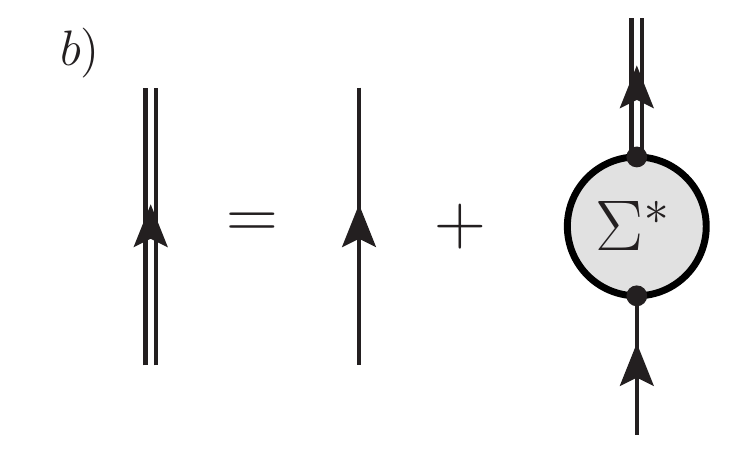}
\caption{Diagrammatic representations of the Dyson equation. The diagram on the left represents Eq.~\eqref{eq:Dyson_a}, while its conjugate equation~\eqref{eq:Dyson_b} is shown to the right. Single lines with an
arrow represent the unperturbed propagator $g^{(0)}(\omega)$ and  double lines are the fully
dressed propagator $g(\omega)$ of Eq.~\eqref{eq:g1Leh}.  Both equations, when expanded in terms
of $g^{(0)}(\omega)$, give rise to the same series of diagrams for the correlated propagator. } 
\label{fig:DysonEq}
\end{center}
\end{figure}

The one-body Green's function~\eqref{eq:g1Leh} is completely determined by solving the Dyson equation:
\begin{subequations}
\label{eq:Dyson}
\begin{align}
  \label{eq:Dyson_a}
  g_{\alpha\beta}(\omega) ={}&g^{(0)}_{\alpha\beta}(\omega) ~+~ \sum_{\gamma\delta} \; g^{(0)}_{\alpha\gamma}(\omega) \, \Sigma_{\gamma\delta}^{\star}(\omega) \, g_{\delta\beta}(\omega) 
   \\   \label{eq:Dyson_b}
  ={}&g^{(0)}_{\alpha\beta}(\omega) ~+~ \sum_{\gamma\delta} \; g_{\alpha\gamma}(\omega) \, \Sigma_{\gamma\delta}^{\star}(\omega) \, g^{(0)}_{\delta\beta}(\omega)  \; ,
\end{align}
\end{subequations}
where we have put in evidence that there exists two different conjugate forms of this equation, corresponding to the first and second lines. 
In Eqs.~\eqref{eq:Dyson},  the unperturbed propagator $g^{(0)}_{\alpha\beta}(\omega)$ is the initial reference state (usually a mean-field or Hartree-Fock state), while $g_{\alpha\beta}(\omega)$ is called the {\em correlated} or {\em dressed} propagator. The quantity $\Sigma_{\gamma\delta}^{\star}(\omega)$ is the {\em irreducible self-energy}
and it is often referred to as the {\em mass operator}.  This operator plays a central role in the GF formalism and can be interpreted as the 
non-local and energy-dependent potential that each fermion feels due to the interactions with the medium.  For frequencies $\omega>0$,  the solution of Eqs.~\eqref{eq:Dyson} yields a continuum spectrum with $E^{A+1}_n > E^A_0$ and the state $\vert\Psi^{A+1}_n\rangle$  describes the elastic scattering of the additional nucleon off the $\vert\Psi^A_0\rangle$ target. It can be show that $\Sigma^{\star}(\omega)$ is an exact optical potential for scattering of a particle  from the many-body target~\cite{ch11_MahauxSartor91,ch11_Capuzzi1996,ch11_Cederbaum2001}.
The Dyson equation is nonlinear in its solution, $g(\omega)$, and thus it corresponds to an all-orders resummation of diagrams involving the self-energy.
 The Feynman diagrams corresponding to both forms of the Dyson equation are shown in Fig.~\ref{fig:DysonEq}.
In both cases, by recursively substituting the exact Green's function (indicated by double lines) that appears on the right hand side with the whole equation, one finds a unique expansion in terms of the unperturbed $g^{(0)}(\omega)$ and the irreducible self-energy. The solution of Eqs.~\eqref{eq:Dyson} is referred to as {\em dressed} propagators since it formally results by `dressing' the free particle by repeated interactions with the system ($\Sigma^{\star}(\omega)$).

 A full knowledge of the self-energy  $\Sigma^{\star}(\omega)$ (see Eqs.~(\ref{eq:Dyson})) would yield the exact solution for $g(\omega)$ but in practice this has to be approximated somehow.
 Standard perturbation theory,  expands $\Sigma^{\star}(\omega)$ in a series of terms that depend on the interactions and on the unperturbed propagator $g^{(0)}(\omega)$.  However, it is also possible to rearrange the 
perturbative expansion in diagrams that depend only on the exact dressed propagator itself (that is, $\Sigma^{\star}=\Sigma^{\star}[g(\omega)]$).  Since any propagator in this diagrammatic expansion is already dressed, one only needs to consider a smaller set of contributions---the so-called  {\em skeleton} diagrams.
These are diagrams that  do not explicitly include any self-energy insertion, as these are already generated by Eqs.~\eqref{eq:Dyson}. 
We will discuss these aspects in more detail in Sec.~\ref{sec:pertexp}.
For the present discussion, we only need to be aware that the functional dependence of $\Sigma^{\star}[g(\omega)]$  requires an iterative procedure in which  $\Sigma^{\star}(\omega)$ and Eqs.~\eqref{eq:Dyson} are calculated several times until they converge to a unique solution.
This approach defines the SCGF method and it is particularly important since it can be shown that full self-consistency allows to exactly satisfy fundamental symmetries and conservations laws~\cite{ch11_Baym1961,ch11_Baym1962}.
In practical applications, and especially in finite systems, this scheme may not be achievable exactly and self-consistency is implemented only partially for the most important contributions. Normally this is still sufficient to obtain highly accurate results.
 We will present suitable approximation schemes to calculate the self-energy in the following sections. In particular, we will focus on the ADC($n$) method   that can be applied with discretized bases in finite and infinite systems in Secs.~\ref{sec:scgf_adc} and~\ref{sec:scgf_comp}. The case of extended systems at finite temperature is discussed in Sec.~\ref{sec:scgf_finiteT}.   Before going into the actual approximation schemes, we need to see how experimental quantities can be calculated  once the one-body propagator is known, as well as to discuss the basic results of  perturbation theory.

\subsection{Spectral function and relation to experimental observations}
\label{sec:scgf_obs}

\begin{figure}[t]
\begin{center}
\includegraphics[height=0.38\textwidth]{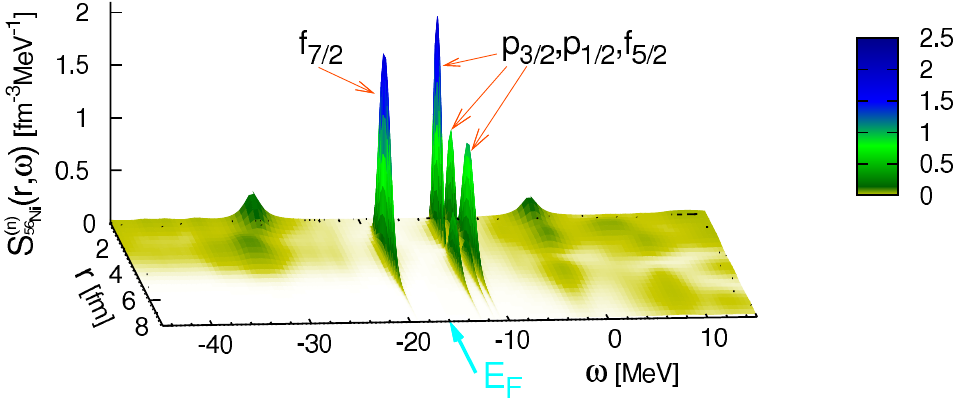}
\caption{
Calculated single-particle spectral function for the addition and removal of a neutron to and form $^{56}$Ni, from Ref.~\cite{ch11_Barbieri2009Ni}. The diagonal part, $S_{r,r}(\omega)$, is
shown in coordinate space. Energies below the Fermi level $E_F$ correspond to the  one-hole spectral function $S^h_{r,r}(\omega)$  which describes
the distribution of nucleons in energy and coordinate space. Integrating over all the quasihole energies yields the matter  density distribution, Eq.~\eqref{eq:rho_r}.
Energies above $E_F$ are for the one-particle spectral function $S^p_{r,r}(\omega)$.}
\label{fig:ScptFnctN56}
\end{center}
\end{figure}

Once the one-body Green's function is known, it can be used to calculate the total binding energy and the expectation values of all one-body observables.
The attractive feature of the SCGF approach is that $g(\omega)$ describes the one-body dynamics completely.  
This information can be recast in the particle and hole spectral functions, which contain the separate responses for the attachment and removal of a nucleon.
They can be obtained directly from Eq.~\eqref{eq:g1Leh}, as follows:
 \begin{align}
S^{p}_{\alpha \beta}(\omega) ={}& -\frac 1 \pi \rm{Im} \; g_{\alpha \beta}(\omega) = \sum_{n} \left(\pazocal{X}^n_\alpha \right)^*  \pazocal {X}^n_\beta ~ \delta\Big(\omega-(E^{A+1}_n - E^A_0)\Big) \; ,  \qquad \rm{for ~ }\omega \geq \varepsilon^+_0 \; ,
\nonumber  \\
 \label{eq:SpSh}
S^{h}_{\alpha \beta}(\omega) ={}& \quad  \frac 1 \pi \rm{Im } \;  g_{\alpha \beta}(\omega) = \sum_{k}  \pazocal{Y}^k_\alpha  (\pazocal{Y}^k_\beta )^* ~\delta\Big(\omega-(E^A_0 - E^{A-1}_k)\Big) \; ,  \qquad \rm{for ~ }\omega \leq \varepsilon^-_0 \;  .
\end{align} 
The diagonal parts of Eqs.~\eqref{eq:SpSh}, have a straightforward physical interpretation~\cite{ch11_FetterWalecka,ch11_Dickhoff2008}. 
The particle part, $S^p_{\alpha \alpha}(\omega)$, is the joint probability of adding a nucleon with quantum numbers $\alpha$ to the A-body ground state, $|\Psi^A_0\rangle$, and then to find the system in a final state with energy $E^{A+1}=E^A_0 + \omega$. Likewise,  $S^h_{\alpha \alpha}(\omega)$ gives the probability of removing a particle from state $\alpha$ while leaving the nucleus in an eigenstate of energy $E^{A-1}=E^A_0 - \omega$.  These are demonstrated in coordinate space in Fig.~\ref{fig:ScptFnctN56} for neutrons around $^{56}$Ni.  Below the Fermi energy, \hbox{$E_F\equiv \frac 1 2 (\varepsilon^+_0 + \varepsilon^-_0)$}, one can see a single dominant quasihole peak corresponding to the $f_{7/2}$ orbit. The states from the $sd$ shell are at lower energies and are instead very fragmented. Just above $E_F$, there are sharp quasiparticles corresponding to the attachment of a neutron to the remaining $pf$ orbits. Finally, for $\omega>0$, one has neutron-$^{56}$Ni elastic scattering states. Remarkably, one can see that dominant quasiparticle peaks persist around the Fermi surface, which confirms the underlying shell structure outside the $^{40}$Ca core for this nucleus.

The existence of isolated dominant peaks  as those shown in Fig.~\ref{fig:ScptFnctN56} indicates that the eigenstates $|\Psi^{A+1}_n\rangle$ and $|\Psi^{A-1}_k\rangle$ are to a very good approximation constructed of a nucleon or a hole independently orbiting the ground state $|\Psi^A_0\rangle$. This is the basic hypothesis at the origin of the nuclear shell-model.
 How much a real nucleus deviates from this assumption can be gauged by the deviations in the values of their {\em spectroscopic factors}. These are defined
 as the normalization overlap of the spectroscopic amplitudes for the attachment or removal of a particle:
\begin{eqnarray}
  SF_n^+=   \sum_\alpha  |{\pazocal X}^n_\alpha|^2 \; ,
 \qquad \qquad && \qquad \qquad 
  SF_k^- =  \sum_\alpha  |{\pazocal Y}^k_\alpha|^2 \; .
\label{eq:SFdef}
\end{eqnarray}
The energy distribution of spectroscopic factors is given by
\begin{eqnarray}
  S(\omega) &=& ~ \sum_\alpha S^p_{\alpha \alpha}(\omega) \quad + \quad \sum_\alpha S^h_{\alpha \alpha}(\omega)  \nonumber \\
          &=& ~ \sum_n  SF_n^+ \, \delta( \omega - E^{A+1}_n + E^A_0) 
          ~+~  \sum_n  SF_k^- \, \delta( \omega - E^A_0 + E^{A-1}_k ) \; ,
\label{eq:SFvsE}
\end{eqnarray}
where each $\delta$-peak corresponds to eigenstates of a neighboring isotope with $A\pm1$ particles. These quasiparticle energies are directly observed in nucleon addition and removal experiments.
Note that the total strength  seen in similar experiments results from a convolution of the spectroscopic amplitudes with the dynamics of the reaction mechanisms. Hence, while the quasiparticle energies appearing in the poles of Eq.~\eqref{eq:g1Leh} are strictly observed, the magnitude of the spectral strength $S(\omega)$ only gives a semi-quantitative description of the strength of the observed cross sections.

Any one-body observable can be calculated via the one-body density matrix $\rho_{\alpha\beta}$, which is obtained from $g_{\alpha\beta}(\omega)$ as follows:
\begin{equation}
 \rho_{\alpha \beta} ~\equiv ~ \langle\Psi_0^A\vert a^{\dag}_{\beta}a_{\alpha} \vert\Psi_0^A\rangle
  ~=~  \int_{-\infty}^{\varepsilon^-_0}S^h_{\alpha\beta}(\omega)~{\rm d}\omega=\sum_k ({\pazocal Y}^k_{\beta})^*{\pazocal Y}^k_{\alpha} .
 \label{eq:OBDM}
\end{equation}
The expectation value of a one-body operator, ${\widehat O}^{1B}$, can then be written in terms of the ${\pazocal Y}$ amplitudes as:
\begin{equation}
\label{eq:den_one}
\langle {\widehat O}^{1B}\rangle  =\sum_{\alpha\beta}  O^{1B}_{\alpha \beta}\,\,\rho_{\beta\alpha}=\sum_k \sum_{\alpha\beta}~ ({\pazocal Y}^k_{\alpha})^*~ O^{1B}_{\alpha\beta} ~  {\pazocal Y}^k_{\beta} \; .
\end{equation}
However, evaluating two- and many-nucleon observables requires the knowledge of many-body propagators. 
Eq.~\eqref{eq:OBDM} also implies that the density profile of the system can be obtained by integrating over the hole  spectral function in coordinate space (cf. Fig.~\ref{fig:ScptFnctN56}):
\begin{equation}
  \label{eq:rho_r}
   \rho({\bf r})  =  \int_{-\infty}^{\varepsilon^-_0}S^h_{{\bf r},{\bf r}}(\omega)~{\rm d}\omega \; .
\end{equation}
Likewise, a second sum (or integration) over the coordinate space yields the total number of particles,
\begin{equation}
  \int {\rm d}\,{\bf r} \,  \int_{-\infty}^{\varepsilon^-_0} S^h_{{\bf r},{\bf r}}(\omega)~{\rm d}\omega \;  ~=~
   \sum_\alpha \int_{-\infty}^{\varepsilon^-_0} \, S^h_{\alpha\alpha}(\omega)~{\rm d}\omega ~=~ A \; .
   \label{eq:part_num}
\end{equation}

A very special case is the Koltun sum-rule that allows calculating the total energy of the system by means of the exact one-body propagator alone, $g(\omega)$~\cite{ch11_Galitskii1958KSR,ch11_Koltun1974KSR}. 
This relation is exact for any Hamiltonian containing at most one- and two body interactions.  When many-particle interactions are present, it is necessary 
to correct for the over countings that arise from these additional terms~\cite{ch11_Carbone2013Nov}. For the specific case in which a three-body interaction $\widehat{W}$ is included, the exact relation for the ground state energy is given by the following modified Koltun rule:
\begin{equation}
  \label{eq:Koltun_hW}
  E^A_0 = \sum_{\alpha\beta} \frac{1}{2} 
            \int_{-\infty}^{\varepsilon^-_0}  [\,T_{\alpha\beta}+\omega\,\delta_{\alpha\beta}\, ]
            \, S^h_{\beta\alpha}(\omega) \; {\rm d}\omega
            ~-~  \frac{1}{2} \langle \widehat W\rangle \, .
\end{equation}
This still relies on the use of a one-body propagator but
 it requires  the additional evaluation of the expectation value of the three-body interaction,~$\langle \widehat W \rangle$ (which in principle requires the knowledge of more complex Green's functions).
Thankfully, in most cases the total strength of $\widehat{W}$ is much smaller than other terms in the Hamiltonian. Thus, one can safely approximate its expectation value at lowest order, in terms of three  correlated density matrices, as
\begin{equation}
   \label{eq:Wddd}
    \langle \widehat W\rangle\simeq\frac{1}{6} \, \sum_{\alpha\beta\mu\gamma\delta\nu} W_{\alpha\beta\mu,\gamma\delta\nu}~\rho_{\gamma\alpha}~\rho_{\delta\beta}~\rho_{\nu\mu} \; .
\end{equation}
As a typical example in finite nuclei, the error from this approximation has been estimated not to exceed 250~keV for the total binding energies for $^{16}$O and $^{24}$O~\cite{ch11_Cipollone2013prl}. However, the accuracy of Eq.~\eqref{eq:Wddd} is not guaranteed and needs to be verified case by case.

\subsection{Perturbation expansion of the Green's function}
\label{sec:pertexp}

In order to understand the following sections and to devise appropriate approximations to the self-energy $\Sigma^{\star}(\omega)$ it is necessary to understand the basic elements of perturbation theory.  These will be also fundamental to derive all-order summation schemes leading to non-perturbative solutions and to discuss the concept of self-consistency. We summarize here the material needed to understand the following sections, while the full set of Feynman rules is reviewed in  Appendix 1.

We work with a system of $A$ non-relativistic fermions interacting by means of two-body and three-body interactions.
We divide the Hamiltonian into two parts, $\widehat H = \widehat H_0 + \widehat H_1$. 
The unperturbed term, $\widehat H_0 = \widehat T + \widehat U$, is given by the sum of the kinetic term and an auxiliary one-body operator~$\widehat U$.
Its choice defines the reference state, $\vert\Phi_0^A\rangle$, and the corresponding
unperturbed propagator $g^{(0)}(\omega)$ that are the starting point for the perturbative expansion\footnote{A typical choice in nuclear physics would be a Slater determinant such as the solution of the Hartree-Fock problem or a set of single-particle harmonic oscillator wave functions.}.  
The perturbative term is then  
$\widehat H_1 = -\widehat U + \widehat V + \widehat W$, where $\widehat V$ denotes the two-body interaction operator and $\widehat W$  is the three-body interaction.
In a second-quantized framework, the full Hamiltonian reads:
\begin{equation}
\label{eq:H}
\widehat H = \sum_{\alpha} \varepsilon^0_\alpha\, a^\dag_\alpha a_\alpha - \sum_{\alpha\beta}U_{\alpha \beta}\, a^\dag_\alpha a_{\beta}
+\frac{1}{4} \sum_{\substack{\alpha\gamma\\\beta\delta}}V_{\alpha\gamma,\beta\delta}\, a_\alpha^\dag a_\gamma^\dag a_{\delta} a_{\beta}
+\frac{1}{36}\sum_{\substack{\alpha\gamma\epsilon \\ \beta\delta\eta}} W_{\alpha\gamma\epsilon,\beta\delta\eta}\,
a_\alpha^\dag a_\gamma^\dag a_\epsilon^\dag a_{\eta} a_{\delta} a_{\beta} \, .
\end{equation}
In Eq.~\eqref{eq:H} we continue to use greek indices $\alpha$,$\beta$,$\gamma$,\ldots to label the single particle basis that defines the model space. But we make the additional assumption that these are the same states which diagonalize the unperturbed Hamiltonian, $\widehat H_0$, with eigenvalues $\varepsilon_\alpha^0$.    This choice is made in most  applications of  perturbation theory but it is not strictly necessary here and it will not affect our discussion in the following sections. 
The matrix elements of the one-body operator $\widehat U$ are given by $U_{\alpha \beta}$. And we  work with
properly antisymmetrized matrix elements of the two-body and three-body forces, $V_{\alpha\gamma,\beta\delta}$ and $W_{\alpha\gamma\epsilon,\beta\delta\eta}$.

In time representation, the many-body Green's functions are defined as the expectation value of  time-ordered products of  annihilation and creation operators  in the Heisenberg picture. This is shown by Eq.~\eqref{eq:g1Time} for the single particle propagator.
Every Green's function can be expanded in a perturbation series in powers of $\widehat{H}_1$.
For the one-body propagator this  reads \cite{ch11_Mattuck1992,ch11_Dickhoff2008}:
\begin{equation}
\label{gpert}
g_{\alpha\beta}(t_\alpha-t_\beta) = (- i) \sum_{n=0}^\infty \left(-  i\right)^n\frac{1}{n!}\int \hspace{-1mm} {\rm d} t_1 \; \ldots \int \hspace{-1mm} {\rm d} t_n \langle\Phi_0^A\vert {\pazocal T} [\widehat H^I_1(t_1) \ldots \widehat H^I_1(t_n)a^I_\alpha(t_\alpha){a_{\beta}^I}^\dag(t_\beta)]\vert\Phi_0^A\rangle_\text{conn} \; ,
\end{equation}
where $\widehat H^I_1(t)$, $a^I_\alpha(t)$ and ${a_{\beta}^I}^\dag(t)$ are now intended as operators in the interaction picture with respect to $H_0$. 
The subscript ``conn'' implies that only \emph{connected} diagrams have to be considered when performing
the Wick contractions of the time-ordered product ${\pazocal T}$.  Each Wick contraction generates an uncorrelated single particle propagator, $g^{(0)}(\omega)$,  which is associated with the system governed by the Hamiltonian~$H_0$.
At order $n=0$, the expansion of Eq.~(\ref{gpert}) simply gives $g^{(0)}(\omega)$.
$H_1$ contains contributions from one-body, two-body and three-body interactions that come from the last three terms on the right hand side of Eq.~\eqref{eq:H}. Thus, for $n\geq1$ the expansion involves terms with individual contributions of each force, or combinations of them, that are linked by uncorrelated propagators.  To each term in the expansion there corresponds a Feynman diagram that gives an intuitive picture of the 
physical process accounted by its contribution. The full set of Feynman diagrammatic rules that stems out of Eq.~(\ref{gpert})
in the presence of three-body interactions is detailed in Appendix 1.

A first reorganization of the contributions generated by Eq.~(\ref{gpert}) is obtained by considering 
\emph{one-particle reducible} diagrams, that is diagrams that can be disconnected by cutting a single fermionic line. 
In general, the reducible diagrams  generated by expansion~(\ref{gpert}) will always have separate structures that are linked together by only one $g^{(0)}(\omega)$ line. These are the same class of diagrams that are created implicitly in the all-orders resummation of the Dyson equation~(\ref{eq:Dyson}). 
 Thus, the irreducible self-energy $\Sigma^\star(\omega)$ is defined as the kernel that collects all the \emph{one-particle irreducible} (1PI) diagrams (with the external legs stripped off).
As already discussed above,  $\Sigma^\star(\omega)$ plays the role of an effective  potential that  is seen by a nucleon inside the system. It splits in static  and frequency dependent terms:
\begin{equation}
  \Sigma^\star_{\alpha \beta}(\omega) ~=~   - U_{\alpha \beta}  ~+~  \Sigma^{(\infty)}_{\alpha \beta}
      ~+~  \widetilde\Sigma_{\alpha \beta}(\omega)  \; ,
\label{eq:SigSplit}
\end{equation}
where we have separated $\widehat{U}$ since this is auxiliary defined and it eventually cancels out when solving the Dyson equation. The term $\Sigma^{(\infty)}$ plays the role of the static mean-field that a nucleon feels due to the average  interactions will all
other particles in the system.
The frequency-dependent part, $\widetilde\Sigma(\omega)$, describes the effects of dynamical excitations of the many-body state that are induced by the nucleon itself. In general, this means the  propagation of (complex) intermediate excitations and therefore it must have  
a Lehmann representation analogous to that of Eq.~\eqref{eq:g1Leh}.
For very large energies ($\omega \rightarrow \pm\infty$) the poles of such Lehmann representation become vanishingly  small and one is left with just $\Sigma^{(\infty)}$ and the auxiliary potential~$\widehat{U}$.

A further level of simplification in the self-energy expansion 
can be obtained if unperturbed propagators, $g^{(0)}(\omega)$, in the internal fermionic lines are replaced by dressed Green's functions, $g(\omega)$. 
This choice further restricts the set of diagrams to the so-called \emph{skeleton} diagrams \cite{ch11_Dickhoff2008}, which are
defined as 1PI diagrams that do not contain  any portion that can be disconnected by cutting any two fermion lines at different points. 
These portions would correspond to self-energy insertions, which are already re-summed into the dressed propagator $g(\omega)$ by Eq.~(\ref{eq:Dyson}).
The SCGF approach is precisely based on expressing the irreducible self-energy in terms of such skeleton diagrams 
with dressed propagators.
The SCGF framework offers great advantages. First, it is intrinsically non-perturbative and completely
independent from any choice of the reference state and auxiliary one-body potential. This is so because $\Sigma^\star(\omega)$ no longer depends on $g^{(0)}(\omega)$ and $\widehat{U}$ always drops out of the Dyson equation (see Eq.~\eqref{eq:DysSchrod} below).
Second, many-body correlations are expanded directly in terms of single particle excitations of the true propagator, which are generally closer to the exact solution than those associated with the unperturbed state, $\vert\Phi_0^A\rangle$. 
Third, given an appropriate  truncation of self-energy, if a full SCGF calculation is possible then it automatically satisfies the basic conservation laws of particle number, angular momentum, etc...~\cite{ch11_Baym1961,ch11_Baym1962,ch11_Dickhoff2008}. 
Finally, the number of diagrams to be considered is vastly reduced to 1PI skeletons one. However, this is not always a simplification since a dressed propagator contains a very large number of poles, which can be much more difficult to deal with than for  the corresponding uncorrelated~$g^{(0)}(\omega)$.

\begin{figure}[t]
\includegraphics[width=0.95\textwidth]{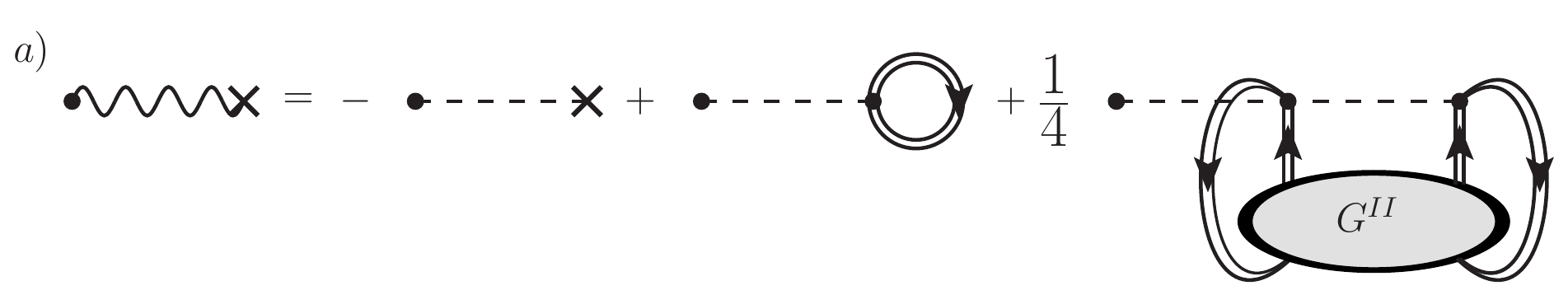}   
\vskip  0.4cm
\includegraphics[width=0.71\textwidth]{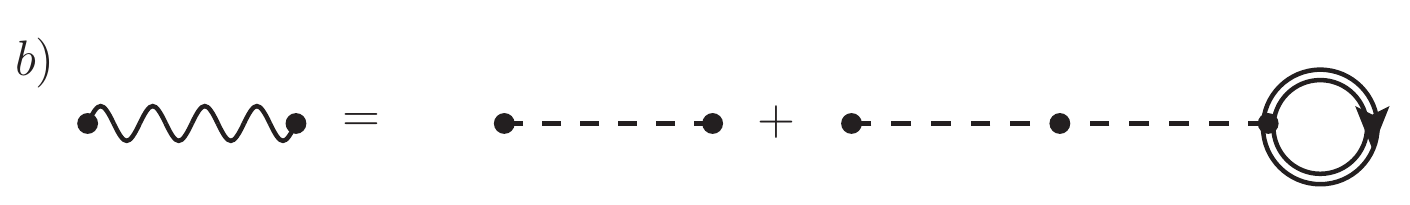}  
\caption{Graphical representation of the effective one-body interaction of Eq.~\eqref{eq:U_eff}, top row, and the effective two-body interaction~\eqref{eq:V_eff}, bottom row. Dashed lines represent the
one-, two,- and three-body interactions entering Eq.~\eqref{eq:H} and wavy lines are the effective operators $\widetilde U$ and  $\widetilde V$.}
\label{fig:EffOps}
\end{figure}

If three- or many-body forces are included in the Hamiltonian, the number of Feynman diagrams that need to be considered at a given order increases very rapidly. In this case it becomes very useful and instructive to restrict the attention to an even smaller class of diagrams that are  {\em interaction-irreducible}~\cite{ch11_Carbone2013Nov}. An interaction vertex is said to be reducible if the whole diagram can be disconnected in two parts by cutting the vertex itself. In general, this happens for an $m$-body interaction when there is a smaller number of $n$ lines ($n < m$) that leave the interaction, may interact only among themselves, and eventually all return to it. The net outcome is that one is left with a ($m-n$)-body operator that results from the average interactions with other $n$-spectator nucleons. This plays the role of a system dependent effective force that is irreducible. Fig.~\ref{fig:EffOps} shows diagrammatically how  $\widehat{V}$ and $\widehat{W}$  can be reduced to  one- and two-body {\em effective interactions} in this way.
Hence, for a system with up to three-body forces, we define an effective Hamiltonian
\begin{equation}
\widetilde H_1= {\widetilde U} + {\widetilde  V} + \widehat W \; ,
\label{eq:Heff}
\end{equation}
where $\widetilde U$ and  $\widetilde V$ are the effective interaction operators. 
The diagrammatic expansion arising from  Eq.~(\ref{gpert}) with the effective Hamiltonian $\widetilde H_1$ is
formed only of (1PI, skeleton) interaction-irreducible diagrams. Note that the three-body interaction, $\widehat W$, remains the same as in Eq.~\eqref{eq:H} but enters only diagrams as an  interaction-irreducible three-body force.  
%
The explicit expressions for the one-body and two-body effective interaction operators can be obtained form the Feynman diagrams of Fig.~\ref{fig:EffOps} and they are given by:  
\begin{subequations}
\label{eq:UV_eff}
\begin{align}
  \label{eq:U_eff}
  \widetilde{U}_{\alpha\beta} ={}& -U_{\alpha \beta}  ~ ~ +  ~\sum_{\delta\gamma} V_{\alpha\gamma,\beta\delta}~\rho_{\delta\gamma}
                        ~ +  ~ \frac{1}{4} \sum_{\mu \nu \gamma \delta}W_{\alpha\mu\nu,\beta\gamma\delta}
                   ~\Gamma_{\gamma \delta, \mu \nu}  \; ,
  \\\label{eq:V_eff} 
  \widetilde{V}_{\alpha\beta,\gamma\delta} ={}& \quad V_{\alpha\beta,\gamma\delta}  ~+ ~ 
                          \sum_{\mu\nu}W_{\alpha\beta \mu ,\gamma \delta \nu}~\rho_{\nu\mu}  \; .
\end{align}
\end{subequations}
where we used the reduced two-body density matrix $\Gamma$, which can be computed from the exact  two-body Green's function:
\begin{equation}
 \Gamma_{\gamma \delta, \mu \nu} = \lim_{\tau \rightarrow 0^-}  -i  \, G^{II}_{\gamma \delta, \mu \nu}(\tau) = 
     \langle\Psi^A_0\vert  a^\dagger_\nu a^\dagger_\mu  a_\gamma  a_\delta  \vert\Psi^A_0\rangle \; .
   \label{eq:2BDM}
\end{equation}

The effective Hamiltonian of Eq.~(\ref{eq:Heff})  not only regroups Feynman diagrams in 
a more efficient way, but also defines the effective one-body and two-body terms from 
higher order interactions. As long as interaction-irreducible diagrams are used together with the 
effective Hamiltonian, $\widetilde{H}_1$, this approach provides a systematic
way to incorporate many-body forces in the calculations and to 
generate effective in-medium interactions. More importantly, the formalism is such that all
symmetry factors are guaranteed to be correct and no diagram is over-counted~\cite{ch11_Carbone2013Nov}.
 Eqs.~\eqref{eq:UV_eff} can be seen as a generalization of the normal
ordering of the Hamiltonian with respect to the reference state $\vert\Phi_0^A\rangle$ discussed in chapter~8. However, these contractions go beyond normal ordering
because they are performed with respect to the exact
correlated density matrices. To some extent, one can intuitively think of 
the effective Hamiltonian $\widetilde{H}_1$  as being ordered with respect
to the interacting many-body ground-state $|\Psi_0^A\rangle$, rather than 
the non-interacting  $\vert\Phi_0^A\rangle$.

\begin{figure}[t]
\begin{center}
\includegraphics[height=0.28\textwidth]{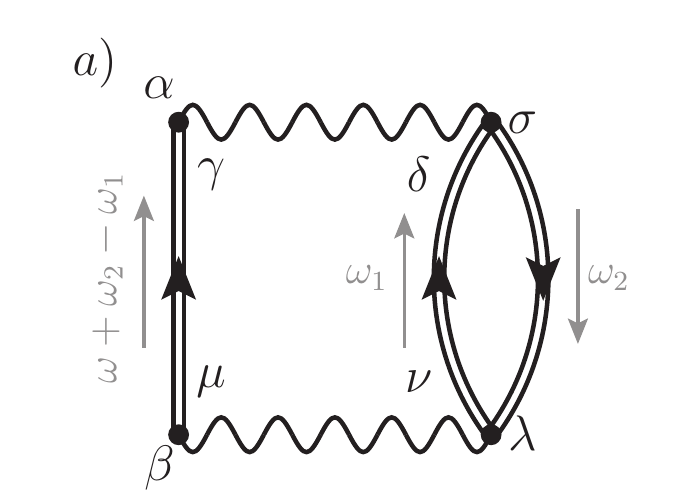}   \hspace{0.06\textwidth} 
\includegraphics[height=0.28\textwidth]{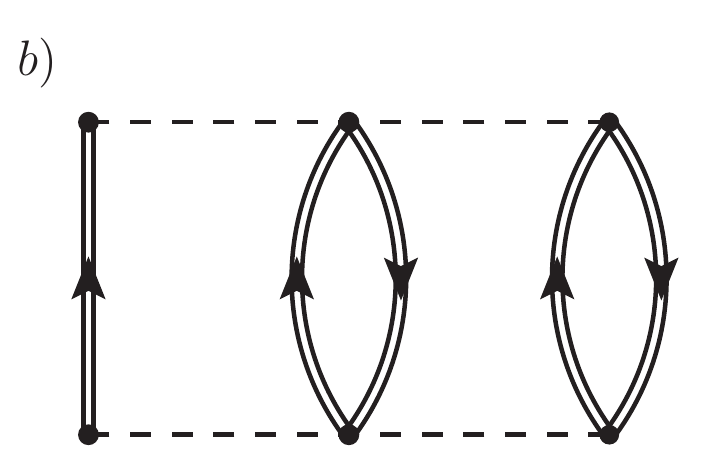}
\caption{Second-order interaction-irreducible contributions to the self-energy arising from both two- and three-nucleon forces. The diagram depending on the effective two-body interactions (left) also shows the indices and labels that are used for calculating its contribution in {\bf Example 11.2}. }
\label{fig:2ndOrd}
\end{center}
\end{figure}

Since the static self-energy does not propagate any intermediate excitations, it can only receive contribution 
when the incoming and outgoing lines of a Feynman diagram are attached to the same interaction vertex.
Thus, by definition, $\Sigma^{(\infty)}$ must include  the one body term in $\widehat{H}_1$ plus any higher order 
interaction that are reduced to effective one-body interactions, hence:
\begin{equation}
   \widetilde{U}_{\alpha\beta} ~=~ - U_{\alpha \beta}  ~+~  \Sigma^{(\infty)}_{\alpha \beta} \; ,
   \label{eq:UeffSig}
\end{equation}
which defines $\Sigma^{(\infty)}$ by comparison with Eq.~\eqref{eq:U_eff}.
The two terms that contribute to $\Sigma^{(\infty)}$ represent extensions of the Hartree-Fock (HF) potentials to correlated 
ground states.  The correlated Hartree-Fock potential from ${\widehat V}$ is the only effective operator
when just two-body forces are present. In this case there is very little gain in using the concept of the effective Hamiltonian~\eqref{eq:Heff}.
However, with three-body interactions, additional effective interaction terms appear in both $\widetilde U$ and $\widetilde V$. 
From Eq.~\eqref{eq:UeffSig} we see that the perturbative SCGF expansion of the $\widetilde{H}_1$ Hamiltonian has only 
one (1PI, skeleton and interaction-irreducible)  term at first order.  The first contributions to $\widetilde\Sigma(\omega)$  appear at second order with the two diagrams in Fig.~\ref{fig:2ndOrd}. Expanding with respect to $\widehat{H}_1$, there would have been five diagrams instead of only the two interaction-irreducible ones shown in  Fig.~\ref{fig:2ndOrd}. These diagrams indeed have a proper Lehmann representation (see Example 11.2 and Exercise 11.2) and propagate {\em intermediate state configurations} (ISCs) of type 2-particle 1-hole (2p1h), 2h1p, 3p2h, etc...
At third order, $\widetilde{H}_1$ generates 17 SCGF diagrams two of which contain only  two-body interactions. The simplest of these, that involve at most 2p1h and 2h1p ISCs, are shown in Fig.~\ref{fig:3rdOrd}.
All interaction-irreducible contributions to the proper self-energy up to third order in perturbation theory are discussed in details in Ref.~\cite{ch11_Carbone2013Nov}. 

\vskip .3 cm
\noindent
{\bf Example 11.1.} Calculate the Feynman-Galitskii propagator, $G^{II,f}(\tau)$, that corresponds to the propagation of two particles or two holes that do not interact with each other.

\vskip 0.2 cm
This is the lowest order approximation to the two-times and two-body propagator
which evolves two particle from states $\alpha$ and $\beta$ to  states  $\gamma$ and  $\delta$ after a time $\tau >0$, or two holes  from $\gamma$ and  $\delta$ to $\alpha$ and $\beta$ when $\tau < 0$.  By applying the perturbative expansion equivalent to Eq.~\eqref{gpert} at order $n=0$, we find:
\begin{eqnarray}
 G^{II\, (0)}_{\alpha \beta, \gamma \delta}(\tau) &=& -i \langle\Phi_0^A\vert {\pazocal T} [
       \, a^I_\beta(\tau)  \, a^I_\alpha(\tau) \, {a_{\gamma}^I}^\dag(0) \, {a_{\delta}^I}^\dag(0) \, ]  \vert\Phi_0^A\rangle
\nonumber \\
 &=& i g^{(0)}_{\alpha\gamma}(\tau) \; \, g^{(0)}_{\beta\delta}(\tau)  ~-~ i  g^{(0)}_{\alpha\delta}(\tau) \; \, g^{(0)}_{\beta\gamma}(\tau)
 ~ \equiv ~  G^{II\, (0),f}_{\alpha \beta, \gamma \delta}(\tau)   - G^{II\, (0),f}_{\alpha \beta, \delta\gamma}(\tau) \; .
\label{eq:FeynGalvsG2}
\end{eqnarray}
The Feynman-Galitskii propagator is precisely defined as the non antisymmetrized part of Eq.~\eqref{eq:FeynGalvsG2}.  We  now transform this to frequency space and apply the Feynman rules of Appendix~1 
to calculate the  $G^{II,f}$ for the more general case of two {\em dressed} propagator lines:
\begin{eqnarray}
G^{II,f}_{\alpha \beta, \gamma \delta}(\omega) &=& \int {\rm d}\tau \; e^{i\omega \tau} \, G^{II,f}_{\alpha \beta, \gamma \delta}(\tau)
~=~ (-i) \int \frac {{\rm d} \omega_1}{2\pi} 
  \; i \,  g_{\alpha\gamma}(\omega-\omega_1) \; i \, g_{\beta\delta}(\omega_1)
\label{eq:GIIf_int} \\ \nonumber
&=& - \int \frac {{\rm d}  \omega_1}{2\pi i}  
  \left\{ \frac{(\pazocal{X}^{n_1}_\alpha)^*  \pazocal{X}^{n_1}_\gamma}{\omega - \omega_1  - \varepsilon^+_{n_1} + i\eta} 
        + \frac{\pazocal{Y}^{k_1}_\alpha  (\pazocal{Y}^{k_1}_\gamma)^*}{\omega - \omega_1  - \varepsilon^-_{k_1} - i\eta}  \right\}
          \left\{ \frac{(\pazocal{X}^{n_2}_\beta)^*  \pazocal{X}^{n_2}_\delta}{\omega_1  - \varepsilon^+_{n_2} + i\eta} 
        + \frac{\pazocal{Y}^{k_2}_\beta  (\pazocal{Y}^{k_2}_\delta)^*}{\omega_1  - \varepsilon^-_{k_2} - i\eta}  \right\} \, ,
\end{eqnarray}
where we have  used the convention that repeated indices are summed over. The integral in the above equation can be performed with the Cauchy theorem by closing an arch on either the positive or the negative imaginary half planes. Hence, contributions where all the poles are  on the same side of the real axis cancel out. Extracting the residues of the other contributions leads to the following result:
\begin{equation}
G^{II,f}_{\alpha \beta, \gamma \delta}(\omega) =
\sum_{n_1, \, n_2} \frac{(\pazocal{X}^{n_1}_\alpha \pazocal{X}^{n_2}_\beta)^*  \pazocal{X}^{n_1}_\gamma \pazocal{X}^{n_2}_\delta}
                      {\omega  - (\varepsilon^+_{n_1}  + \varepsilon^+_{n_2}) + i\eta} 
~-~ \sum_{k_1, \, k_2} \frac{\pazocal{Y}^{k_1}_\alpha \pazocal{Y}^{k_2}_\beta \, (\pazocal{Y}^{k_1}_\gamma \pazocal{Y}^{k_2}_\delta)^*}
                     {\omega  - (\varepsilon^-_{k_1} + \varepsilon^-_{k_2}) - i\eta}   \; .
\label{eq:GIIf}
\end{equation}

\vskip 0.3 cm
\noindent
{\bf Exercise 11.1.} Calculate the contribution of the three-body force $\widehat W$ to the  effective one body potential, in the approximation of 
two dressed but non interacting spectator nucleons.

\vskip 0.2 cm
\noindent
{\bf Solution.}  This is the last term in Fig.~\ref{fig:EffOps}a) and Eq.~\eqref{eq:U_eff} but with $G^{II}(\tau)$ approximated by two independent fermion lines, as for the dressed Feynman-Galitskii propagator. Using Eq.~\eqref{eq:2BDM} and re-expressing the second line of~\eqref{eq:FeynGalvsG2} in terms of $g(\tau)$, we arrive at:
\begin{equation}
\label{eq:ueff_3b_first}
 \widetilde{U}^{(W)}_{\alpha\beta} ~=~ \frac{1}{2} \sum_{\mu \nu \gamma \delta}W_{\alpha\mu\nu,\beta\gamma\delta}
               ~\rho_{\gamma\mu}~\rho_{\delta\nu}  \; .
\end{equation}

\vskip 0.3 cm
\noindent
{\bf Example 11.2.} Calculate the expression for the second-order contribution to $\Sigma^\star(\omega)$ from two-nucleon interactions only.

\vskip 0.2 cm
This is the  diagram of Fig.~\ref{fig:2ndOrd}a). By applying the Feynman rules of Appendix 1 we have:
\begin{eqnarray}
 \Sigma^{(2,2N)}_{\alpha\beta}(\omega) &=&
  - \frac{(i)^2}{2} \int \frac {{\rm d} \omega_1}{2\pi} \frac {d \omega_2}{2\pi}  V_{\alpha\sigma,\gamma\delta} 
         \,  g_{\gamma\mu}(\omega+\omega_2-\omega_1) \, g_{\delta\nu}(\omega_1) \, g_{\lambda\sigma}(\omega_2) 
           \, V_{\mu \nu, \beta \lambda}
\nonumber \\
   &=& + \frac{1}{2} \int \frac {{\rm d} \omega_2}{2\pi i} V_{\alpha\sigma,\gamma\delta} \; 
     G^{II,f}_{\gamma \delta, \mu \nu}(\omega+\omega_2)   \, g_{\lambda\sigma}(\omega_2) 
           \, V_{\mu \nu, \beta \lambda}
\nonumber \\
  &=& \frac{1}{2}   \int \frac {{\rm d} \omega_2}{2\pi i} \, V_{\alpha\sigma,\gamma\delta} \,
  \left\{
    \frac{(\pazocal{X}^{n_1}_\gamma \pazocal{X}^{n_2}_\delta)^*  \pazocal{X}^{n_1}_\mu \pazocal{X}^{n_2}_\nu}
                      {\omega  + \omega_2  - (\varepsilon^+_{n_1}  + \varepsilon^+_{n_2}) + i\eta} 
 -  \frac{ \pazocal{Y}^{k_1}_\gamma \pazocal{Y}^{k_2}_\delta \, (\pazocal{Y}^{k_1}_\mu \pazocal{Y}^{k_2}_\nu)^*}
                     {\omega  + \omega_2  - (\varepsilon^-_{k_1} + \varepsilon^-_{k_2}) - i\eta}
  \right\}
\nonumber \\
&& \qquad \qquad \quad \times  \left\{ \frac{(\pazocal{X}^{n_3}_\lambda )^* \pazocal{X}^{n_3}_\sigma}{\omega_2  - \varepsilon^+_{n_3} + i\eta} 
        + \frac{\pazocal{Y}^{k_3}_\lambda  (\pazocal{Y}^{k_3}_\sigma)^*}{\omega_2  - \varepsilon^-_{k_3} - i\eta}  \right\} 
          \, V_{\mu \nu, \beta \lambda}
 \; ,
\end{eqnarray}
where we have used the two-body interaction $\widehat V$, but it could have been equally calculated with the effective interaction $\widetilde V$. 
Note that the  integration over $\omega_1$ is exactly the same as in Eq.~\eqref{eq:GIIf_int}. Thus, we can directly substitute the expression for the Feynman-Galitskii propagator~\eqref{eq:GIIf}  in the last two lines above. By performing the last Cauchy integral we find that only two out of four possible terms survive. The final result for the second-order irreducible self-energy is: 
\begin{equation}
\Sigma^{(2,2N)}_{\alpha\beta}(\omega) = \frac{1}{2}  V_{\alpha\sigma,\gamma\delta} \left\{
   \sum_{\substack{n_1, \, n_2 \\  k_3}} \frac{(\pazocal{X}^{n_1}_\gamma   \pazocal{X}^{n_2}_\delta    \pazocal{Y}^{k_3}_\sigma)^*
                                                      \pazocal{X}^{n_1}_\mu \pazocal{X}^{n_2}_\nu  \pazocal{Y}^{k_3}_\lambda}
                      {\omega  - (\varepsilon^+_{n_1}  + \varepsilon^+_{n_2}- \varepsilon^-_{k_3}) + i\eta} 
+ \sum_{\substack{k_1, \, k_2 \\ n_3}} \frac{\pazocal{Y}^{k_1}_\gamma     \pazocal{Y}^{k_2}_\delta \pazocal{X}^{n_3}_\sigma \, 
                                                       (\pazocal{Y}^{k_1}_\mu \pazocal{Y}^{k_2}_\nu \pazocal{X}^{n_3}_\lambda )^*}
                     {\omega  - (\varepsilon^-_{k_1} + \varepsilon^-_{k_2}- \varepsilon^+_{n_3}) - i\eta}  
       \right\}  V_{\mu \nu, \beta \lambda} \; ,
\label{eq:Sig_2nd}
\end{equation}
where repeated greek indices are summed over implicitly but we show the explicit summation over the poles corresponding to 2p1h and 2h1p ISCs.

\vskip 0.3 cm
\noindent
{\bf Exercise 11.2.} Calculate the expression for the  other second-order contribution to $\Sigma^\star(\omega)$ arising from three-nucleon interactions (diagram of Fig.~\ref{fig:2ndOrd}b). Show that this contains ISCs of 3p2h and 3h2p.

\vskip 0.2 cm
\noindent
{\bf Solution.} Upon performing the four frequency integrals, one obtains:
\begin{eqnarray}
\Sigma^{(2,3N)}_{\alpha\beta}(\omega) &=& \frac{1}{12}  W_{\alpha\gamma\delta, \mu\nu\lambda} \left\{
   \sum_{\substack{n_1, \, n_2 ,\, n_3 \\  k_4,\, k_5}} 
       \frac{(\pazocal{X}^{n_1}_\mu   \pazocal{X}^{n_2}_\nu   \pazocal{X}^{n_3}_\lambda    \pazocal{Y}^{k_4}_\gamma  \pazocal{Y}^{k_5}_\delta)^*
                 \pazocal{X}^{n_1}_{\mu'} \pazocal{X}^{n_2}_{\nu'}  \pazocal{X}^{n_3}_{\lambda'}    \pazocal{Y}^{k_4}_{\gamma'} \pazocal{Y}^{k_5}_{\delta'}}
                      {\omega  - (\varepsilon^+_{n_1}  + \varepsilon^+_{n_2} + \varepsilon^+_{n_3}- \varepsilon^-_{k_4}- \varepsilon^-_{k_5}) + i\eta} 
 \right. \nonumber \\
 &&\qquad \qquad \quad  \left. +
\sum_{\substack{k_1, \, k_2 ,\, k_3 \\ n_4 ,\, n_5}} 
         \frac{\pazocal{Y}^{k_1}_\mu     \pazocal{Y}^{k_2}_\nu    \pazocal{Y}^{k_3}_\lambda \pazocal{X}^{n_4}_\gamma \pazocal{X}^{n_5}_\delta \, 
               (\pazocal{Y}^{k_1}_{\mu'} \pazocal{Y}^{k_2}_{\nu'}     \pazocal{Y}^{k_3}_{\lambda'} \pazocal{X}^{n_4}_{\gamma'} \pazocal{X}^{n_5}_{\delta'} )^*}
                     {\omega  - (\varepsilon^-_{k_1} + \varepsilon^-_{k_2}+ \varepsilon^-_{k_3}- \varepsilon^+_{n_4}- \varepsilon^+_{n_5}) - i\eta}  
       \right\}  W_{\mu' \nu' \lambda' , \beta \gamma' \delta'} \, . \qquad \qquad
\label{eq:Sig_2nd_3b}
\end{eqnarray}

\section{The Algebraic Diagrammatic Construction method}
\label{sec:scgf_adc}

The most general form of the irreducible self-energy is given by Eq.~\eqref{eq:SigSplit}. The $\Sigma^{(\infty)}$  is defined by the mean-field diagrams of Fig.~\ref{fig:EffOps}a) and Eq.~\eqref{eq:U_eff}, while $\widetilde\Sigma(\omega)$ has a Lehmann representation as seen in the examples of Eqs.~\eqref{eq:Sig_2nd} and~\eqref{eq:Sig_2nd_3b}. Similarly to the case of a propagator, the  pole structure of the energy-dependent part is dictated by the principle of causality with the correct boundary conditions coded by the  $\pm i\eta$ terms in the denominators.  This implies a dispersion relation that  can link the real and imaginary parts of the self-energy~\cite{ch11_MahauxSartor91,ch11_Dickhoff2008}.  Correspondingly, the direct coupling of single particle orbits to  ISCs (of 2p1h and 2h1p character or more complex) imposes  the separable structure of the residues. In this section we consider the case of a  finite system, for which it is useful to use a discretized single particle basis $\{\alpha\}$ as the model space. 
From now on we will use the Einstein convention that repeated indices ($n$, $k$, $\alpha$...) are summed over even if not explicitly stated.
Thus, the above constraints impose the following  analytical form the self-energy operator:
\begin{equation}
\Sigma^{(\star)}_{\alpha\beta}(\omega) =  - U_{\alpha \beta}  ~+~ \Sigma^{(\infty)}_{\alpha\beta} ~+~ M^\dagger_{\alpha, r} \left[\frac1{\omega - (E^> +C) + i  \eta}  \right]_{r,r'} M_{r', \beta} 
       ~+~ N_{\alpha, s} \left[\frac1{\omega - (E^< +D) - i \eta} \right]_{s,s'} N^\dagger_{s', \beta}  \; ,
\label{eq:ADC_SE_form}
\end{equation}
where, here and in the following, $\omega$ and $\pm i\eta$ are to be intended as multiplication operators (that is, with matrix elements $[\omega+i\eta]_{s,s'}=(\omega+i\eta)\delta_{s,s'}$) and the fraction means a matrix inversion.
In Eq.~\eqref{eq:ADC_SE_form}, the $E^>$ and $E^<$ are the unperturbed energies for the forward and backward ISCs and $r$ and $s$ are collective indices that label sets of configurations beyond single particle structure. Specifically, $r$ is for particle addition and will label 2p1h, 3p2h, 4p3h, ... states, in the general case. Likewise, $s$ is for particle removal and we will use it to label 2h1p states (or higher configurations). However, for the approximations presented in this chapter and for our discussion below we will only be limited to 2p1h and 2h1p ISCs.

The expansion of the self-energy at second order in perturbation theory trivially satisfies Eq.~\eqref{eq:ADC_SE_form}. In the results  of Eq.~\eqref{eq:Sig_2nd}, the sums over $r$ and $s$ can be taken to run over ordered configurations $r\equiv\{n_1 < n_2,k_3\}$ and  $s\equiv\{k_1 < k_2, n_3\}$. Because of the Pauli principle, the half residues of each pole are antisymmetric with respect to exchanging two quasiparticle or two quasihole indices. Therefore the constraints  $n_1 < n_2$ and  $k_1 < k_2$  can be imposed to avoid counting the same configurations twice. Thus, we can identify the expressions for the residues and poles as follows:
\begin{subequations}
 \label{eq:ADC2_MEC}
\begin{align}
 M_{r , \alpha} ={}&  \pazocal{X}^{n_1}_\mu \pazocal{X}^{n_2}_\nu \pazocal{Y}^{k_3}_\lambda \, V_{\mu \nu, \alpha \lambda}
  \label{eq:ADC2_M}  \\
 E^>_{r,r'} ={}& \mathrm{diag} \left( \,\varepsilon^+_{n_1} + \varepsilon^+_{n_2} - \varepsilon^-_{k_3}  \, \right)
 \label{eq:ADC2_Efw}  \\
 C_{r,r'} ={}& 0
 \label{eq:ADC2_C}
 \end{align}
\end{subequations}
and 
\begin{subequations}
\label{eq:ADC2_NED}
\begin{align}
 N_{\alpha,s} ={}& V_{\alpha \lambda, \mu \nu} \, \pazocal{Y}^{k_1}_\mu \pazocal{Y}^{k_2}_\nu \pazocal{X}^{n_3}_\lambda  
   \label{eq:ADC2_N}  \\
 E^<_{s,s'} ={}& \mathrm{diag} \left( \, \varepsilon^-_{k_1} + \varepsilon^-_{k_2} - \varepsilon^+_{n_3} \, \right)
   \label{eq:ADC2_Ebk}  \\
 D_{s,s'}  ={}& 0  \; ,
  \label{eq:ADC2_D}
\end{align}
\end{subequations}
where the factor $1/2$ from Eq.~\eqref{eq:Sig_2nd} disappears because we restricted the sums to triplets of indices where  $n_1<n_2$ and $k_1<k_2$.
As we will discuss in the next section, Eqs.~\eqref{eq:ADC2_MEC} and \eqref{eq:ADC2_NED} define the algebraic diagrammatic method at second order [ADC(2)].

\begin{figure}[t]
\begin{center}
\includegraphics[height=0.18\textheight]{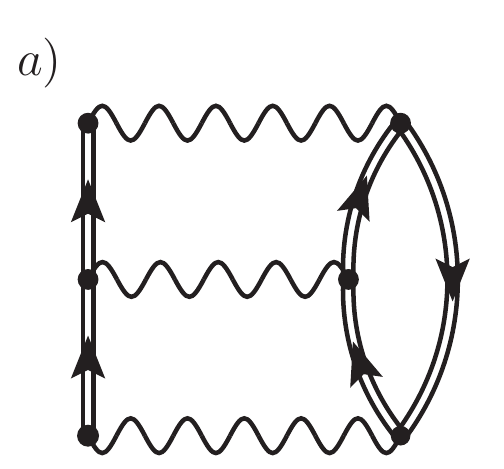}   \hspace{0.03\textwidth} 
\includegraphics[height=0.18\textheight]{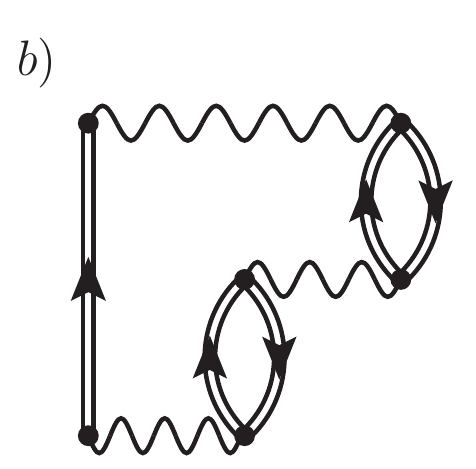}   \hspace{0.03\textwidth} 
\includegraphics[height=0.18\textheight]{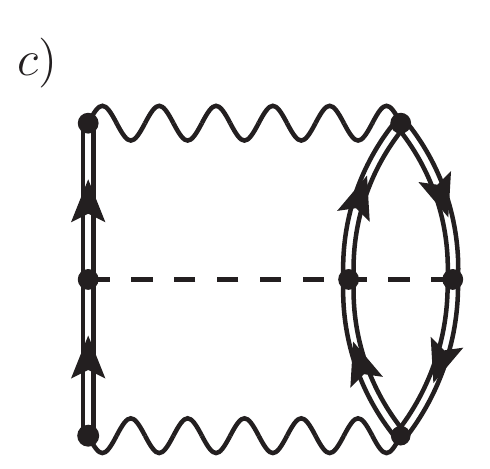}  
\caption{The three simplest skeleton and interaction-irreducible diagrams contributing to the self-energy at third order.  All these terms involve intermediate state configurations of at most 2p1h and 2h1p. The first two contain only two-nucleon interactions and are the first terms in the  resummation of ladders~[diagram~a)] and rings [diagram b)]. The diagram c) is the first contribution containing an irreducible three-nucleon interaction. All the remaining 14 diagrams at third order require explicit three-body interactions and ISCs with 3p2h and 3h2p excitations~\cite{ch11_Carbone2013Nov,ch11_Raimondi_inprep}.  }
\label{fig:3rdOrd}
\end{center}
\end{figure}

Unfortunately, $\Sigma^\star(\omega)$ loses its analytical form of Eq.~\eqref{eq:ADC_SE_form} as soon as one moves to higher orders in perturbation theory. To demonstrate this, let us calculate the contribution of the third-order `ladder' diagram of Fig.~\ref{fig:3rdOrd}a). By exploiting the Feynman rules and Eq.~\eqref{eq:GIIf_int} we obtain:
\begin{align}
  \Sigma^{(3,ld)}_{\alpha\beta}(\omega)={}& - \frac{i^3}{4} \int \frac{{\rm d}\omega_1}{2\pi} \int \frac{{\rm d}\omega_2}{2\pi} \int \frac{{\rm d}\omega_3}{2\pi} 
  V_{\alpha\sigma,\gamma\delta} 
         \,  g_{\gamma\gamma'}(\omega+\omega_3-\omega_1) \, g_{\delta\delta'}(\omega_1) 
           \, V_{\gamma' \delta', \mu' \nu'}
  \nonumber \\ {}& \qquad \qquad \qquad \qquad \qquad \qquad \qquad \qquad \times
    \,  g_{\mu'\mu}(\omega+\omega_3-\omega_2) \, g_{\nu'\nu}(\omega_2) \, V_{\mu \nu, \beta \lambda} \,  g_{\lambda\sigma}(\omega_3) 
\nonumber \\
  ={}&  \frac{1}{4}  \int \frac{{\rm d}\omega_3}{2\pi i} 
  V_{\alpha\sigma,\gamma\delta} 
         \,  G^{II,f}_{\gamma\delta, \gamma' \delta'}(\omega+\omega_3) \,  
           \, V_{\gamma' \delta', \mu' \nu'}     \,  G^{II,f}_{\mu' \nu', \mu \nu}(\omega+\omega_3) \, V_{\mu \nu, \beta \lambda}  \, g_{\lambda\sigma}(\omega_3) 
\nonumber \\
  ={}&  \frac{1}{4}  \int \frac{{\rm d}\omega_3}{2\pi i} 
    V_{\alpha\sigma,\gamma\delta} \,
  \left\{
    \frac{(\pazocal{X}^{n_1}_\gamma \pazocal{X}^{n_2}_\delta)^*  \pazocal{X}^{n_1}_{\gamma'} \pazocal{X}^{n_2}_{\delta'}}
                      {\omega+\omega_3  - (\varepsilon^+_{n_1}  + \varepsilon^+_{n_2}) + i\eta} 
 -  \frac{ \pazocal{Y}^{k_1}_\gamma \pazocal{Y}^{k_2}_\delta \, ( \pazocal{Y}^{k_1}_{\gamma'} \pazocal{Y}^{k_2}_{\delta'})^*}
                     {\omega+\omega_3  - (\varepsilon^-_{k_1} + \varepsilon^-_{k_2}) - i\eta}
  \right\}
\nonumber \\
  & \qquad \qquad \times    \, V_{\gamma' \delta', \mu' \nu'} \,  \left\{
    \frac{( \pazocal{X}^{n_4}_{\mu'} \pazocal{X}^{n_5}_{\nu'})^*  \pazocal{X}^{n_4}_\mu \pazocal{X}^{n_5}_\nu}
                      {\omega+\omega_3  - (\varepsilon^+_{n_4}  + \varepsilon^+_{n_5}) + i\eta} 
 -  \frac{ \pazocal{Y}^{k_4}_{\mu'} \pazocal{Y}^{k_5}_{\nu'} \, (\pazocal{Y}^{k_4}_\mu \pazocal{Y}^{k_5}_\nu)^*}
                     {\omega+\omega_3  - (\varepsilon^-_{k_4} + \varepsilon^-_{k_5}) - i\eta}
  \right\}
 \nonumber \\
  & \qquad \qquad \qquad \qquad \times   V_{\mu \nu, \beta \lambda}
   \left\{ \frac{(\pazocal{X}^{n_3}_\lambda )^* \pazocal{X}^{n_3}_\sigma}{\omega_3  - \varepsilon^+_{n_3} + i\eta} 
           + \frac{\pazocal{Y}^{k_3}_\lambda  (\pazocal{Y}^{k_3}_\sigma)^*}{\omega_3  - \varepsilon^-_{k_3} - i\eta}  \right\}   \; .
\label{eq:LaddEg1}
 \end{align}
Performing the Cauchy integrals, only six terms out of the eight combinations of poles survive. To simplify the discussion 
we will focus on the three integrals that contribute to the forward propagation of the self-energy (third term on the r.h.s.
 of \eqref{eq:ADC_SE_form}). This is done by retaining only the poles $(\omega_3  - \varepsilon^-_{k_3} - i\eta)^{-1}$ in the
 last propagator of Eq.~\eqref{eq:LaddEg1}, which lie above the real axis with respect to the integrand $\omega_3$. Thus,
 we have:
 \begin{align}
  \Sigma^{(ld,>)}_{\alpha\beta}(\omega) ={}& 
 \frac{1}{4}  \int \frac{{\rm d}\omega_3}{2\pi i} 
    V_{\alpha\sigma,\gamma\delta} \,
  \left\{
 -  \frac{ \pazocal{Y}^{k_1}_\gamma \pazocal{Y}^{k_2}_\delta \, ( \pazocal{Y}^{k_1}_{\gamma'} \pazocal{Y}^{k_2}_{\delta'})^*}
                     {\omega+\omega_3  - (\varepsilon^-_{k_1} + \varepsilon^-_{k_2}) - i\eta}
  \right\}
\nonumber \\
 & \qquad \qquad \times
   \, V_{\gamma' \delta', \mu' \nu'} \,  \left\{
    \frac{( \pazocal{X}^{n_4}_{\mu'} \pazocal{X}^{n_5}_{\nu'})^*  \pazocal{X}^{n_4}_\mu \pazocal{X}^{n_5}_\nu}
                      {\omega+\omega_3  - (\varepsilon^+_{n_4}  + \varepsilon^+_{n_5}) + i\eta} 
  \right\}
  V_{\mu \nu, \beta \lambda}
   \left\{  \frac{\pazocal{Y}^{k_3}_\lambda  (\pazocal{Y}^{k_3}_\sigma)^*}{\omega_3  - \varepsilon^-_{k_3} - i\eta}  \right\}
     \nonumber \\
     &+
 \frac{1}{4}  \int \frac{{\rm d}\omega_3}{2\pi i} 
    V_{\alpha\sigma,\gamma\delta} \,
  \left\{
    \frac{(\pazocal{X}^{n_1}_\gamma \pazocal{X}^{n_2}_\delta)^*  \pazocal{X}^{n_1}_{\gamma'} \pazocal{X}^{n_2}_{\delta'}}
                      {\omega+\omega_3  - (\varepsilon^+_{n_1}  + \varepsilon^+_{n_2}) + i\eta} 
  \right\}
\nonumber \\
 & \qquad \qquad \times
   \, V_{\gamma' \delta', \mu' \nu'} \,  \left\{
 -  \frac{ \pazocal{Y}^{k_4}_{\mu'} \pazocal{Y}^{k_5}_{\nu'} \, (\pazocal{Y}^{k_4}_\mu \pazocal{Y}^{k_5}_\nu)^*}
                     {\omega+\omega_3  - (\varepsilon^-_{k_4} + \varepsilon^-_{k_5}) - i\eta}
  \right\}
  V_{\mu \nu, \beta \lambda}
   \left\{  \frac{\pazocal{Y}^{k_3}_\lambda  (\pazocal{Y}^{k_3}_\sigma)^*}{\omega_3  - \varepsilon^-_{k_3} - i\eta}  \right\}
     \nonumber \\
 &+
 \frac{1}{4}  \int \frac{{\rm d}\omega_3}{2\pi i} 
    V_{\alpha\sigma,\gamma\delta} \,
  \left\{
    \frac{(\pazocal{X}^{n_1}_\gamma \pazocal{X}^{n_2}_\delta)^*  \pazocal{X}^{n_1}_{\gamma'} \pazocal{X}^{n_2}_{\delta'}}
                      {\omega+\omega_3  - (\varepsilon^+_{n_1}  + \varepsilon^+_{n_2}) + i\eta} 
  \right\}
\nonumber \\
 & \qquad \qquad \times
   \, V_{\gamma' \delta', \mu' \nu'} \,  \left\{
    \frac{( \pazocal{X}^{n_4}_{\mu'} \pazocal{X}^{n_5}_{\nu'})^*  \pazocal{X}^{n_4}_\mu \pazocal{X}^{n_5}_\nu}
                      {\omega+\omega_3  - (\varepsilon^+_{n_4}  + \varepsilon^+_{n_5}) + i\eta} 
  \right\}
  V_{\mu \nu, \beta \lambda}
   \left\{  \frac{\pazocal{Y}^{k_3}_\lambda  (\pazocal{Y}^{k_3}_\sigma)^*}{\omega_3  - \varepsilon^-_{k_3} - i\eta}  \right\}
\nonumber \\
\nonumber \\
  ={}&   
   \frac{  \frac{1}{2} V_{\alpha\sigma,\gamma\delta} \,  \pazocal{Y}^{k_1}_\gamma \pazocal{Y}^{k_2}_\delta 
     ~  ( \pazocal{Y}^{k_1}_{\gamma'} \pazocal{Y}^{k_2}_{\delta'})^* \;  V_{\gamma' \delta', \mu' \nu'}  \;
        ( \pazocal{X}^{n_4}_{\mu'} \pazocal{X}^{n_5}_{\nu'} \pazocal{Y}^{k_3}_\sigma)^* }
                     {[\varepsilon^-_{k_1} + \varepsilon^-_{k_2} -\varepsilon^+_{n_4}  - \varepsilon^+_{n_5} ]}
    \frac{1}{2}
    \frac{  \pazocal{X}^{n_4}_\mu \pazocal{X}^{n_5}_\nu \pazocal{Y}^{k_3}_\lambda}
                      {\omega  - (\varepsilon^+_{n_4}  + \varepsilon^+_{n_5}  - \varepsilon^-_{k_3}) + i\eta} 
  V_{\mu \nu, \beta \lambda}
      \nonumber \\
 &+
 V_{\alpha\sigma,\gamma\delta} \,
    \frac{(\pazocal{X}^{n_1}_\gamma \pazocal{X}^{n_2}_\delta \pazocal{Y}^{k_3}_\sigma)^* }
                      {\omega  - (\varepsilon^+_{n_1}  + \varepsilon^+_{n_2}  - \varepsilon^-_{k_3}) + i\eta} 
                 \frac{1}{2} 
 \frac{   \pazocal{Y}^{k_3}_\lambda \pazocal{X}^{n_1}_{\gamma'} \pazocal{X}^{n_2}_{\delta'} \;  V_{\gamma' \delta', \mu' \nu'} \;
 \pazocal{Y}^{k_4}_{\mu'} \pazocal{Y}^{k_5}_{\nu'} \, (\pazocal{Y}^{k_4}_\mu \pazocal{Y}^{k_5}_\nu)^*  \frac{1}{2} V_{\mu \nu, \beta \lambda} }
                  {[\varepsilon^-_{k_4} + \varepsilon^-_{k_5} - \varepsilon^+_{n_1}  - \varepsilon^+_{n_2}]}
\nonumber \\
  &+
    \frac{     V_{\alpha\sigma,\gamma\delta}  ~ (\pazocal{X}^{n_1}_\gamma \pazocal{X}^{n_2}_\delta \pazocal{Y}^{k_3}_\sigma)^* }
                      {\omega  - (\varepsilon^+_{n_1}  + \varepsilon^+_{n_2} - \varepsilon^-_{k_3}) + i\eta} 
 \frac{1}{2}  
   \pazocal{X}^{n_1}_{\gamma'} \pazocal{X}^{n_2}_{\delta'}  \, V_{\gamma' \delta', \mu' \nu'} \, ( \pazocal{X}^{n_4}_{\mu'} \pazocal{X}^{n_5}_{\nu'})^*
 \frac{1}{2}  
    \frac{  \pazocal{X}^{n_4}_\mu \pazocal{X}^{n_5}_\nu \pazocal{Y}^{k_3}_\lambda  ~ V_{\mu \nu, \beta \lambda}}
                      {\omega  - (\varepsilon^+_{n_4}  + \varepsilon^+_{n_5} - \varepsilon^-_{k_3}) + i\eta} 
 \nonumber \\
\nonumber \\
  \equiv{}& \quad M^{(2,ld) \, \dagger}\frac1{\omega - E^> + i \eta}M^{(1)}
  \nonumber \\
    &+~  M^{(1) \, \dagger}\frac1{\omega - E^> + i \eta}M^{(2,ld)}
  \nonumber \\
    & +~  M^{(1) \, \dagger}\frac1{\omega - E^> + i \eta}C^{(ld)}\frac1{\omega - E^> + i \eta}M^{(1)}  \; ,
\label{eq:LaddEg2}
 \end{align}
where $M^{(1)}$ and $E^>$ are the same as in Eqs.~\eqref{eq:ADC2_MEC} and the factors 1/2 are again absorbed by summing over the ordered configurations for $r$ and $r'$. The 2p1h ladder interaction $C^{(ld)}$ is at first order in $V$, while the coupling matrix $ M^{(2,ld)}$ is at second order. These can be read from the previous lines of Eq.~\eqref{eq:LaddEg2} and turn out to be (showing all summations explicitly):
\begin{align}
M^{(2,ld)}_{r,\alpha} ={}&  \sum_{k_4, \, k_5} \quad \sum_{\substack{ \sigma ,\, \zeta ,\, \gamma, \,  \delta \\ \mu ,\, \nu , \, \lambda}} 
  \frac{   \pazocal{X}^{n_1}_{\gamma} \pazocal{X}^{n_2}_{\delta} \;  V_{\gamma \delta, \sigma \zeta} \;
 \pazocal{Y}^{k_4}_{\sigma} \pazocal{Y}^{k_5}_{\zeta} \, (\pazocal{Y}^{k_4}_\mu \pazocal{Y}^{k_5}_\nu)^*  \pazocal{Y}^{k_3}_\lambda  }
                  {[\varepsilon^-_{k_4} + \varepsilon^-_{k_5} - \varepsilon^+_{n_1}  - \varepsilon^+_{n_2}]} \frac{1}{2} V_{\mu \nu, \alpha \lambda}
  \nonumber \\
  C^{(ld)}_{r,r'} ={}&  \sum_{ \alpha ,\, \beta ,\, \gamma ,\, \delta} \pazocal{X}^{n_1}_{\alpha} \pazocal{X}^{n_2}_{\beta}  \, V_{\alpha \beta, \gamma \delta} \, ( \pazocal{X}^{n_1'}_{\gamma} \pazocal{X}^{n_2'}_{\delta})^* \; \delta_{k_3, k_3'} \; .
\end{align}

Eq.~\eqref{eq:LaddEg2}  clearly breaks the known Lehmann representation for the self-energy and would even lead to inconsistent 
results unless its contribution is very small compared to the second-order contribution of Eq.~\eqref{eq:Sig_2nd}. That is, Eq.~\eqref{eq:LaddEg2} would
invalidate the perturbative expansion unless $V$ is small.
Therefore, we need to identify proper corrections that allow to retain these third order contributions but at the same time let
us recover the correct analytical form~\eqref{eq:ADC_SE_form}.  For the first two terms on the right hand side of Eq.~\eqref{eq:LaddEg2}, this
issue can be easily solved by remembering that the corresponding diagram from $\Sigma^{(2)}(\omega)$ (see Eq.~\eqref{eq:Sig_2nd}) is to be included.
If then one adds an extra term that is quadratic in  $M^{(2,ld)}$, this leads to:
\begin{equation}
  \Sigma^{(2)}(\omega)  + \Sigma^{(3, ld)}(\omega) + M^{(2,ld) \, \dagger}\frac 1 {\omega  - E^> + i\eta} M^{(2,ld)}
 \longrightarrow  \left[ M^{(1)} +  M^{(2,ld)}\right]^\dagger \frac1{\omega  - E^> + i\eta}\left[ M^{(1)} + M^{(2,ld)}\right]  \, ,
\end{equation}
which resolves the issue of obtaining the residues in separable form. Note that this new correction is just one specific Goldstone diagram among
the many that contribute to the self-energy at {\em fourth order}. On the other hand, adding all of the fourth-order diagrams would lead to 
new terms that break the Lehmann representation themselves and that in turn would call for the inclusions of selected Goldstone terms at even higher orders.  In other words, we have achieved to recover the structure of Eq.~\eqref{eq:ADC_SE_form}  but at the price of giving up a systematic 
perturbative expansion that is complete at each order in $\widetilde V$. Given that the Lehmann representation is dictated by physical properties, this is 
a more satisfactory rearrangement of the perturbation series.

The last term in Eq.~\eqref{eq:LaddEg2} is more tricky to correct since it contains second-order poles as $(\omega  - E - i\eta)^{-2}$, which cannot be 
canceled by single contributions at higher order. Instead, we are forced to perform a non-perturbative resummation of Goldstone diagrams to all orders that results in a geometric series. This is done by considering the relation
\begin{equation}
 \frac1{A-B}  ~=~  \frac1 A ~+~ \frac1 A  \, B \,  \frac1{A-B}  ~=~  \frac1 A  ~+~  \frac1 A  \, B \,  \frac1 A  ~+~  \frac1 A  \, B \,   \frac1 A   \, B \,   \frac1 A
             ~+~ \frac1 A  \, B \,  \frac1 A  \, B \,  \frac1 A  \, B \,  \frac1 A  ~+ \ldots
  \label{eq:1overAB}
\end{equation}
for two operators $A$ and $B$. If we chose  $A\equiv{\omega  - E^> + i\eta}$ and $B\equiv C^{(ld)}$,  the first and second term on the right hand 
side can then be  identified respectively with the contribution from $\Sigma^{(2)}(\omega)$ and the  last term of Eq.~\eqref{eq:LaddEg2}. 
Also in this case, all perturbative terms up to third order have been kept unchanged  but we are forced to select a series of Goldstone diagrams up to infinite order.   

If then one adds an extra term that is quadratic in  $M^{(2,ld)}$, this lead to:
\begin{equation}
  \Sigma^{(2)}(\omega)  ~+~ \Sigma^{(3, ld)}(\omega) ~+~ \begin{array}{c} \hbox{terms beyond} \\ \hbox{ 3$^{rd}$ order } \end{array} 
 \longrightarrow  \left[ M^{(1)} +  M^{(2,ld)}\right]^\dagger \frac1{\omega  - E^> - C^{(ld)} + i\eta}\left[ M^{(1)} + M^{(2,ld)}\right]  \ ,
 \label{eq:pp_ladder}
\end{equation}
which now contains {\em all}  the perturbation theory terms at second~\eqref{eq:Sig_2nd} and third order~\eqref{eq:LaddEg2} while preserving the expected analytical form for~$\widetilde\Sigma(\omega)$. 

It can be shown that the summation implicit in Eq.~\eqref{eq:pp_ladder} is equivalent to a full resummation of two-particle ladder diagrams in the Tamn-Dancoff approximation (TDA)~\cite{ch11_RingSchuck}. In this sum the remaining quasi hole state appearing in the 2p1h ISC remains uncoupled from the ladder series, as it can be seen in Fig.~\ref{fig:3rdOrd}a), which is the first term in the series.  Likewise, one would find that the remaining backward-going terms in Eq.~\eqref{eq:LaddEg1} would lead to resumming the two-hole TDA ladders within the 2h1p configurations. Instead, diagram in Fig.~\ref{fig:3rdOrd}b) involves a resummation of ph ring diagrams. Extensions of these series to random-phase approximation (RPA) is also possible, this would introduce a larger set of high-order Goldstone diagrams but it would not be required to enforce consistency with perturbation theory at third order.

\vskip 0.3 cm
\noindent
{\bf Exercise 11.3.}  Complete  the calculation of Eq.~\eqref{eq:LaddEg1} and derive the remaining corrections to the 2h1p interaction $D^{(ld)}$  and the 1h-2h1p coupling term  $N^{(2,ld)}$.

\subsection{The \texorpdfstring{ADC($n$)}{ADC(n)} approach and working equations at third order}

 The  procedure discussed above to devise reliable approximations for the self-energy is at the heart of the
 ADC method, originally introduced by J.~Schirmer and collaborators~\cite{ch11_Schirmer1982ADC2,ch11_Schirmer1983ADCn}.
 This approach  generates  a hierarchy of approximations of increasing accuracy such that,
 at a given order $n$, the ADC($n$) equations will maintain the analytic form of Eq.~\eqref{eq:ADC_SE_form} and will be consistent with perturbation theory up to order~$n$. Note that this does not mean that ADC($n$) is a perturbative truncation but that it must contain at least all the Feynman diagrams for $\Sigma^\star(\omega)$ up to order~$n$, among higher terms. In fact, we will see below that for $n>2$ it always involve an infinite resummation of diagrams (see also Eqs.~\eqref{eq:1overAB} and~\eqref{eq:pp_ladder}). 
To implement this scheme for the dynamic self-energy, $\widetilde\Sigma(\omega)$, we expand its Lehmann representation in powers of the perturbation interaction $\widehat{H}_1$ (or, equivalently, $\widetilde{H}_1$). The interaction matrices 
$C$ and $D$ appearing in the denominators of Eq.~\eqref{eq:ADC_SE_form} can only be of first order in either $\widehat{U}$,  $\widehat{V}$ or $\widehat{W}$. However, the coupling matrices can contain terms of any order:
\begin{align}
  M =& M^{(1)} +  M^{(2)} +  M^{(3)} +  \ldots
   \nonumber  \\
  N =&  N^{(1)} +  N^{(2)} +  N^{(3)} + \ldots  
 \label{eq:MNexp}
\end{align}
Using  Eqs.~\eqref{eq:1overAB}  and~\eqref{eq:MNexp} one finds the following expansion
for Eq.~\eqref{eq:ADC_SE_form}:
\begin{align}
  \Sigma^{\star}(\omega) ={}&  - \widehat{U}  ~+~   \Sigma^{(\infty)} 
  \nonumber \\
   +~& M^{(1) \, \dagger}\frac1{\omega - E^>  + i \eta}M^{(1)} +N^{(1)}\frac1{\omega - E^<  - i \eta}N^{(1) \, \dagger}
    \nonumber \\
  +~& M^{(2) \, \dagger}\frac1{\omega - E^> + i \eta}M^{(1)}  +   M^{(1) \, \dagger}\frac1{\omega - E^> + i \eta}M^{(2)}
 +{}  M^{(1) \, \dagger}\frac1{\omega - E^> + i \eta} C \frac1{\omega - E^> + i \eta}M^{(1)}  
    \nonumber \\
  +~& N^{(2)}\frac1{\omega - E^< - i \eta}N^{(1) \, \dagger} ~ +   N^{(1)}\frac1{\omega - E^< - i \eta}N^{(2) \, \dagger}
 ~+{}  N^{(1)}\frac1{\omega - E^< - i \eta} D \frac1{\omega - E^< - i \eta}N^{(1) \, \dagger}  
 \nonumber \\
 +~&   {\pazocal O}({ \widehat{H}_1}^4)  \; ,
 \label{eq:ADC_SE_form_exp}
\end{align}
where all terms up to third order in $\widehat H_1$ are shown explicitly.
The ADC procedure is then to simply calculate all possible diagrams up
to order $n$. By comparing them to Eq.~\eqref{eq:ADC_SE_form_exp}, one then reads
the minimum expressions for the coupling and interaction matrices, $M$, $N$, $C$ and $D$ that are needed to
retain all the $n$-order diagrams for $\widetilde\Sigma(\omega)$. Correspondingly, the energy-independent self-energy $\Sigma^{(\infty)}$ needs to be computed at least up to order $n$ as well.
 Note that the dynamic part of the self-energy, which propagates ISCs, appears only starting from 
 second order. This is so because any such diagram needs at least one perturbing interaction $V$ to generate an ISC
 and a second one to annihilate it back to a single particle state. In general, if the Hamiltonian contains
 up to $m$-body forces and $i$ is an integer, then the ADC($2i$) and ADC($2i+1$) will require
 ISCs up to  ($k$+1)-particle--$k$-hole and \hbox{($k$+1)-hole--$k$-particle}, where~\hbox{$k$=($m$-1)*$i$}.
 Thus, with two-nucleon forces ADC(2) and ADC(3) include  2p1h and 2h1p states,  ADC(4) and ADC(5) need up to 
 3p2h and 2h3p states, and so on. However, the full ADC(2/3) sets with three-nucleon forces already
 includes 3p2h and 3h2p configurations~\cite{ch11_Raimondi_inprep}.

At first order, ADC(1) requires to only calculate diagram(s) that contribute to $\widetilde{U}=-\widehat{U}+\Sigma^{(\infty)}$,
see Fig.~\ref{fig:EffOps}a), and thus the scheme reduces to Hartree-Fock theory.
At second order and with at most two-body interactions, there is only one diagram contributing to $\widetilde\Sigma(\omega)$
which is already in the proper Lehmann form. Hence, Eqs.~\eqref{eq:Sig_2nd},~\eqref{eq:ADC2_MEC} and~\eqref{eq:ADC2_NED}
fully define the ADC(2) approximation. In this case, $\Sigma^{(\infty)}$ also requires a second-order non-skeleton term.

Higher order cases are more complicated. For a two-body Hamiltonian, the only skeleton diagrams at third order
are the ladder and ring diagrams shown in Figs.~\ref{fig:3rdOrd}a) and ~\ref{fig:3rdOrd}b).
  As long as one works with a Hartree-Fock reference state or a fully self-consistent (dressed)
propagators, no other diagram is needed because the additional  non-skeleton terms either vanish or
must not be included (see {\bf Exercise 11.5}).
In these cases, one obtains the following working expressions the for the ADC(3) approximation:
\begin{subequations}
\label{eq:ADC3_MEC}
\begin{align}
M_{r , \alpha} ={}&  \pazocal{X}^{n_1}_\mu \pazocal{X}^{n_2}_\nu \pazocal{Y}^{k_3}_\lambda \, V_{\mu \nu, \alpha \lambda} ~+~
  \frac{   \pazocal{X}^{n_1}_{\gamma} \pazocal{X}^{n_2}_{\delta} \;  V_{\gamma \delta, \sigma \zeta} \;
 \pazocal{Y}^{k_4}_{\sigma} \pazocal{Y}^{k_5}_{\zeta}  }
                  {2 \, [\varepsilon^-_{k_4} + \varepsilon^-_{k_5} - \varepsilon^+_{n_1}  - \varepsilon^+_{n_2}]} \,
                  \, (\pazocal{Y}^{k_4}_\mu \pazocal{Y}^{k_5}_\nu)^*  \, \pazocal{Y}^{k_3}_\lambda \,  V_{\mu \nu, \alpha \lambda}
 \label{eq:ADC3_M} \\
 &{}+ \frac{
{\pazocal X}^{n_2}_{\rho} {\pazocal Y}^{k_3}_{\sigma}
V_{\rho \delta, \sigma \gamma}
{\pazocal Y}^{k_5}_{\gamma} {\pazocal X}^{n_6}_{\delta}
}{[\varepsilon^-_{k_3} - \varepsilon^+_{n_2} + \varepsilon^-_{k_5} - \varepsilon^+_{n_6}]}
  ({\pazocal Y}^{k_5}_{\nu} {\pazocal X}^{n_6}_{\lambda})^* \, {\pazocal X}^{n_1}_{\mu} \,
V_{\mu \nu, \alpha \lambda}
 ~-~
\frac{
{\pazocal X}^{n_1}_{\rho} {\pazocal Y}^{k_3}_{\sigma}
V_{\rho \delta, \sigma \gamma}
{\pazocal Y}^{k_5}_{\gamma} {\pazocal X}^{n_6}_{\delta}
}{[\varepsilon^-_{k_3} - \varepsilon^+_{n_1} + \varepsilon^-_{k_5} - \varepsilon^+_{n_6}]}
  ({\pazocal Y}^{k_5}_{\nu} {\pazocal X}^{n_6}_{\lambda})^* \, {\pazocal X}^{n_2}_{\mu} \,
V_{\mu \nu, \alpha \lambda}
 \nonumber \\
 \nonumber \\
  E^>_{r,r'} ={}& \mathrm{diag} \left( \,\varepsilon^+_{n_1} + \varepsilon^+_{n_2} - \varepsilon^-_{k_3}  \, \right)
  \label{eq:ADC3_Efw}  \\
 \nonumber \\
C_{r,r'} ={}&  \pazocal{X}^{n_1}_{\alpha} \pazocal{X}^{n_2}_{\beta}  \, V_{\alpha \beta, \gamma \delta} \, ( \pazocal{X}^{n_1'}_{\gamma} \pazocal{X}^{n_2'}_{\delta})^* \; \delta_{k_3, k_3'}
 \nonumber \\
 &{} +  {\pazocal X}^{n_1}_\alpha {\pazocal Y}^{k_3}_\beta \, V_{\alpha \delta, \beta \gamma} \,
 ({\pazocal X}^{n_1'}_\gamma {\pazocal Y}^{k_3'}_\delta )^*  \; \delta_{n_2, n_2'}
  -  {\pazocal X}^{n_2}_\alpha {\pazocal Y}^{k_3}_\beta \, V_{\alpha \delta, \beta \gamma} \,
 ({\pazocal X}^{n_1'}_\gamma {\pazocal Y}^{k_3'}_\delta )^*  \; \delta_{n_1, n_2'}
  \label{eq:ADC3_C} \\
&{}  -  {\pazocal X}^{n_1}_\alpha {\pazocal Y}^{k_3}_\beta \, V_{\alpha \delta, \beta \gamma} \,
 ({\pazocal X}^{n_2'}_\gamma {\pazocal Y}^{k_3'}_\delta )^*  \; \delta_{n_2, n_1'}
  +  {\pazocal X}^{n_2}_\alpha {\pazocal Y}^{k_3}_\beta \, V_{\alpha \delta, \beta \gamma} \,
 ({\pazocal X}^{n_2'}_\gamma {\pazocal Y}^{k_3'}_\delta )^*  \; \delta_{n_1, n_1'}
\nonumber
\end{align}
\end{subequations}
and
\begin{subequations}
\label{eq:ADC3_NED}
\begin{align}
 N_{\alpha,s} ={}& V_{\alpha \lambda, \mu \nu} \, \pazocal{Y}^{k_1}_\mu \pazocal{Y}^{k_2}_\nu \pazocal{X}^{n_3}_\lambda 
~+~
V_{\alpha \lambda, \mu \nu}  \,  {\pazocal X}^{n_3}_{\lambda} \, ( {\pazocal X}^{n_7}_{\mu} {\pazocal X}^{n_8}_{\nu} )^*
\frac{
 {\pazocal X}^{n_7}_{\gamma} {\pazocal X}^{n_8}_{\delta}
V_{\gamma \delta ,\sigma \rho}
 {\pazocal Y}^{k_1}_{\sigma} {\pazocal Y}^{k_2}_{\rho}
}{2 \, [\varepsilon^-_{k_1} + \varepsilon^-_{k_2} - \varepsilon^+_{n_7} - \varepsilon^+_{n_8}]}
  \label{eq:ADC3_N} \\
&{} +
V_{\alpha \lambda, \mu \nu}
\, {\pazocal Y}^{k_1}_{\mu} \,  ({\pazocal X}^{n_5}_{\nu} {\pazocal Y}^{k_6}_{\lambda})^* \,
\frac{
{\pazocal X}^{n_5}_{\gamma} {\pazocal Y}^{k_6}_{\delta} V_{\gamma \rho, \delta \sigma }
{\pazocal Y}^{k_2}_{\sigma} {\pazocal X}^{n_3}_{\rho}
}{[\varepsilon^-_{k_2} - \varepsilon^+_{n_3} + \varepsilon^-_{k_6} - \varepsilon^+_{n_5}]}
-
V_{ \alpha \lambda, \mu \nu}
\, {\pazocal Y}^{k_2}_{\mu} \, ({\pazocal X}^{n_5}_{\nu} {\pazocal Y}^{k_6}_{\lambda})^* \,
\frac{
{\pazocal X}^{n_5}_{\gamma} {\pazocal Y}^{k_6}_{\delta} V_{\gamma \rho, \delta \sigma  }
{\pazocal Y}^{k_1}_{\sigma} {\pazocal X}^{n_3}_{\rho}
}{[\varepsilon^-_{k_1} - \varepsilon^+_{n_3} + \varepsilon^-_{k_6} - \varepsilon^+_{n_5}]}
 \nonumber \\
 \nonumber \\
 E^<_{s,s'} ={}& \mathrm{diag} \left( \, \varepsilon^-_{k_1} + \varepsilon^-_{k_2} - \varepsilon^+_{n_3} \, \right)
   \label{eq:ADC3_Ebk}  \\
 \nonumber \\
 D_{s,s'}={}&  - (\pazocal{Y}^{k_1}_{\alpha} \pazocal{Y}^{k_2}_{\beta})^*  \, V_{\alpha \beta, \gamma \delta} \, \pazocal{Y}^{k_1'}_{\gamma} \pazocal{Y}^{k_2'}_{\delta} \; \delta_{n_3, n_3'}
\nonumber \\
&{} - ({\pazocal Y}^{k_1}_\alpha    {\pazocal X}^{n_3}_\beta )^* V_{\alpha \delta, \beta \gamma} \, {\pazocal Y}^{k_1'}_\gamma    {\pazocal X}^{n_3'}_\delta   \; \delta_{k_2, k_2'}
 + ({\pazocal Y}^{k_2}_\alpha    {\pazocal X}^{n_3}_\beta )^* V_{\alpha \delta, \beta \gamma} \, {\pazocal Y}^{k_1'}_\gamma    {\pazocal X}^{n_3'}_\delta  \; \delta_{k_1, k_2'}
\label{eq:ADC3_D} \\
&{} + ({\pazocal Y}^{k_1}_\alpha    {\pazocal X}^{n_3}_\beta )^* V_{\alpha \delta, \beta \gamma} \, {\pazocal Y}^{k_2'}_\gamma    {\pazocal X}^{n_3'}_\delta   \; \delta_{k_2, k_1'}
 - ({\pazocal Y}^{k_2}_\alpha    {\pazocal X}^{n_3}_\beta )^* V_{\alpha \delta, \beta \gamma} \, {\pazocal Y}^{k_2'}_\gamma    {\pazocal X}^{n_3'}_\delta  \; \delta_{k_1, k_1'}  \; ,
  \nonumber
\end{align}
\end{subequations}
where  only  ordered configurations $r$=$\{n_1 < n_2, k_3 \}$ and  $s$=$\{k_1 < k_2, n_3 \}$ need to be considered, in accordance
with the Pauli principle.  Note that  these equations apply to the case of  two-body interactions  but they remain unchanged for  an effective operator $\widetilde{V}$ that is derived from {\hbox{three-body} forces.  However the full inclusion of $\widehat{W}$ would require the inclusion of the diagram of Fig.~\ref{fig:2ndOrd}b) at the ADC(2) level and several other interaction-irreducible diagrams for ADC(3).  The non-skeleton contributions to $\widetilde\Sigma(\omega)$ that arise at third order when the reference propagator is not dressed are shown in Fig.~\ref{fig:SEins_3ndOrd}. The case of three-nucleon forces is discussed in full detail in Ref.~\cite{ch11_Raimondi_inprep}.

\begin{figure}[t]
\begin{center}
\includegraphics[height=0.22\textwidth]{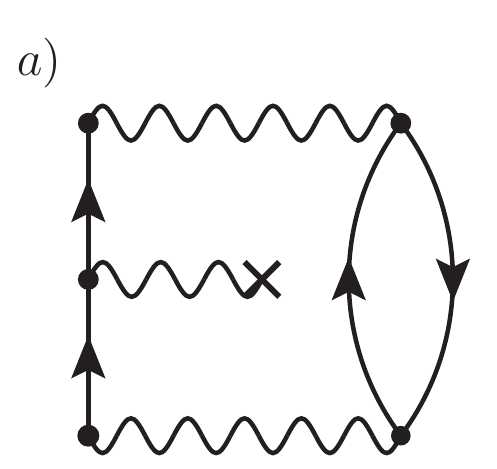}   \hspace{0.25\textwidth}
\includegraphics[height=0.22\textwidth]{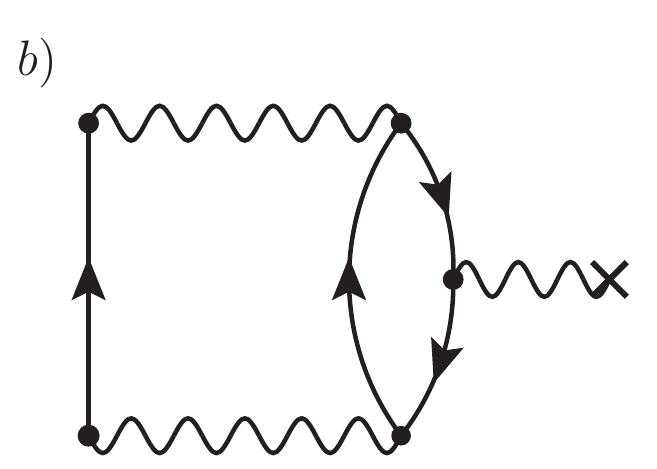}   \hspace{0.20\textwidth}
\vskip 0.8 cm
\includegraphics[height=0.21\textwidth]{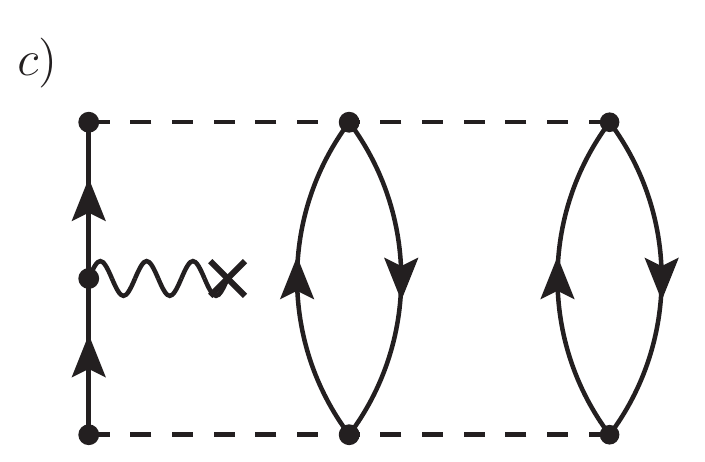}   \hspace{0.1\textwidth}
\includegraphics[height=0.21\textwidth]{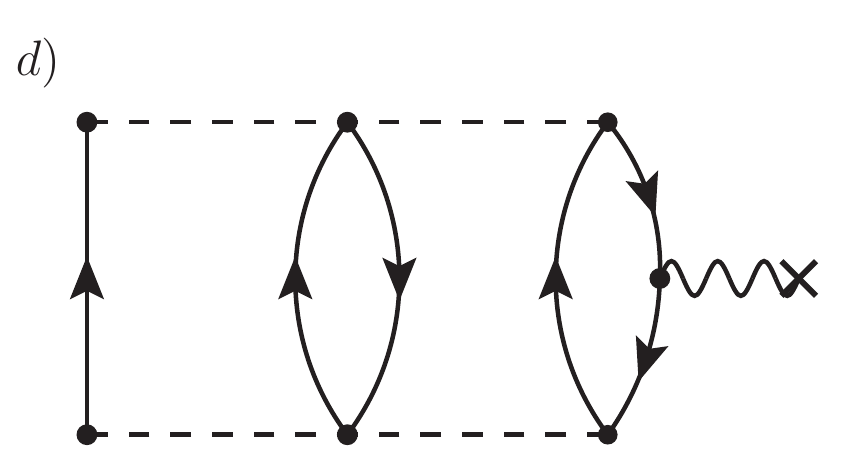}
\caption{Self-energy insertion diagrams that appear, at third order, in the perturbative expansion for $\widetilde\Sigma(\omega)$ with two- and three-nucleon interactions. These non-skeleton diagrams need to be considered  when the reference propagators are not self-consistent.  Diagrams a) and b) involve only one- and two-body interactions and results from self-energy insertion into the diagram of Fig.~\ref{fig:2ndOrd}a).  With the inclusion of three-nucleon interactions, the diagrams c) and d) arise from the one of Fig.~\ref{fig:2ndOrd}b).  When a Hartree-Fock reference state is used all these contributions cancel out (see {\bf Exercise 11.5}). }
\label{fig:SEins_3ndOrd}
\end{center}
\end{figure}

To remain consistent with the ADC($n$) formulation,  the static self-energy $\Sigma^{(\infty)}$ must also be computed at least to the same order $n$. However, this involves a large number of non-skeleton diagrams when self-consistency is not implemented. In practice, it is relatively inexpensive to compute it directly from dressed propagators, as given by 
\eqref{eq:U_eff} and therefore if can be iterated to self-consistency.
This prescription, in which $\widetilde\Sigma(\omega)$ is calculated from an unperturbed reference state $g^{0)}(\omega)$ but $\Sigma^{(\infty)}$
is obtained self-consistently, is often used in nuclear physics applications and we refer to it as the {\em sc0} approximation~\cite{ch11_Soma2014Lanc}.
When dealing with the Coulomb force in molecular systems, the dynamic self-energy can be simply calculated in terms of a Hartree-Fock reference state.  In nuclear physics, a Hartree-Fock reference state is adequate only if the chosen Hamiltonian is particularly soft.  Otherwise, it is necessary to optimize the reference state by choosing a $\widehat{H}_0$ and $g^{0)}(\omega)$  that better represent the correlated single particle energies in the dressed propagator. In all cases, at least the {\em sc0} approach to $\Sigma^{(\infty)}$ is always required when computing finite nuclei and infinite nucleonic matter.

The standard ADC($n$) prescription is to identify the {\em minimal} matrices  $M$, $N$, $C$ and $D$  that make the self-energy  consistent with perturbation theory up to order $n$. However, other intermediate approximations are also possible and have been exploited in the past.
The so-called \hbox{\em 2p1h-TDA} method is an extension of the second order scheme of Eqs.~\eqref{eq:ADC2_MEC} and~\eqref{eq:ADC2_NED}  where the matrices $C$ and $D$ are  calculated at first order instead, as given by  Eqs.~\eqref{eq:ADC3_C} and~\eqref{eq:ADC3_D}. As a rule of thumb, the ADC(2)  approximation yields roughly 90\% of the total correlation energy in most applications, while the ADC(3) can account for about 99\% of it---hence, with a~1\% error in binding energies. The 2p1h-TDA  contains the ADC(2) in full but it further resums the full set of two-particle~(pp), two-holes~(hh) and particle-hole~(ph) diagrams. This can result in a sensible improvement in the accuracy of binding energies but without the price of computing corrections to the $M$ and $N$ coupling matrices. Nevertheless, the 2p1h-TDA misses the second order terms from Eqs.~\eqref{eq:MNexp} that are known to contribute strongly to quasiparticle energies. As a consequence the one nucleon addition and separation energies (or, equivalently, ionization potentials and electron affinities in molecules) would be predicted poorly in 2p1h-TDA and in general they require full ADC(3) calculations~\cite{ch11_VonNiessen1984ConPhysRep}.
In nuclear physics applications, the description of collective excitations often requires that  particle-hole configurations are diagonalized at least in the  
RPA scheme. While this is similar to the TDA all-order summations included in 2p1h-TDA and in ADC(3), extra ground state correlations effects from the RPA series are deemed important to reproduce collective modes typical of nuclear systems~\cite{ch11_RingSchuck}.
To account for these effects on needs to separate the partial summations in the pp, hh and ph channels, substitute them with equivalent RPA series and  recouple these through a Faddeev-like expansion, in order to eventually reconstruct the self-energy~\cite{ch11_Danielewicz1994OMP,ch11_Barbieri2001frpa,ch11_Barbieri2003ExO16,ch11_Barbieri2006plbO16}.
The Faddeev-RPA (FRPA)  method contains the ADC(3) in full but it also generates additional ground state correlation terms that are induced by the RPA summation and are at fourth and higher order in the perturbative expansion of the self-energy. The working implementation of the FRPA approach has been formulated in Refs.~\cite{ch11_Barbieri2001frpa,ch11_Barbieri2007Atoms,ch11_Degroote2011frpa}.

  Another important extension of the ADC(3) framework comes from the realization that Eqs.~\eqref{eq:MNexp} still imply a   perturbative truncation for $M$ and~$N$. This causes the energy denominators in Eqs.~\eqref{eq:ADC3_M} and~\eqref{eq:ADC3_N} to become unstable if the system is close to being degenerate. The way out from this situation is again to perform an all-orders summation. Since the coupling matrices correspond to specific energy-independent parts  of Goldstone diagrams, they can be resummed in the same way as for the coupled cluster (CC) technique~\cite{ch11_Barb_unp}. We show this for the second term on the right hand side of~\eqref{eq:ADC3_M}, which can be rewritten as follows:
 \begin{align}
  \frac{   \pazocal{X}^{n_1}_{\gamma} \pazocal{X}^{n_2}_{\delta} \;  V_{\gamma \delta, \sigma \zeta} \;
 \pazocal{Y}^{k_4}_{\sigma} \pazocal{Y}^{k_5}_{\zeta} \, (\pazocal{Y}^{k_4}_\mu \pazocal{Y}^{k_5}_\nu)^*  \pazocal{Y}^{k_3}_\lambda  }
                  {2 \, [\varepsilon^-_{k_4} + \varepsilon^-_{k_5} - \varepsilon^+_{n_1}  - \varepsilon^+_{n_2}]} \, V_{\mu \nu, \alpha \lambda}
   \quad \longrightarrow {}& \quad
  \frac{1}{2}  \;  t^{(0)}{}^{n_1 \, n_2}_{k_4 \, k_5} \; (\pazocal{Y}^{k_4}_\mu \pazocal{Y}^{k_5}_\nu)^*  \pazocal{Y}^{k_3}_\lambda   \, V_{\mu \nu, \alpha \lambda}  \; ,
 \label{eq:ADC3CCM}
 \end{align}
 where the amplitude
 \begin{align}
    t^{(0)}{}^{n_1 \, n_2}_{k_4 \, k_5} \equiv {}& 
    \frac{   \pazocal{X}^{n_1}_{\gamma} \pazocal{X}^{n_2}_{\delta} \;  V_{\gamma \delta, \sigma \zeta} \; \pazocal{Y}^{k_4}_{\sigma} \pazocal{Y}^{k_5}_{\zeta}   }
     { \varepsilon^-_{k_4} + \varepsilon^-_{k_5} - \varepsilon^+_{n_1}  - \varepsilon^+_{n_2} }
 \label{eq:dressed_t2}
 \end{align}
generalizes the zeroth approximation to the CC operator~$\hat{T}_2$ (see Sec. 8.7 or Ref.~\cite{chapter8}).  In case of a dressed propagator, the spectroscopic amplitudes $\pazocal{X}$~($\pazocal{Y}$) account for the fragmentation of single particle strength. However, for a standard mean-field reference, they simply select the particle~(hole) reference orbits and $t^{(0)}$ is exactly the same as for the CC approach. In order to mitigate effects of the perturbative truncation in Eqs.~\eqref{eq:ADC3_M} and ~\eqref{eq:ADC3_N} (and to resum the 2p-2h ISCs) one simply substitutes $t^{(0)}$ with the corresponding CC solution.  In general, when $t$ is computed using the CC doubles (CCD) approach we refer to the whole self-energy as being in the ADC(3)-D approximation, when $t$ is obtained by resumming both singles and doubles (CCSD) it will give the ADC(3)-SD approximation, and so on.  In Sec.~\ref{sec:pairing_model}, we  will see a case when these corrections are important.

The working equations for the self-energy at the ADC(4) level and beyond are discussed in Ref.~\cite{ch11_Schirmer1983ADCn}.

\vskip .3 cm
\noindent
{\bf Exercise 11.4.}  Calculate the ladder and ring diagrams in Fig.~\ref{fig:3rdOrd}  and prove Eqs.~\eqref{eq:ADC3_MEC} and~\eqref{eq:ADC3_NED} in full.
[Hint: for the ring diagrams it is simpler to first perform integrations for the free polarization propagator,
$\Pi^f_{\alpha\beta,\gamma\delta}(\omega)=\int \frac{{\rm d} \, \omega_1}{2\pi i} g_{\alpha\gamma}(\omega+\omega_1)g_{\delta\beta}(\omega_1)$,
which describes  non interacting particle-hole states.]

\vskip .3 cm
\noindent
{\bf Exercise 11.5.}  In case of a reference propagator that is not fully self-consistent, it is necessary to also include non-skeleton diagrams. For $\widetilde\Sigma(\omega)$ these first appear at third order with the diagrams shown in Fig.~\ref{fig:SEins_3ndOrd}).
Calculate the expressions for  diagrams in a) and~b), then:
\begin{itemize}
\item Deduct the corresponding corrections to Eqs.~\eqref{eq:ADC3_MEC} and~\eqref{eq:ADC3_NED}. These will be the complete ADC(3) working equations.
\item Show that they cancel out exactly if the reference propagator is of Hartree-Fock type. Hence these corrections do not need to
be taken into account even tough the Hartree-Fock reference state is {\em not} a dressed---and fully self-consistent---input in this case.
\end{itemize}
[Hint: In Hartee-Fock theory, the static self-energy $\Sigma^{(\infty)}$ reduces to the Hartree-Fock potential. The reference state in this case is given by $\widehat{H}_0=\widehat{T}+\widehat{U}^{HF}\equiv\widehat{H}^{HF}$, which is also the Hartree-Fock Hamiltonian. Additionally, in the notation of Eqs.~\eqref{eq:Z_ampl} below,
the (orthogonal) single particle wave functions are the solutions of $\{T + \Sigma^{HF}\} \pazocal{Z}^i = \varepsilon^i \pazocal{Z}^i$.]

\subsection{Solving the Dyson equation}
\label{sec:DysonDiag}

 Once we have  a suitable approximation to the self-energy, it is necessary to solve the Dyson equation~\eqref{eq:Dyson}
 to obtain the single particle propagator, the  associated  observables and the  spectral function. The latter will also yield spectroscopic amplitudes and their spectroscopic factor for the addition and removal of a nucleon form the correlated state $|\Psi^A_0\rangle$.  In doing this, Eqs.~\eqref{eq:Dyson} take the form of a one-body Schr\"odinger equation for the scattering of a particle or a hole inside the medium. Given that all the Cauchy integrals associated with Feynman diagrams have been carried out, we can safely take the limit $\pm i \eta \rightarrow 0$ in all denominators for simplicity. The same equation applies to states both above and below the Fermi surface.
 Thus, it is convenient to take a general index $i$ and using $\varepsilon_i$ and $\pazocal Z^i$ to label energies and spectroscopic amplitudes for all quasiparticle and quasihole states. Specifically,
 \begin{equation}
\varepsilon_i \longrightarrow \left\{
\begin{array}{lcl}
\varepsilon_n^+ & \quad & \hbox{for $i$=$n$, particle,}  \\ ~ \\
\varepsilon_k^- &  & \hbox{for $i$=$k$, hole,}
\end{array} \right.
\qquad \hbox{and} \qquad
\pazocal Z^i_\alpha  \longrightarrow \left\{
\begin{array}{lcl}
(\pazocal X^n_\alpha)^* & \quad & \hbox{for $i$=$n$, particle,} \\ ~ \\
 \pazocal Y^k_\alpha &  & \hbox{for $i$=$k$, hole.}
\end{array} \right.
\label{eq:Z_ampl}
\end{equation}

In order to extract the solution for the pole $i$ in the Lehmann representation, we extract the corresponding residue on both the left and right hand side of Eq.~\eqref{eq:Dyson_a}:
 \begin{equation}
   \lim _{\omega \rightarrow \varepsilon_i}  \; (\omega - \varepsilon_i)
   \left\{
     g_{\alpha\beta}(\omega) = g^{(0)}_{\alpha\beta}(\omega) + g^{(0)}_{\alpha\gamma}(\omega)  \Sigma^\star_{\gamma\delta}(\omega) g_{\delta\beta}(\omega)
   \right\} \; ,
 \end{equation}
which gives
 \begin{equation}
    \pazocal Z^i_\alpha (\pazocal Z^i_\beta)^*  =  \left. g^{(0)}_{\alpha\gamma}(\omega)  \Sigma^\star_{\gamma\delta}(\omega) \pazocal Z^i_\delta (\pazocal Z^i_\beta)^*
    \right|_{\omega = \varepsilon_i}
 \; .
 \end{equation}
By dividing out $(\pazocal Z^i_\beta)^*$ and using the fact that $[g^{(0)}(\omega)]^{-1} = \omega - \widehat{H}_0$ 
we finally obtain the  eigenvalue equation
 \begin{eqnarray}
   \varepsilon_i   \pazocal Z^i_\alpha &=&   \left. \left\{ \widehat{T} + \widehat{U} +   \Sigma^\star(\omega)   \right\}_{\alpha \, \delta}
  \pazocal Z^i_\delta  \right|_{\omega = \varepsilon_i}
  \nonumber \\
  &=&  \left. \left\{  \widehat{T}  +  \Sigma^{(\infty)} +M^\dagger\frac1{\omega - E^> -C + i \eta}M +N\frac1{\omega - E^< -D - i \eta}N^\dagger     \right\}_{\alpha \, \delta}
  \pazocal Z^i_\delta  \right|_{\omega = \varepsilon_i}  \; ,
\label{eq:DysSchrod}
 \end{eqnarray}
where the potential $\widehat{U}$ defining the unperturbed state completely cancels out. From here we see that the true irreducible self-energy $\Sigma^{(\infty)}+\widetilde\Sigma(\omega)$
acts as a non-local and energy dependent potential that accounts for the motion of both particles and holes inside the system and for their
coupling intermediate excitations.
At positive energies ($\omega > 0$) this equation describes the elastic scattering of a nucleon off the $|\Psi^A_0\rangle$ ground
state and the self-energy can be identified with a fully microscopic optical potential~\cite{ch11_Capuzzi1996,ch11_Cederbaum2001,ch11_Barbieri2005}.
In this case the spectroscopic amplitudes $\pazocal Z^i$ correspond to scattering wave functions with the usual asymptotic
normalization.
Instead, at $\omega < 0$, Eq.~\eqref{eq:DysSchrod}  describes the transition to states of $|\Psi^{A\pm1}_i\rangle$ with bound amplitudes.
The norm of each $\pazocal Z^i$ gives the corresponding spectroscopic factor and it is obtained as
\begin{equation}
 SF_i = \sum_\alpha |\pazocal Z^i_\alpha|^2 =  \frac 1 {1 - (\overline{\pazocal Z}^i_\beta)^*
   \left. \frac{d \, \Sigma^\star_{\beta \gamma}(\omega)}{d \, \omega} \right|_{\omega = \varepsilon_i}
    \overline{\pazocal Z}^i_\gamma}   \; ,
\label{eq:SFnorm1}
\end{equation}
where $\overline{\pazocal Z}^i \equiv {\pazocal Z}^i / \sqrt{SF_i}$ is the spectroscopic amplitude normalized to 1.

Equations~\eqref{eq:DysSchrod} and~\eqref{eq:SFnorm1} are the central equations of the Green's function formalism and show how the
single-particle propagator is the solution of an effective one-body Schr\"odinger equation for a nucleon or a
hole propagating inside the correlated system. The energy dependence of  $\Sigma^\star(\omega)$  and its non-locality are a consequence of the underlying  many-body dynamics. Eq.~\eqref{eq:SFnorm1} also shows that the reduction of spectral strength commonly observed in correlated systems arises from the dispersion
properties of the self-energy.

 In spite of its beauty, Eq.~\eqref{eq:DysSchrod} is also the worst starting point to solve the Dyson equation in a discretized finite basis. Unless one is interested in just a few solutions near the Fermi surface or the model space is extremely small, this approach will require high computational times due to the large amounts of diagonalizations required to extract the correct eigenvalues. The reason is that root-finding algorithms are needed to match the eigenvalues $\varepsilon_i$ with the argument of  $\Sigma^\star(\varepsilon_i)$, but simple searching algorithms may miss a large amount of solutions. The consequences of missing a large portion of spectral strength are that wrong results would be obtained for the ground state observables computed as in Sec.~\ref{sec:scgf_obs}. This can also deteriorate the self-consistency already at the level of the static self-energy, $\Sigma^{(\infty)}=\widetilde{U}$.  If Eq.~\eqref{eq:DysSchrod} must be used, it is possible to gather all the necessary solutions by starting from extremely fine energy meshes to be sure that all eigenvalues are bracketed first. However, this easily becomes suicidal in terms of the increase of computing time.
We discuss here a different approach that is not affected by these problems and that will also give some further insight into the physics content of the Dyson equation.

First, for each solution of the Dyson equation we define two new vectors $\pazocal{W}^i$ and  $\pazocal{V}^i$ which live in the ISCs space as follows:
\begin{align}
 [\omega - E^> -C ]_{r,r'}  \, \pazocal{W}^i_{r'}  ~\equiv{}&~
M_{r,\delta}   Z^i_\delta  \; ,
 \nonumber \\
  [\omega - E^< -D ]_{s,s'} \, \pazocal{V}^i_{s'} ~\equiv{}&~
   N^\dagger_{s,\delta}    Z^i_\delta \; ,
 \label{eq:defWV}
\end{align}
where we have let $i\eta\rightarrow 0$ as this is no longer needed in a finite and discretized basis. With these definitions, Eq.~\eqref{eq:DysSchrod}  is easily rearranged into a single eigenvalue problem of larger dimensions but where the corresponding matrix is energy independent:
\begin{equation}
\left( \begin{array}{ccccc}
 \widehat{T} + \Sigma^{(\infty)}  &~&   M^\dagger   &~&  N~  \\
&&\\
    M   &&  E^>+C  && \\
    &&\\
    N^\dagger    &&      &&  E^<+D
\end{array} \right)
\left( \begin{array}{c}
\pazocal{Z}^i \\ ~ \\ \pazocal{W}^i \\~ \\ \pazocal{V}^i
\end{array} \right)
=\left( \begin{array}{c}
  \pazocal{Z}^i \\ ~\\ \pazocal{W}^i \\~\\ \pazocal{V}^i
\end{array} \right)
  \varepsilon_i
\label{eq:DysMtx}
\end{equation}
and the normalization condition~\eqref{eq:SFnorm1} becomes
\begin{equation}
\sum_\alpha  |\pazocal Z^i_\alpha|^2 + \sum_r  |\pazocal W^i_r|^2 + \sum_s  |\pazocal V^i_s|^2 = 1   \; .
\label{eq:SFnorm2}
\end{equation}

The advantage of this approach  is that it linearizes the Dyson equation and yields all solutions in one single diagonalization. Although the dimension of the Dyson matrix in Eq.~\eqref{eq:DysMtx} is much larger than a one-body Schr\"odinger problem and that it requires a substantial amount of memory storage, it typically provides the full spectral strength 100 times faster than using Eq.~\eqref{eq:DysSchrod} directly. Furthermore, it is possible to reduce the dimensionality of the eigenvalue problem by projecting matrices $[E^>+C]$ and $[E^<+D]$  (separately!) onto smaller Lanczos/Krylov subspaces~\cite{ch11_Schirmer1989BlkLanc,ch11_Soma2014Lanc}. In this way one reduces the number of poles of $g(\omega)$ far away from the Fermi surface---where only their average is physically meaningful---but conserves the overall strength needed to compute ground state observables.

Eq.~\eqref{eq:DysMtx} also puts in evidence how the Dyson equation is very closely related to a configuration interaction (CI) approach. For solutions ($\varepsilon^+_n$,$\pazocal{X}^n$) in the single particle spectrum,  the eigenstates of $|\Psi^{A+1}_n\rangle$ are expanded in terms of 1p configurations (from the $\widehat{T} + \Sigma^{(\infty)}$ sector) and 2p1h or larger configurations, which is evident from the matrix $C$, in Eq.~\eqref{eq:ADC3_C}. However, additional 2h1p configurations are included through matrix $D$. This is in spirit very similar to how ground state correlations are included in the random phase approximation approach~\cite{ch11_RingSchuck}. Furthermore,  the  matrices that couples these subspaces are the same as in CI only at first order ($M^{(1)}$ and $N^{(1)}$). The eigenstates of Eq.~\eqref{eq:DysMtx} will approach the exact solution as the approximation of the self-energy is systematically improved. Similarly, the propagation of hole states that correspond to the eigenstates of $|\Psi^{A-1}_k\rangle$ are obtained in a CI fashion.
Eq.~\eqref{eq:SFnorm2} is then the natural normalization condition for the CI expansion and shows that the spectroscopic amplitudes are the projection of more complex many-body wave functions onto a single-particle space.

\vskip .3 cm
\noindent
{\bf Exercise 11.6.}  Perform a Taylor expansion of the propagator $g(\omega)$ at zeroth order around a given pole $\varepsilon_i^\pm$. Then, use this and the
conjugate Dyson equation~\eqref{eq:Dyson_b} to obtain the normalization condition for spectroscopic factors given in  Eq.~\eqref{eq:SFnorm1}.

\vskip .3 cm
\noindent
{\bf Exercise 11.7.}  Based on the definitions of vectors $\pazocal{W}^i$ and $\pazocal{V}^i$, Eqs.~\eqref{eq:defWV}, show that~\eqref{eq:SFnorm1} and~\eqref{eq:SFnorm2} are equivalent.

\subsection{A simple pairing model}
\label{sec:pairing_model}

\begin{figure}[ht]
\begin{center}
\includegraphics[width=0.8\textwidth]{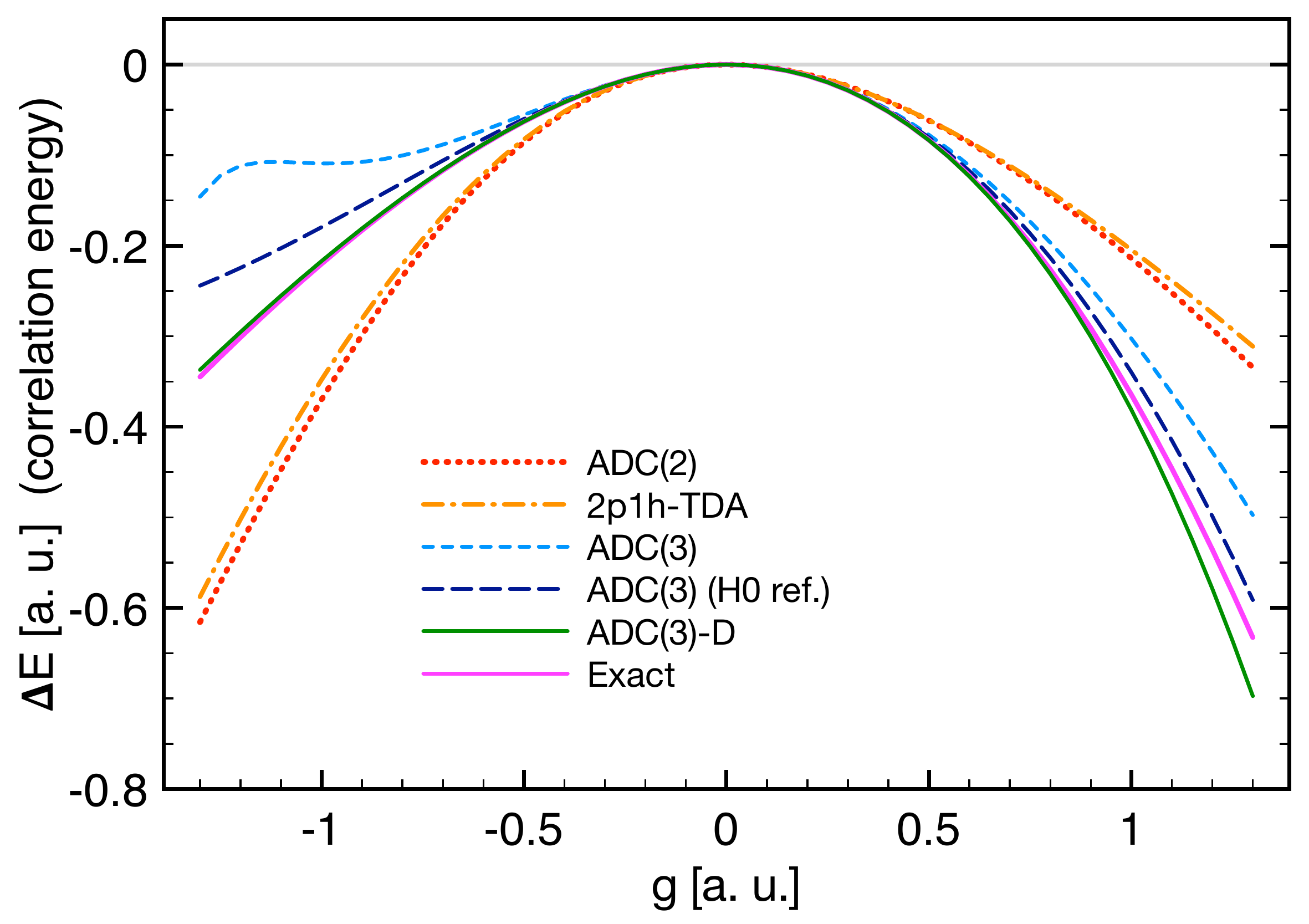}
\caption{Correlation energy for the pairing Hamiltonian of Eq.~\eqref{eq:H_pair} as a function of the coupling $g$, obtained for different  ADC($n$) approximations to the self-energy and in the {\em sc0} scheme. The dotted, dot-dashed, short dashed and full lines are all obtained from the HF reference of Eq.~\eqref{eq:pair_gHF} and show successive approximations of the ADC($n$) hierarchy [respectively: ADC(2), 2p1h-TDA, ADC(3) and ADC(3)-D]. The long dashed line is the same ADC(3) truncation but based on the unperturbed reference propagator of Eq.~\eqref{eq:pair_h0}. The purple line shows the exact result calculated  from a full configuration interaction diagonalization.  }
\label{fig:pairing_adc}
\end{center}
\end{figure}

\begin{figure}[ht]
\begin{center}
\includegraphics[width=0.8\textwidth]{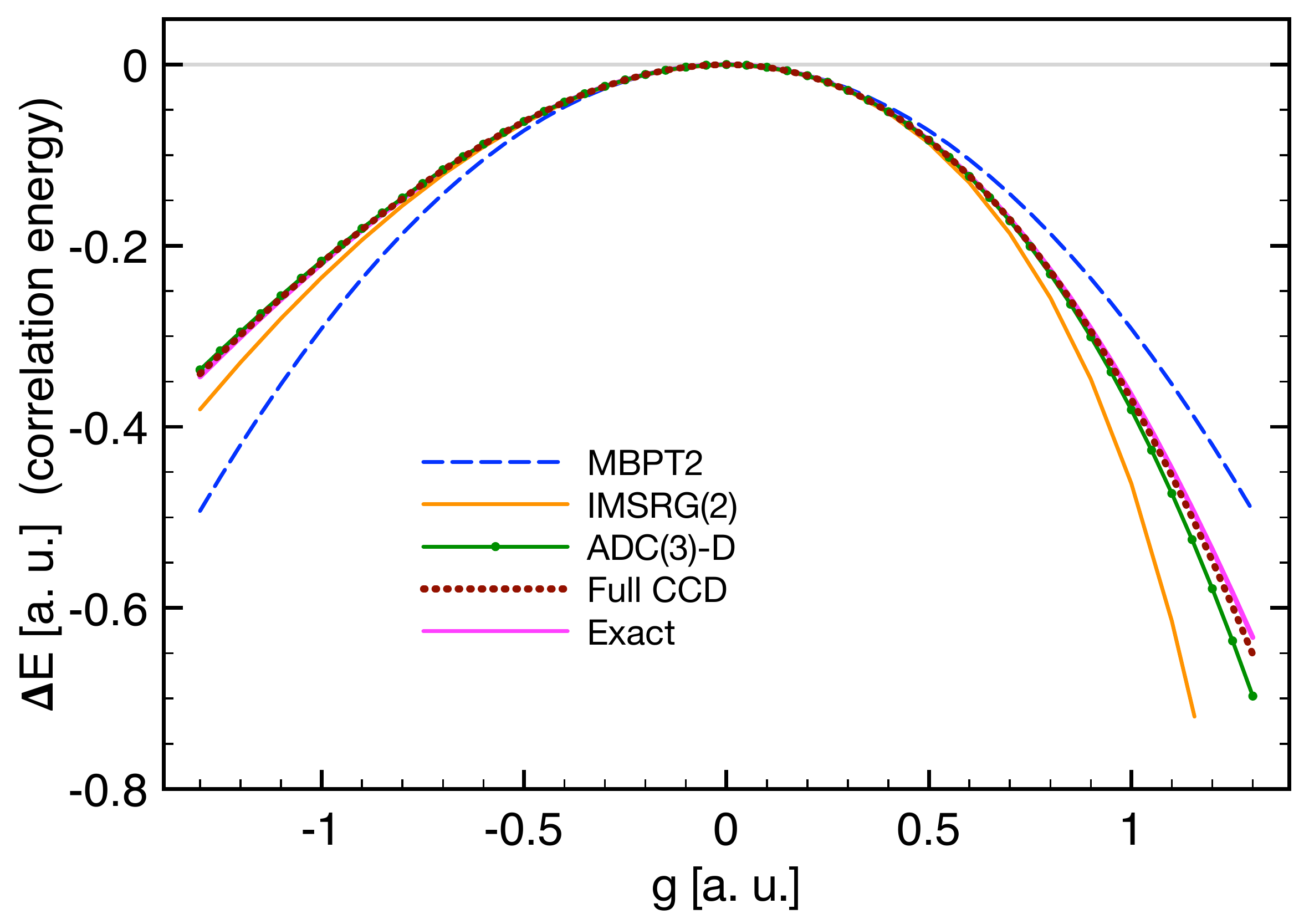}
\caption{Correlation energy for the pairing Hamiltonian of Eq.~\eqref{eq:H_pair} as a function of the coupling $g$,  for  different
many-body methods discussed in this book. The purple line is the exact results from  configuration interaction theory. 
The results for  second-order perturbation theory (MBPT2), for IMSRG(2), for the  CC-corrected ADC(3)-D and  for the standard CC with doubles (CCD)  are compared. See also Section 10.3.6 of Ref.~\cite{chapter10} for higher truncations of MBPT.}
\label{fig:pairing_all}
\end{center}
\end{figure}

As a first demonstration  of the ADC formalism, we consider the pairing Hamiltonian already discussed in chapter~8. This is a system of  four spin-1/2 fermions in a 4-level model space that interact through a pairing force:
  \begin{align}
   \widehat{H} ={}& \widehat{H}_0 +  \widehat{V} = \xi \sum_{p=1}^4 ~ \sum_{\sigma=+, -} (p-1) a^{\dagger}_{p \sigma} a_{p \sigma}
 ~-~ \frac{g}{2} \sum_{p, q=1}^4 a^{\dagger}_{p+}a^{\dagger}_{p-}  a_{q-}a_{q+} \; .
\label{eq:H_pair}
\end{align}

In spite of its simplicity,  this model poses a particularly difficult test for many-body approximations based on ISRs because the Hamiltonian~\eqref{eq:H_pair} does not allow for admixtures of leading order excitations, that is of the particle-hole type. The ground state contains only 2p2h and higher excitations. Correspondingly, the pairing interactions $\widehat{V}$ cannot couple particle states to 2p1h configurations, neither hole states with  2h1p ones.  This is obvious looking at  the  leading terms, Eqs.~\eqref{eq:ADC2_M} and~\eqref{eq:ADC2_N}, that would involve interactions between a particle and a hole (which  cannot be connected by pairing)  but it applies to the full ADC(3) couplings~\eqref{eq:ADC3_M} and~\eqref{eq:ADC3_N} as well. It follows  that  the spectra for  particle attachment and removal are dominated  by 3p2h and 3h2p ISCs. These are partially included in the Dyson equation by couplings between particles and backward going, 2h1p, terms in the self-energy (or between holes and the forward 2p1h terms). However, a complete account of them would require many-body truncations at the ADC(4) level and higher.  Remarkably, it is still possible to reach rather accurate results as demonstrated by Figs.~\ref{fig:pairing_adc} and~\ref{fig:pairing_all}.

The unperturbed propagator, associated with the $\widehat{H}_0$ term of Eq.~\eqref{eq:H_pair}, is given by
  \begin{align}
   g^{(0)}_{p \sigma_p \, ,\, q \sigma_q}(\omega) ={}& \delta_{p q} \delta_{\sigma_p \,  \sigma_q} 
   \left\{ \sum_{n=3,4}  \;  \frac {\delta_{n \,  p} }{ \omega - \varepsilon^{(0)}_n + i\eta}
   ~+~ \sum_{k=1,2 }  \;  \frac {\delta_{k \, p} }{ \omega - \varepsilon^{(0)}_k - i\eta}
\right\}  
\label{eq:pair_h0}
\end{align}
where $\varepsilon^{(0)}_p = \xi (p-1)$ are the unperturbed single particle energies and the gap at the Fermi surface is \hbox{$E^{(0)}_{ph} = \varepsilon^{(0)}_3 - \varepsilon^{(0)}_2 = \xi$}.  
For this particular model, the unperturbed state is also the same state that solves the HF equations. Thus, the HF propagator is written exactly in the
same way but with only a shift in the single particle energies of the hole states (see also Section 8.7.4 and Tab.8.11 of Ref.~\cite{chapter8}):
  \begin{align}
   g^{HF}_{p \sigma_p \, ,\, q \sigma_q}(\omega) ={}& \delta_{p q} \delta_{\sigma_p \,  \sigma_q} 
   \left\{ \sum_{n=3, 4}  \;  \frac {\delta_{n \,  p} }{ \omega - \varepsilon^{HF}_n + i\eta}
   ~+~ \sum_{k=1, 2}  \;  \frac {\delta_{k \, p} }{ \omega - \varepsilon^{HF}_k - i\eta}
\right\}  
\label{eq:pair_gHF}
\end{align}
where
  \begin{align}
  \varepsilon^{HF}_p ={}& \left\{ 
  \begin{array}{lcl}
    \xi (p-1) \; , &   \qquad & \hbox{for~} p=3,4 \\
    ~\\
    \xi (p-1) - g/2 \; , &   \qquad & \hbox{for~} p=1,2    
  \end{array}
  \right.
  \label{eq:pair_ehf}
\end{align}
and the particle-hole gap now depends on the coupling constant, \hbox{$E^{HF}_{ph} = \varepsilon^{HF}_3 - \varepsilon^{HF}_2 = \xi + g/2$}.  One may chose either  of these propagators as the reference state for calculating the ADC($n$) self-energy. However,  $g^{(0)}(\omega)$ will also require  additional corrections terms for the interactions matrices $C$ and $D$, as seen in {\bf Exercise 11.5}.  In practice, these corrections are already included in the shifts of Eq.~\eqref{eq:pair_ehf} and the HF reference is normally a better starting point for calculating the self-energy.

We now set $\xi=1$ and  perform calculations at different levels of approximations in the ADC($n$) approach, by using the $g^{HF}(\omega)$ as reference (except when indicated)  and by calculating $\Sigma^{(\infty)}$ self-consistently in the {\em sc0} scheme.  After solving the Dyson equation, we extract the ground state energy from the Koltun sum rule~\eqref{eq:Koltun_hW}  and calculate the correlation energy $\Delta E=E_{g.s.} - (2\xi - g)$. 
 The result of the ADC(2) equations~\eqref{eq:ADC2_MEC} and~\eqref{eq:ADC2_NED}  is shown by the dotted line in Fig.~\ref{fig:pairing_adc}. 
 The 2p1h-TDA approximation improves upon this by  using  the interaction matrices from Eqs.~\eqref{eq:ADC3_C} and~\eqref{eq:ADC3_D},
 which resums infinite ladders of 2p and 2h states. However, this brings only a very small improvement to this system.
The ADC(3) approximation gives better results and it is shown for  both the $g^{(0)}(\omega)$ and $g^{HF}(\omega)$ choices of the reference state  with long dashed and short dashed lines, respectively.  Remarkably,  these results depend strongly on the reference state and  are much closer to the exact solution for the $g^{(0)}(\omega)$ case, which would have been expected to be a poorer choice. Furthermore, $g^{HF}(\omega)$ behaves erratically for negative values of $g$, corresponding to a  repulsive pairing interaction $\widehat{V}$.  These two calculations differ only in the single particle energies used to calculate the 
coupling matrices $M$ and $N$. Such behavior is simply explained by the dependence of $E^{HF}_{ph}$ on $g$, which can make the denominator in Eqs.~\eqref{eq:ADC3_C} and~\eqref{eq:ADC3_D}  very small and causes the breakdown of the perturbation expansion~\eqref{eq:MNexp}.
To resolved his problem we substitute the $t^{(0)}$ of Eq.~\eqref{eq:dressed_t2} with the converged solution from the CCD equations. The resulting ADC(3)-D is now completely independent of the choice between the two reference states and it also reproduces the exact result closely. This is shown by the two solid lines in Fig.~\ref{fig:pairing_adc}.

Fig.~\ref{fig:pairing_all}  compares the ADC approach with CC,  in-medium similarity renormalization group~(\hbox{IMSRG})  and second-order perturbation theory.  
The ADC(3)-D, the two-body truncation of  IMSRG~(\hbox{IMSRG(2)}, see chapter 10)) and the CC methods perform similarly at $g<0$, where they are all close to the exact solution all the way to $g\approx-1.3$. For smaller values of the coupling the CCD iterations stop converging. At large positive values of $g$ (corresponding to attractive pairing) the various approaches eventually deviate from the exact result but with CCD being slightly better. 
Clearly the full ADC(3)-D is a more complex calculation than CCD but leads to  similar results for the binding energy. On the other hand, this does not only yield the ground state energy but also the whole spectral function for the addition and removal of a particle is generated when solving the Dyson equation~\eqref{eq:DysSchrod} or~\eqref{eq:DysMtx}.  The next section will demonstrate examples of the self-energy and the spectral distribution obtained when calculating the single particle propagator.

The FORTRAN code that generated these results is available online at  {\sloppy  \url{https://github.com/ManyBodyPhysics/LectureNotesPhysics/blob/master/doc/src/Chapter11-programs/Pair_Model}}. We do not examine this code here but we will give a detailed discussion of how to structure a complex ADC($n$) code for infinite matter computations in the next section.

\section{Numerical solutions for infinite matter}
\label{sec:scgf_comp}

In this section we discuss how to implement the calculation of the self-energy and the single particle propagator
in the ADC($n$) formalism. We will demonstrate this for the case of infinite nucleonic matter and use our
results to discuss  general  features of the spectral function.
A general code that can solve for both symmetric and pure neutron matter up to ADC(3) is provided with  this chapter at the {\sloppy
URL  \url{https://github.com/ManyBodyPhysics/LectureNotesPhysics/blob/master/doc/src/Chapter11-programs/Inf_Matter}. } 
We will use the  C++ programming language and will refer to this code for describing the technical details of the implementation.
We then show results based on the Minnesota nuclear potential from Ref.~\cite{ch11_minnesota}. This is a very simplified model of the nuclear interaction that allows for an easy implementation. On the other hand, it still retains some physical properties of the nuclear Hamiltonian that will allow us to discuss the basic features of the spectral function of  nucleonic matter (and of infinite fermionic systems in general).  The reader interested in these physics aspects could refer directly to Sec.~\ref{sec:scgf_comp_results}.

\subsection{Computational details for \texorpdfstring{ADC($n$)}{ADC(n)}}
\label{sec:scgf_comp_code}

The first fundamental step  to set up a SCGF computation is the choice of the model space. For
infinite matter, translational invariance imposes that the Dyson equation is diagonal in momentum  and
therefore it becomes much easier to solve the problem in momentum space. However, there remain two possible
choices for how to encode single particle degrees of freedom.
The first one is to subdivide the infinite space in  boxes of finite size  and to impose periodic boundary conditions
(see also chapter 8). In this way, the number of fermions included in each box is finite and determined by the particle
density of the system. The resulting model space is naturally expressed by a set of discretized single particle
states and one solves the working equations in the form of Eqs.~\eqref{eq:ADC3_MEC}, \eqref{eq:ADC3_NED} and \eqref{eq:DysMtx}.
This path requires  the same technical steps needed to calculate  finite systems in a box. 
Numerical results then need to be converged with respect to the truncation of the k-space (and, for an infinite system, with respect to the
number of  nucleons inside each periodic box).   We will follow this approach for the present
computational project.
The other approach is to retain the full momentum space and write the SCGF equations already in the full
thermodynamic limit. This choice is best suited to solve the Dyson equation at finite temperatures and
in a full SCGF fashion and will be discussed further in Sec.~\ref{sec:scgf_finiteT}. 

\vskip 0.5 cm
{\bf Construction of the model space.}
For simplicity, we assume a total number $A$ of nucleons in each (cubic) periodic box. For 
boxes of  length $L$, the density and the Fermi momentum are expressed, respectively as ($\hbar$=1):
\begin{align}
  \rho = \frac{A}{L}  \qquad  \qquad  & \hbox{and}  \qquad  \qquad  p_F =  \sqrt[\leftroot{-1}\uproot{2}\scriptstyle 3]{ \frac{6 \pi^2 \rho}{\nu_d} } \, ,
  \label{eq:pf_vs_rho}
\end{align}
where the degeneracy $\nu_d$ is twice the  number of different  spin-$\frac 1 2$ fermions
and the basis states are defined by the cartesian quantum numbers $n_x$, $n_y$, $n_z$=\,0, 1, 2...  with momentum
\begin{align}
  {\bf p} =  \frac{2 \pi}{L}  \left(  \begin{array}{c}  n_x \\n_y  \\ n_z   \end{array} \right) \, .
\end{align}
The kinetic energies, and hence the unperturbed single particle energies, will depend \hbox{on $|{\bf p}|^2$} and hence the values
of $N_{sq}=n_x^2+n_y^2+n_z^2$ define a set of separate shells. Since we need closed shell reference states, only certain
values for the number of nucleons in each box,  $A$, are possible. The size of the model space is given by 
$N_{sq}^{\rm max} = \max \{n_x^2+n_y^2+n_z^2\}$.
The construction of the single particle model space is then straightforward. We will do it constructing a specific class 
with pointers to arrays for each relevant quantum number and additional arrays for the kinetic energies or any
other useful quantity associated with each state.

\lstset{language=c++}
\begin{lstlisting}
class SpBasisK {

  public:
    int SpNmax, SpNAlloc;      // total number of s.p. states and allocated space
    int *nx, *ny, *nz, *spin;  // quantum numbers
    double *e_kin;  // kinetic energy
    
    double Lbox;  // side length of the periodic box 
    int N_holes;   // number of nucleons in a box (# of occupied states)  
    
    // grouping s.p. states of equal symmetry
    int N_grps;  //  number of different groups
    int *gr_mlt, *gr_rep;
    
    // functions
  public:
    void Build_sp_basis(int, double, int);
    
    int Build_groups_table(void );
    
  };
\end{lstlisting}

The constructor for the model space will be necessary to order the basis with increasing values of $N_{sq}$, so that 
the orbits corresponding to the $A$ hole states come first. This becomes useful later to construct ISCs.
We first count the total number of possible $(n_x, n_y, n_z)$ configurations. Once it is known how many single particle $\vec k$ states there are, we can allocate arrays in memory to store the relevant quantum numbers of each of them:

\lstset{language=c++}
\begin{lstlisting}
const double PI          = 3.141592653589793; 
const double hbarc       = 197.326968;  // [MeV*fm]
const double NUCLEONmass = 939.565;     // [MeV]

void SpBasisK::Build_sp_basis(int Nsq_max, double Lbox, int A) {

  int imax = int( sqrt(double Nsq_max) + 1 ); // max value of |n_x|, |n_y| or |n_z|
  
  int i_count = 0;  // counts the number of basis states:
  for (int ix=-imax; ix<=imax; ++ix)
    for (int iy=-imax; iy<=imax; ++iy)
      for (int iz=-imax; iz<=imax; ++iz)
        if (ix*ix + iy*iy + iz*iz <= Nsq_max) ++i_count;
  
  
  SpNAlloc = 2 * i_count; //  2 is the spin-1/2 degeneracy; we assume PNM here
  
  cout << "\n Allocating space for "<< SpNAlloc << " sp states... \n";
  
  nx    = new int[SpNAlloc];  // Allocate basis' arrays
  ny    = new int[SpNAlloc];
  nz    = new int[SpNAlloc];
  spin  = new int[SpNAlloc];
  e_kin = new double[SpNAlloc];
  
  double xek;
  
  cout << "\n Single particle basis:\n ----------------------";
  cout << "\n  orbit    n_x   n_y   n_z    Nsq     E_kin\n";
  
  i_count = 0;
  for (int isq=0; isq<=Nsq_max; ++isq) {
    for (int ix=-imax; ix<=imax; ++ix)
      for (int iy=-imax; iy<=imax; ++iy)
        for (int iz=-imax; iz<=imax; ++iz) {
          if ((ix*ix + iy*iy + iz*iz) != isq) continue;
          
          xek = double(isq) * pow((hbarc * 2.0 * PI / Lbox),  2.0) / 2.0 / NUCLEONmass;
          cout <<i_count <<"  " <<ix <<"  " <<iy <<"  " <<iz <<"  " <<isq <<"  " <<xek <<endl;
          
          for (int is=-1; is<2; is+=2) {
            nx[i_count] = ix;
            ny[i_count] = iy;
            nz[i_count] = iz;
            spin[i_count] = is;
            e_kin[i_count] = xek;
            ++i_count;
          }
          
        }  // end of ix, iy, iz loop
  } // end of isq loop
  SpNmax = i_count;

  this->N_holes = A; // very important! Must set the # of occupied states

  return;}
\end{lstlisting}

\vskip 0.5 cm
{\bf Construction of the ISCs.}
Due to translational invariance the Dyson equation~\eqref{eq:Dyson} separates in a set of uncoupled equations
for each  values of $\{{\bf p}_i, s^i_z \}$ in the model space (where $s_z$ is the spin projection and $i$ labels
the basis states):
\begin{align}
   g({\bf p}_i, s^i_z; \omega) = g^{(0)}({\bf p}_i, s^i_z; \omega) + g^{(0)}({\bf p}_i, s^i_z; \omega) \, \Sigma^\star({\bf p}_i, s^i_z; \omega) \,  g({\bf p}_i, s^i_z; \omega) \; .
\end{align}
This diagonal equation can be formally inverted as shown in Eqs.~\eqref{eq:G_algebraic}  and~\eqref{eq:S_self_all} below. However, we will solve for all of its eigenstates instead and this is better done by diagonalizing Eq.~\eqref{eq:DysMtx}.  For each state $i$, we need to generate
tables for the relevant 2p1h and 2h1p ISCs and then calculate the elements of  the Dyson matrix. 
One can build a class whose objects are associated to  a particular orbit of the given model space  and then construct the ISCs  
in accordance with the conservation  of momentum and other symmetries of the Hamiltonian, which are implicit in the matrix elements for the coupling ($M$ and $N$) and interaction ($C$ and $D$) matrices.
Schematically, looking only at the 2p1h configurations for simplicity, this will be:
\lstset{language=c++}
\begin{lstlisting}
class ADC3BasisK {
  
  public:
    int *Bas_2p1h, *Bas_2h1p; // pointers to 2p1h/2h1p bases
    int Nbas_2p1h, Nbas_2h1p; // dimensions of the 2p1h/2h1p bases
    
    int iSpLoc; // {p,s_z} state in the s.p. basis associated with the 2p1h/2h1p 
    SpBasisK *SpBasLoc;
    
    // functions
  public:
    void Build_2p1h_basis(SpBasisK*, int );
  };

void ADC3BasisK::Build_2p1h_basis(SpBasisK *InBasis, int isp) {

  this->SpBasLoc = InBasis; // keep track of the basis and the s.p. states associated
  this->iSpLoc = isp;       // to this 2p1h ICSs, for use by other functions

  Nbas_2p1h = ... ;  // must compute the number of expected 2p1h configurations
  
  if (NULL != Bas_2p1h) delete [] Bas_2p1h;
  this->Bas_2p1h = new int[3*(Nbas_2p1h)];  // need 3 indices for each config (n1, n2, k3)
  
  int    k3_x, k3_y, k3_z, k3_sp;

  i_count = 0;
  for (int n1=SpBasLoc->N_holes; n1<SpBasLoc->SpNmax; ++n1) {
    for (int n2=n1+1; n2<SpBasLoc->SpNmax; ++n2) { // n1 < n2 due to Pauli

      // expected q.#s for 3rd index (k3), imposed by the Hamiltonian's symmetries:
      k3_x  = SpBasLoc->nx[n1]   + SpBasLoc->nx[n2]   - SpBasLoc->nx[isp];
      k3_y  = SpBasLoc->ny[n1]   + SpBasLoc->ny[n2]   - SpBasLoc->ny[isp];
      k3_z  = SpBasLoc->nz[n1]   + SpBasLoc->nz[n2]   - SpBasLoc->nz[isp];
      k3_sp = SpBasLoc->spin[n1] + SpBasLoc->spin[n2] - SpBasLoc->spin[isp];
      
      for (int k3=0; k3<SpBasLoc->N_holes; ++k3) {
        if ( (k3_x != SpBasLoc->nx[k3])  ||  (k3_y        != SpBasLoc->ny[k3]      ) ||
               (k3_z != SpBasLoc->nz[k3]) || (k3_sp != SpBasLoc->spin[k3]) ) continue;
        
        this->Bas_2p1h[3*i_count    ] = n1;
        this->Bas_2p1h[3*i_count + 1] = n2;
        this->Bas_2p1h[3*i_count + 2] = k3;
        ++i_count;

      } // end k3 loop
    } // end n2 loop
  } // end n1 loop
  if (i_count > Nbas_2p1h) {/* This is a trouble */} else {Nbas_2p1h = i_count;}

  return;}
\end{lstlisting}

\vskip 0.5 cm
{\bf Spectral representation.}
Both the propagator and the self-energy have spectral representations in terms of poles, with residues in separable form. Hence, we can devise a general class that could store both objects. Specifically, by using the conservation of spin and the fact that the propagator is diagonal in momentum space, one can write the Lehmann representation~\eqref{eq:g1Leh} as
\begin{align}
   g({\bf p}_i, s^i_z; \omega) ={}& \sum_n \frac{~~ S^{p}({\bf p}_i, s^i_z; \varepsilon^{{\bf p}_i +}_n ) ~~}{\omega - \varepsilon^{{\bf p}_i +}_n + i \Gamma}
        ~ + ~                                     \sum_k \frac{~~ S^{h}({\bf p}_i, s^i_z; \varepsilon^{{\bf p}_i -}_k ) ~~}{\omega - \varepsilon^{{\bf p}_i -}_k - i \Gamma} \; ,
\label{eq:MattK_gsp_Leh}
\end{align}
where $S^{p(h)}({\bf p}_i, s^i_z; \omega )$ are the particle and hole parts of the spectral function (see Eqs.~\eqref{eq:SpSh}). Hence, it is simpler and more
efficient to store the full residues rather than separate spectroscopic amplitudes. The self-energy can be casted in the same simple pole structure 
by diagonalizing the interactions matrices. Assuming that \hbox{$U_C \,(E^>+C) \, U_C^\dagger = \mathrm{diag}(\lambda_C^r)$} and \hbox{$U_D \, (E^<+D) \, U_D^\dagger = \mathrm{diag}(\lambda_D^q)$},
with $\lambda_{C,D}$ being the eigenvalues, we  rewrite Eq.~\eqref{eq:ADC_SE_form} as follows:
\begin{align}
   \Sigma^\star({\bf p}_i, s^i_z; \omega) ={}& \Sigma^{(\infty)}({\bf p}_i, s^i_z) 
       ~ + ~      \sum_r\frac{ | \widetilde{M}_{r\, ; \, {\bf p}_i, s^i_z} |^2 }{ ~~ \omega -  \lambda_C^r + i \Gamma ~~}
       ~ + ~      \sum_q \frac{ | \widetilde{N}_{{\bf p}_i, s^i_z \, ; \, q} |^2 }{ ~ \omega -  \lambda_D^q - i \Gamma ~} \; ,
\label{eq:MattK_Sig_Leh}
\end{align}
where $\widetilde{M} = U_C M$  and $\widetilde{N} = N U_D^\dagger$.  
A full pre-diagonalization of the interaction matrices $C$ and $D$ is not needed to construct the Dyson matrix. Thus, storing the self-energy in the form of Eq.~\eqref{eq:MattK_Sig_Leh} is worth only if  self-energy is to be calculated for specific values of its arguments (for example to plot it). However, in most cases, a reduction of these matrices through a Lanczos algorithm is still necessary to reduce the dimensionality of the problem, as discussed below here. The resulting tridiagonal matrices can be accommodated in the same structure as for the propagator by simply adding an extra array for the sub-diagonal elements.
Thus, the class for the Lehmann representation has  the following structure:
\lstset{language=c++}
\begin{lstlisting}
class SpctDist {

  public:
    SpBasisK *SpBasLoc;  // associated s.p. basis
    
    int N_LEH_ALLOC; // number of Lehmann representations to store
  
    int     *N_fw_pls,     *N_bk_pls,  *N_PLS_ALLOC;
    double **ek_fw,       **ek_bk; // - poles of the propagator/self-energy
    double **eb_fw,       **eb_bk; // - eb_xx Lanczos subdiagonal for storing self-energy
    double **Sk_fw,       **Sk_bk; // - this is the FULL residue (not the amplitude X,Y)
    double *Sig_inf;  // static self-energy

  // functions
  public:
    SpctDist(SpBasisK* ); // constructor
    int add_k_channel(int, int, double*, double*, int, double*, double*,
                            double in_Sig_inf=0.0, double *B_fw_in=NULL, double *B_bk_in=NULL);
  };

void SpctDist::SpctDist(SpBasisK *InBasis ) {
  //
  //  Use constructor to initialize the object with a table
  // of pointers for all basis states
  
  this->SpBasLoc = InBasis; // keeps track of the associated model space

  N_LEH_ALLOC = this->SpBasLoc->SpNmax;


  Sig_inf = new double[N_LEH_ALLOC];

  N_fw_pls    = new int[N_LEH_ALLOC];        N_bk_pls = new int[N_LEH_ALLOC];
  N_PLS_ALLOC = new int[N_LEH_ALLOC];
  ek_fw       = new double*[N_LEH_ALLOC];    ek_bk    = new double*[N_LEH_ALLOC];
  Sk_fw       = new double*[N_LEH_ALLOC];    Sk_bk    = new double*[N_LEH_ALLOC];
  eb_fw       = new double*[N_LEH_ALLOC];    eb_bk    = new double*[N_LEH_ALLOC];
  
  for (int isp=0; isp<N_LEH_ALLOC; ++isp) {
    Sig_inf[isp] = 0.0;
    
    N_fw_pls  [isp] = -100;    N_bk_pls  [isp] = -100;
    N_PLS_ALLOC[isp] = -100;
    ek_fw[isp]      = NULL;    ek_bk[isp]      = NULL;
    Sk_fw[isp]      = NULL;    Sk_bk[isp]      = NULL;
    eb_fw[isp]      = NULL;    eb_bk[isp]      = NULL;
  }
  
  return;}

void SpctDist::add_k_channel(int i_Leh, int N_fw_in, double *A_fw_in, double *E_fw_in,
                                        int N_bk_in, double *A_bk_in, double *E_bk_in,
                                        double in_Sig_inf /*=0.0*/,
                                        double *B_fw_in/*=NULL*/, double *B_bk_in/*=NULL*/){
  //
  //  This function is to load and store the spectral representation of a s.p. propagator
  // or a self-energy, if the additional array for the subdiagonal elements the self-energy
  // are not provided, they are set automatically to zero.

  //  Allocate memory for the basis' state  i_Leh; only one array is allocate for both hole
  // and particle poles, the xx_fw[] arrays will just point to where the particles begin
  N_PLS_ALLOC[i_Leh] = N_bk_in + N_fw_in;
  ek_bk[i_Leh] = new double[N_PLS_ALLOC[i_Leh]];  ek_fw[i_Leh] = ek_bk[i_Leh] + N_bk_in;
  eb_bk[i_Leh] = new double[N_PLS_ALLOC[i_Leh]];  eb_fw[i_Leh] = eb_bk[i_Leh] + N_bk_in;
  Sk_bk[i_Leh] = new double[N_PLS_ALLOC[i_Leh]];  Sk_fw[i_Leh] = Sk_bk[i_Leh] + N_bk_in;
  
  // store hole poles
  N_bk_pls[i_Leh] = N_bk_in;
  for (int ibk=0; ibk<N_bk_in; ++ibk) {
    ek_bk[i_Leh][ibk] = E_bk_in[ibk];
    Sk_bk[i_Leh][ibk] = A_bk_in[ibk];
    eb_bk[i_Leh][ibk] = 0.0;
    if (NULL != B_bk_in) eb_bk[i_Leh][ibk] = B_bk_in[ibk];
  }
  
  // store particle pole
  N_fw_pls[i_Leh] = N_fw_in;
  for (int ifw=0; ifw<N_fw_in; ++ifw) {
    ek_fw[i_Leh][ifw] = E_fw_in[ifw];
    Sk_fw[i_Leh][ifw] = A_fw_in[ifw];
    eb_fw[i_Leh][ifw] = 0.0;
    if (NULL != B_fw_in) eb_fw[i_Leh][ifw] = B_fw_in[ifw];
  }

  Sig_inf[i_Leh] = in_Sig_inf;  // stores the static self-energy;  == 0.0 if default
  
  return;}
   
\end{lstlisting}

\vskip 0.5 cm
The above classes simplify the calculation of quantities related to SCGF. For example, let us assume a function, {\verb Vpotential(ia,ib,ic,id) }, that returns the matrix elements  of the two-body interaction. 
The ADC(2) coupling matrix~\eqref{eq:ADC2_M} could be calculated using the following code:
\lstset{language=c++}
\begin{lstlisting}
// Configurations for s.p. state iL:
ADC3BasisK ISC2p1h(); ISC2p1h.Build_2p1h_basis(SpBasis, iL);

// Array to store the coupling matrix M:
double M_rp = new double[ISC2p1h.Nbas_2p1h];

for (int ir = 0; ir<ISC2p1h.Nbas_2p1h; ++ir) {
  // no need to loop over s.p. states since we are diagonal in the channel ia
  
  // Single particle states for the ir-th 2p1h configuration:
  im = Bas_2p1h[3*ir ];
  iv = Bas_2p1h[3*ir + 1 ];
  iL = Bas_2p1h[3*ir + 2 ];
  
  // Apply Eq. (11.28a) [a HF ref. state is assumed here... X=Y=1]
  M_rp[ir] = V_potential(im,iv,ia,iL);
}
\end{lstlisting}

Likewise, the correlated HF diagram that contributes to $\Sigma^{(\infty)}$ [second term on the right hand side of Eq.~\eqref{eq:U_eff}] could be obtained
as follows:
\lstset{language=c++}
\begin{lstlisting}
// To calculate the HF potential (V_HF) between states ia and ib we do:

double Sh, Vhf_ab;
int nHoles;
SpBasisK  *Bas = ;  // point to some object containing the model space
SpctDist  SpProp(Bas);  // sp propagator, contains spectral distribution of every (p_i,s_z)

Vhf_ab = 0.0;
for (ic = 0; ic<Bas->SpNmax; ++ic) {
  nHoles = SpProp.N_bk_poles[ic];
  Sh = 0.0;
  for (int k=0; k<nHoles; ++k) Sh += SpProp[ic].Sh[k];
    Vhf_ab += V_potential(ia,ic,ib,ic) * Sh;
    }
\end{lstlisting}

\vskip 0.5 cm
{\bf Reducing the computational load.}
  Practical applications often require rather large model spaces to achieve convergence.  This poses a major hindrance 
since the number of ISCs can grow very fast with the size of the space. The strongest constraint comes from
2p1h configurations (that is, the dimension of the $C$ matrix), which increases quadratically with the 
number of unoccupied states and linearly with the number of occupied ones. As a consequence, it is almost never possible
to attempt a fully self-consistent calculations of the dynamic self-energy because these would be based on the huge number of poles in 
Eqs.~\eqref{eq:g1Leh} or~\eqref{eq:MattK_gsp_Leh}. 
In fact, the dimensionality wall not only  prohibits going beyond a {\em sc0} calculation but the dimensions of the Dyson matrix
can become prohibitive even for a mean-field reference state and  models spaces of moderate size.

As already mentioned in Sec.~\ref{sec:DysonDiag},
the way out from this situation is to substitute the denominators in the Lehmann representation of the self-energy~\eqref{eq:MattK_Sig_Leh}
with a much smaller numbers of effective poles.  This is done by projecting the sub-matrices $E^>+C$ and $E^<+D$ onto Krylov spaces of much smaller dimensions by using a Lanczos algorithm (or Block Lanczos, in the general case when the self-energy is not diagonal in ${\bf p}_i$)~\cite{ch11_MatrixComputations}.  This approach is usually more efficient if the vectors corresponding to the columns of $M$ and $N^\dagger$ are taken as the pivots.
For example, if $\pazocal{L}$ is the \hbox{$N_{\rm red}\times N_{\rm 2p1h}$}  matrix that projects from the full space of 2p1h configurations to the Krylov space of dimension $N_{\rm red}$~($<< N_{\rm 2p1h}$), then the third term on the right hand side of Eq.~\eqref{eq:ADC_SE_form} is modified as follows:
\begin{align}
   M^\dagger \, \frac1{\omega - [E^> +C] + i \eta} \, M    {}&  \quad  \longrightarrow \quad
   M^\dagger  \pazocal{L}^\dagger \,  \frac1{\omega -  \, \pazocal{L}[E^> +C] \pazocal{L}^\dagger \, + i \eta} \, \pazocal{L} M
      \label{eq:sigma_Lanc}
\end{align}
and similarly for the 2h1p sector.  In most cases, a number of Lanczos vectors between $N_{\rm red}=$~50 and 300 is sufficient, depending on model space size and the  accuracy required.  The reason for choosing a Krylov type of projection to reduce the dimensionality of the Dyson eigenvalue problem is that this allows to preserve two  crucial properties of the spectral distribution of $\Sigma^\star(\omega)$. First, the lowest $2N_{\rm red}$ moments of the spectral distribution are conserved, which guarantees to reproduce well the average spectral function at medium and large energies. Second, the eigenvectors at the extremes of the (2p1h or the 2h1p) spectrum converge first in the Lanczos algorithm. This implies that the self-energy and the particle attachment or removal distributions converge fast to the exact one near the Fermi energy. For this reason it is crucial that both the $E^>+C$ and $E^<+D$ matrices are projected and that they are handled separately.
See eRef.~\cite{ch11_Soma2014Lanc} for details of the implementation in the SCGF approach.

In addition to the dimensions problem, one also needs to diagonalize Eq.~\eqref{eq:DysMtx} for each separate channel (${\bf p}_i, \, s^i_z$) in the basis.
On the other hand, some single particle states are equivalent. For example, the momentum states with $n_x$=3, $n_y$=2 and $n_z$=1 is the same
as $n_x$=2, $n_y$=-3 and $n_z$=1 except for a rotation around the $z$-axis. Likewise, $n_x$=3, $n_y$=2 and $n_z$=-1 differs only by a parity inversion.
The diagonalization of each of these channel would yield exactly the same results and needs to be performed only once.  The obvious procedure is that of grouping the model space states according to the same symmetries of the Hamiltonian. In this way, Eq.~\eqref{eq:DysMtx} is typically solved a few tens of times even when the model space is two orders of magnitude larger. 
For an Hamiltonian that is invariant under rotation, parity inversion and spin flipping, the algorithm to separate the basis in groups of the same symmetry is as follows:
\lstset{language=c++}
\begin{lstlisting}
int SpBasisK::Build_groups_table(void ) {
  
  int AbsN_mx = ...  // Maximum absolute value of n_x, n_y or n_z
  
  int N_ALLOC_GRPS = ...  //Max number of different groups expected
  
  gr_rep = new int[N_ALLOC_GRPS]; // for each group, keep track of a representative state
  gr_mlt = new int[N_ALLOC_GRPS]; // number of basis states belonging to a group
  
  int  i_mult, i_rep, n1, n2, n3, itmp;
  
  
  int count=0;
  for (int i1=0; i1<=AbsN_mx; ++i1)
    for (int i2=i1; i2<=AbsN_mx; ++i2)
      for (int i3=i2; i3<=AbsN_mx; ++i3) {
        
        i_mult = 0;
        i_rep = -100;
        for (int isp=0; isp<this->SpNmax; ++isp) {
          
          n1 = abs(nx[isp]);   n2 = abs(ny[isp]);   n3 = abs(nz[isp]);
          
          if (n1 > n2) {itmp=n1; n1=n2; n2=itmp;}  //  order the q.#s of the orbit isp in
          if (n1 > n3) {itmp=n1; n1=n3; n3=itmp;}  // increasing values, according 
          if (n2 > n3) {itmp=n2; n2=n3; n3=itmp;}  // to i1 < i2 < i3
          
          if ((n1==i1) && (n2==i2) && (n3==i3)) {
            ++i_mult;
            if (i_rep < 0) i_rep = isp;
          }
          
        }  // end loop over isp
        
        if (i_rep >= 0) {
          gr_mlt[count] = i_mult;
          gr_rep[count] = i_rep;
          ++count;
        }
        
      }
  
  this->N_grps = count;
  
  cout << "\n\n A total of " << N_grps << " independent groups of single particle basis \n";
  cout <<        "states has been found. All states within one group are equivalent \n";
  cout <<        "under rotation, spin and/or parity inversion.\n";
  
  return N_grps;}
\end{lstlisting}

\subsection{Spectral function in pure neutron and symmetric nuclear matter}
\label{sec:scgf_comp_results}

We test the ADC approach for pure neutron matter (PNM) and symmetric nuclear matter (SNM) using the Minnesota nuclear force~\cite{ch11_minnesota}. This is a simple semi-realistic potential that contains only central terms, for different spin and isospin, but no tensor force. It has often been used in structure studies of light neutron-rich nuclei, although it  fails to  predict any saturation of infinite nuclear matter up to very high densities.  Nevertheless, it is a good toy model for describing certain salient features of nucleonic matter and of quantum liquids in general.
In pure neutron matter, we computed  A=N=66 neutrons in a model space truncated at $N_{sq}^{\rm max}$=36, which is enough to converge the total energy per particle. For symmetric nuclear matter, we fill the same unperturbed orbits with Z=66 protons and N=66 neutron. Thus, we have a total of A=132 nucleons and truncate the model space at $N_{sq}^{\rm max}$=26. This requires up to 30~Gb of memory but it 
is still small enough to be computed on a high-end desktop.  In both cases,  the Dyson equation is solved for each value of the momentum ${\bf p}_i$ as discussed above. We retained $N_{\rm red}$=300 Lanczos vectors in every channel, which is even more than necessary for converging the binding energies and spectral functions with respect to the Krylov projection.

Total energies per particle are shown in Fig.~\ref{fig:minn_adc_eos}, for the reference state (which is HF) and for different approximations that show the convergence with respect to the many-body truncation: in order ADC(2), 2p1h-TDA and ADC(3).
These plots already demonstrate one general feature of infinite nucleonic matter: PNM is relatively weakly correlated and may allow for solutions in MBPT, while SNM is more correlated and  requires more sophisticated  all-orders methods. The correlations energy with respect to the HF reference, $E_{\rm corr.}=Eg.s. - E_{HF}$, varies between 0.5 and 2~MeV for neutrons but it is twice as much ($\approx$4 MeV) for symmetric matter and independent of the density (note the different scales in the two panels). 
Furthermore, the ADC(2) energies for PNM are already very close to the full ADC(3) results, showing that the calculation is extremely well converged. In SNM, the situation is different and  truncations beyond the second  order contribute to  the calculated  correlation energy.  The difference between 2p1h-TDA and the ADC(3) is always about 300~keV/A and the trend shows convergence with respect to the many-body truncation.

\begin{figure}[t]
\begin{center}
\includegraphics[width=0.46\textwidth]{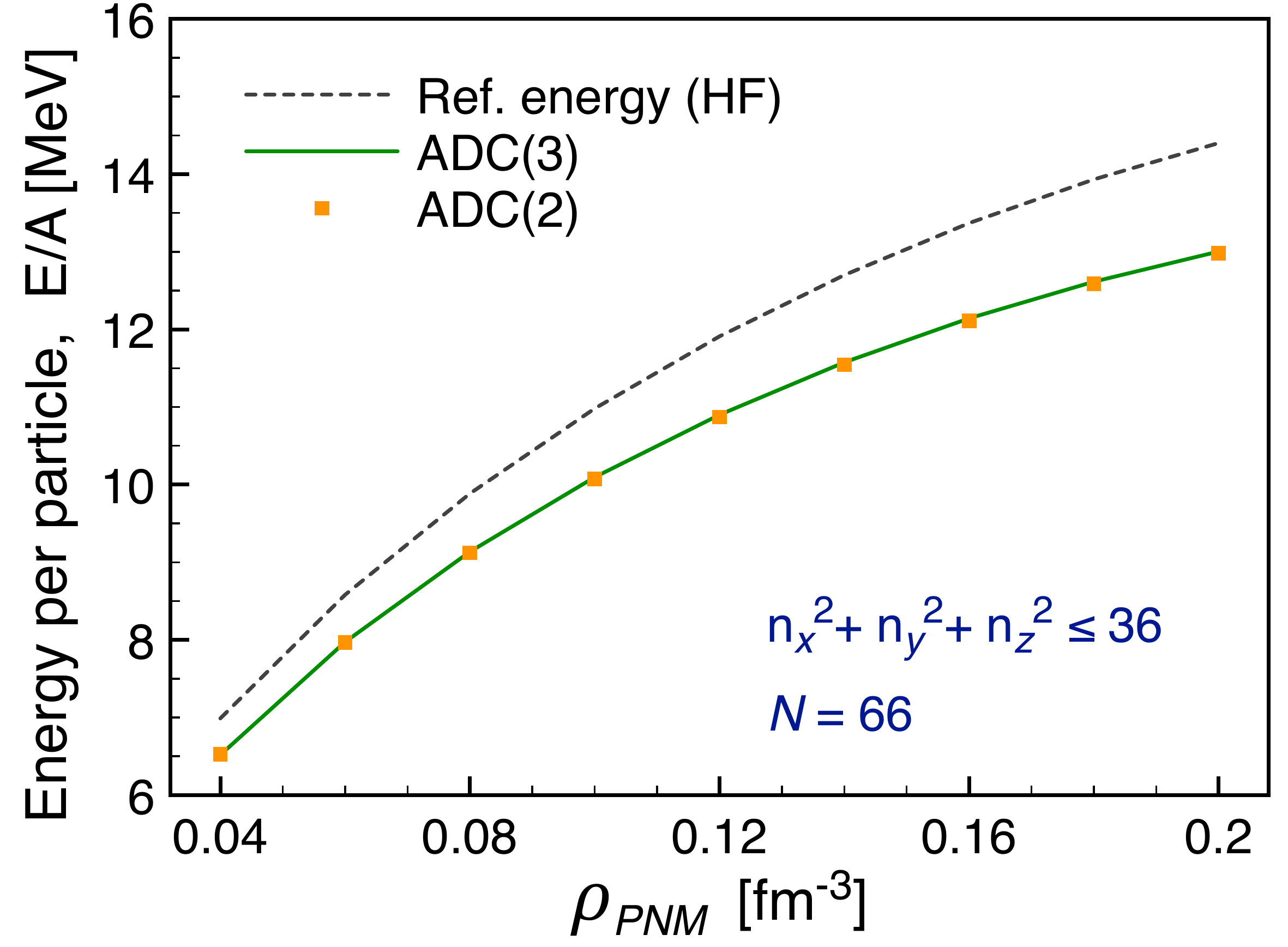}  \hspace{0.06\textwidth}
\includegraphics[width=0.46\textwidth]{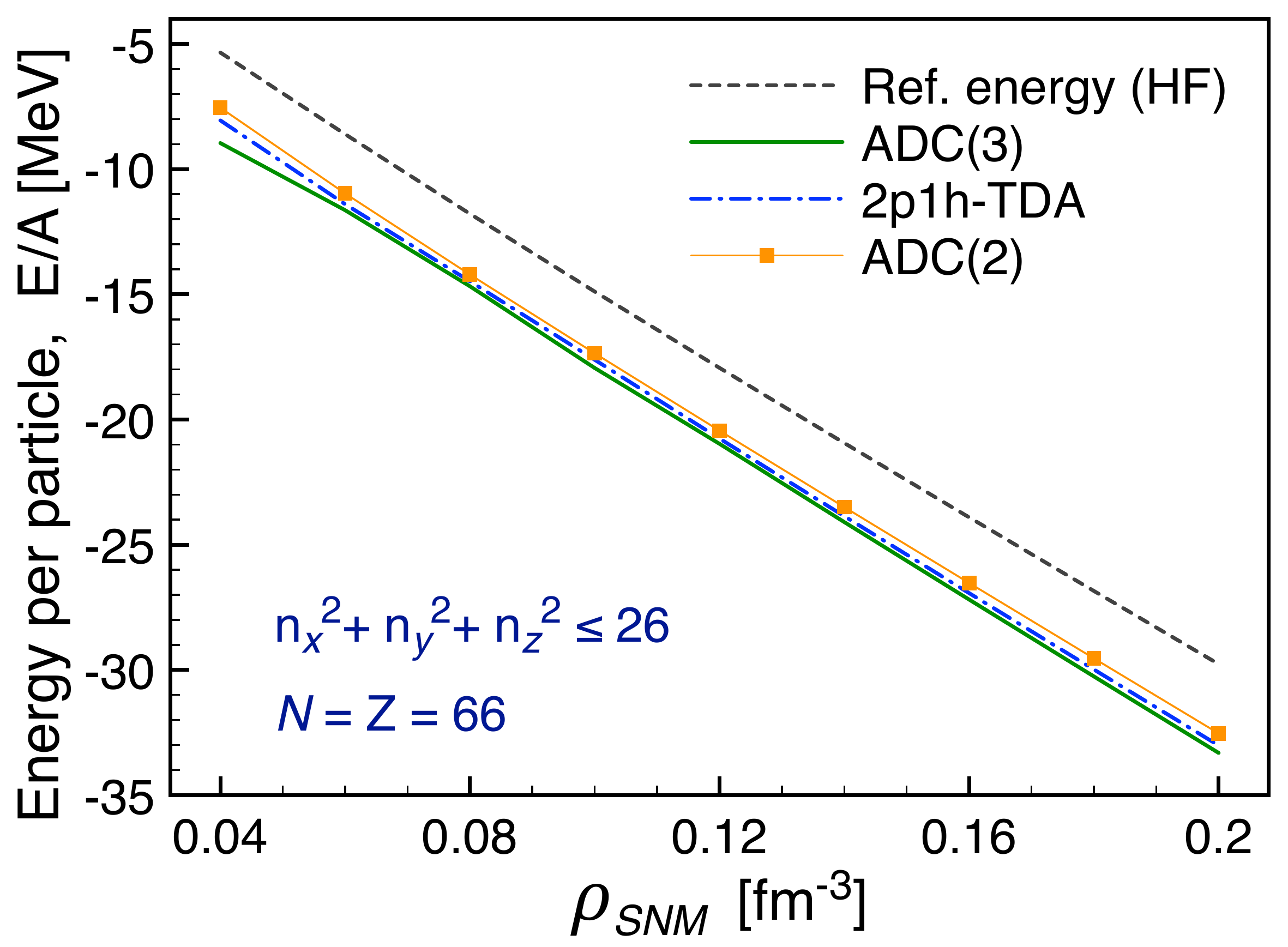}
\caption{Equation of state for PNM (left) and SNM (right) as predicted by the Minnesota two-nucleon interaction. Different curves show results for different ADC approximations. The ADC(2) (filled squares), 2p1h-TDA (dot-dashed line) and full ADC(3) (full lines) are calculated using a Hartree-Fock reference state and unperturbed single particle energies.}
\label{fig:minn_adc_eos}
\end{center}
\end{figure}

\begin{figure}[ht]
\begin{center}
\includegraphics[width=0.86\textwidth]{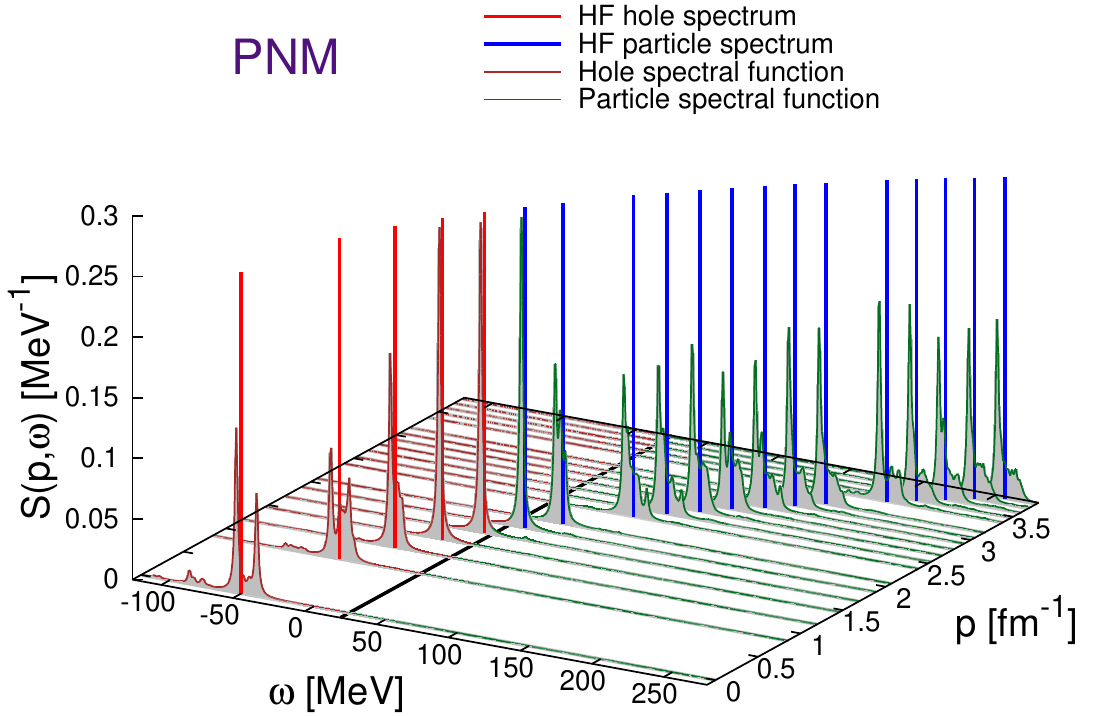} \\  \vspace{0.06\textwidth}
\includegraphics[width=0.86\textwidth]{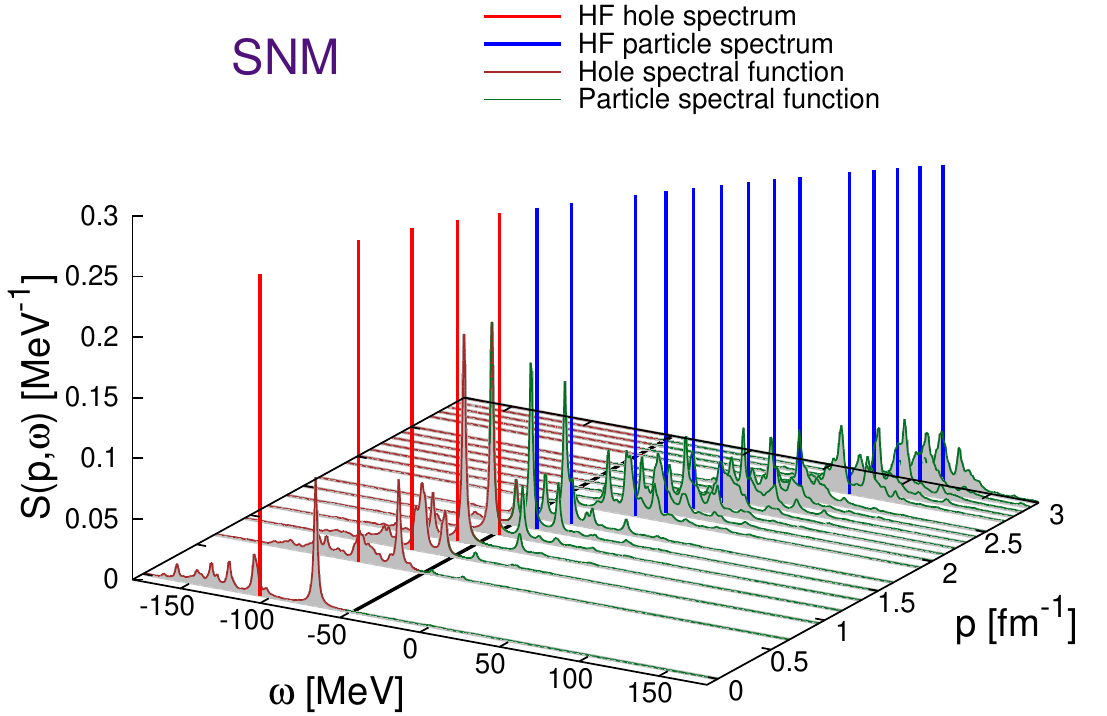} \\
\caption{Spectral function of PNM (top) and SNM (bottom) at nominal saturation density ($\rho=0.16$~fm$^{-3}$) from ADC(3).
The correlated strength distribution is folded with Lorentzians along the energy axis. The isolated vertical lines mark the unperturbed HF spectrum and are normalized to the same height assumed for the Lorentzians, so that a visual comparison with the correlated distribution is meaningful.
The thick line at constant $\omega$ marks the Fermi energy, $E_F$, for the correlated ADC(3) results, which separates the quasihole from the quasiparticle spectrum.
}
\label{fig:minn_adc_sfnct}
\end{center}
\end{figure}

The resulting spectral functions from ADC(3) are shown in Fig.~\ref{fig:minn_adc_sfnct} and compared to the unperturbed (HF) reference state.
Since we are working in a discrete basis, the results are given for the cartesian momenta ${\bf p}_i$ and only  discrete quasiparticle energies 
are obtained from  Eq.~\eqref{eq:DysMtx} [also compare Eqs.~\eqref{eq:SpSh} and~\eqref{eq:MattK_gsp_Leh}].
In order to give a clearer visualization of the  spectral distribution, we fold each state along the energy axis with Lorentzians of width $\Gamma$=1.2~MeV near the Fermi energy and $\Gamma$=7~MeV otherwise.
The corresponding expression of the spectral function in the HF approximation has no fragmentation and displays only isolated \hbox{$\delta$-peaks} for each momenta:
 \begin{align}
S^{HF}({\bf p}, s_z; \, \omega) ={}& S^{h, HF}({\bf p}, s_z; \, \omega) +  S^{p, HF}({\bf p}, s_z; \, \omega) =  \delta\Big(\omega- \varepsilon^{HF}({\bf p})\Big) \; ,   \label{eq:Spw_HF}
\end{align} 
where $\varepsilon^{HF}({\bf p})=\frac{p^2}{2m}+v_{HF}({\bf p})$ are the HF single particle energies. Eq.~\eqref{eq:Spw_HF} is plotted as separate spikes in Fig.~\ref{fig:minn_adc_sfnct}, with their height  taken to be  the same as for the (normalized) Lorentzians near the Fermi surface. Thus, the unperturbed spectral function can be visually compared to the fragmented distribution plotted for the ADC(3).

 Fig.~\ref{fig:minn_adc_sfnct} shows all the general characteristics of the spectral distribution for infinite systems. 
At the HF level, each nucleon has an energy spectrum $\varepsilon^{HF}({\bf p})$ that follows the parabolic trend of its kinetic energy but it is otherwise shifted in energy due to the mean-field HF potential.  The density $\rho$ determines the momentum $p_F$ of the last occupied state according to Eq.~\eqref{eq:pf_vs_rho}, which in turn sets the Fermi energy, $E_F^{(HF)}=\varepsilon^{HF}(p_F)$.
When correlations are included the spectrum becomes  fragmented. Again, it is seen that PNM (top panel) is only weakly correlated  and the quasiparticle peaks are almost unchanged near the Fermi surface. Only deeply bound neutrons, at the smallest momenta, are sensibly fragmented.  On the other hand, the correlated spectral function of  SNM is much more fragmented, some particle strength is visible for small momenta $p<p_F$ and likewise there is a small occupation of states with $p>p_F$. Integrating $S({\bf p}, s_z; \, \omega)$ over the energy interval $]-\infty, E_F]$ yields the momentum distribution (per unit volume), while further integrating over momenta gives the total nucleon density $\rho$ (see Eq.~\eqref{eq:part_num}).

\begin{figure}[tb]
\begin{center}
\includegraphics[width=0.46\textwidth]{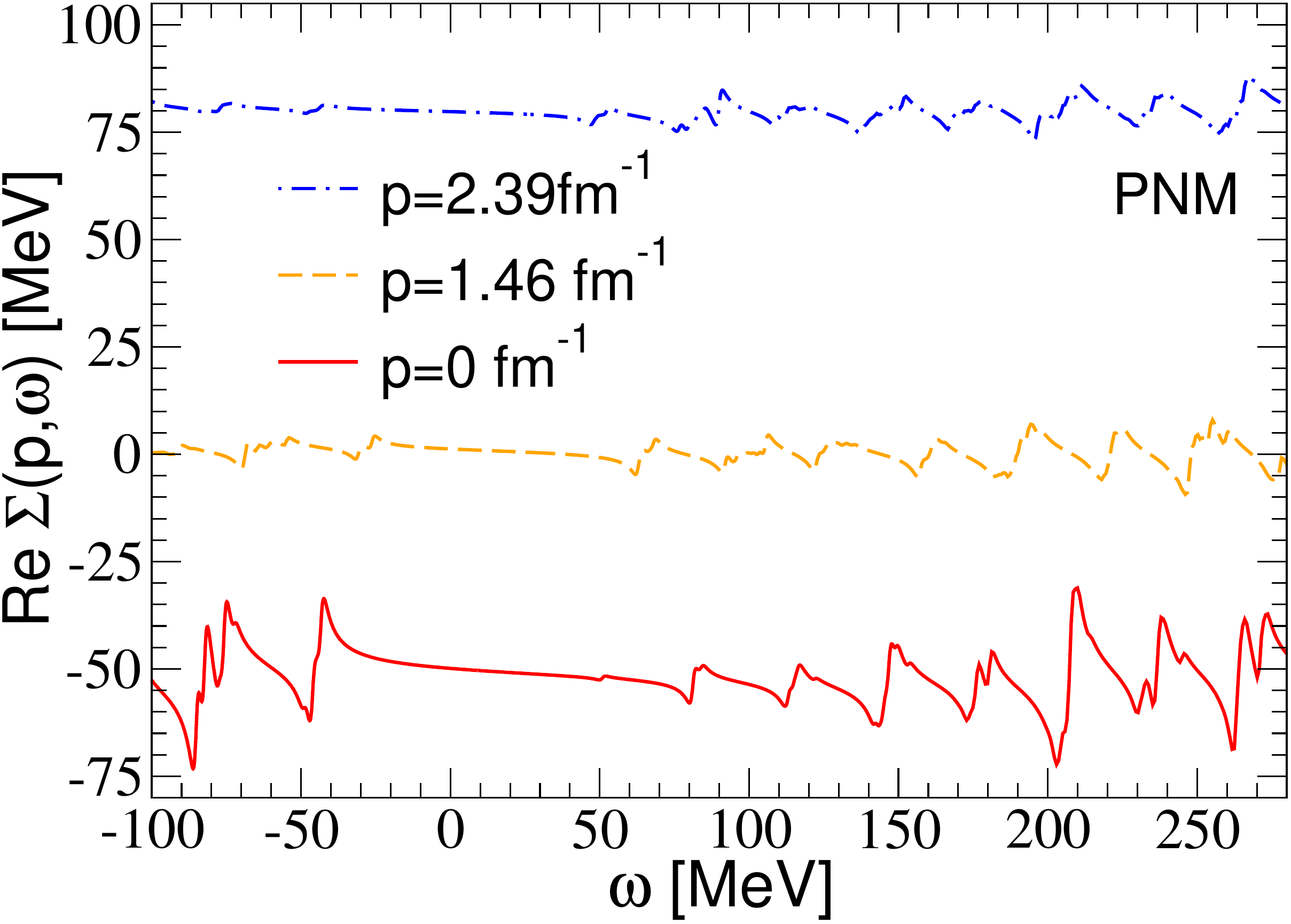}  \hspace{0.05\textwidth}
\includegraphics[width=0.472\textwidth]{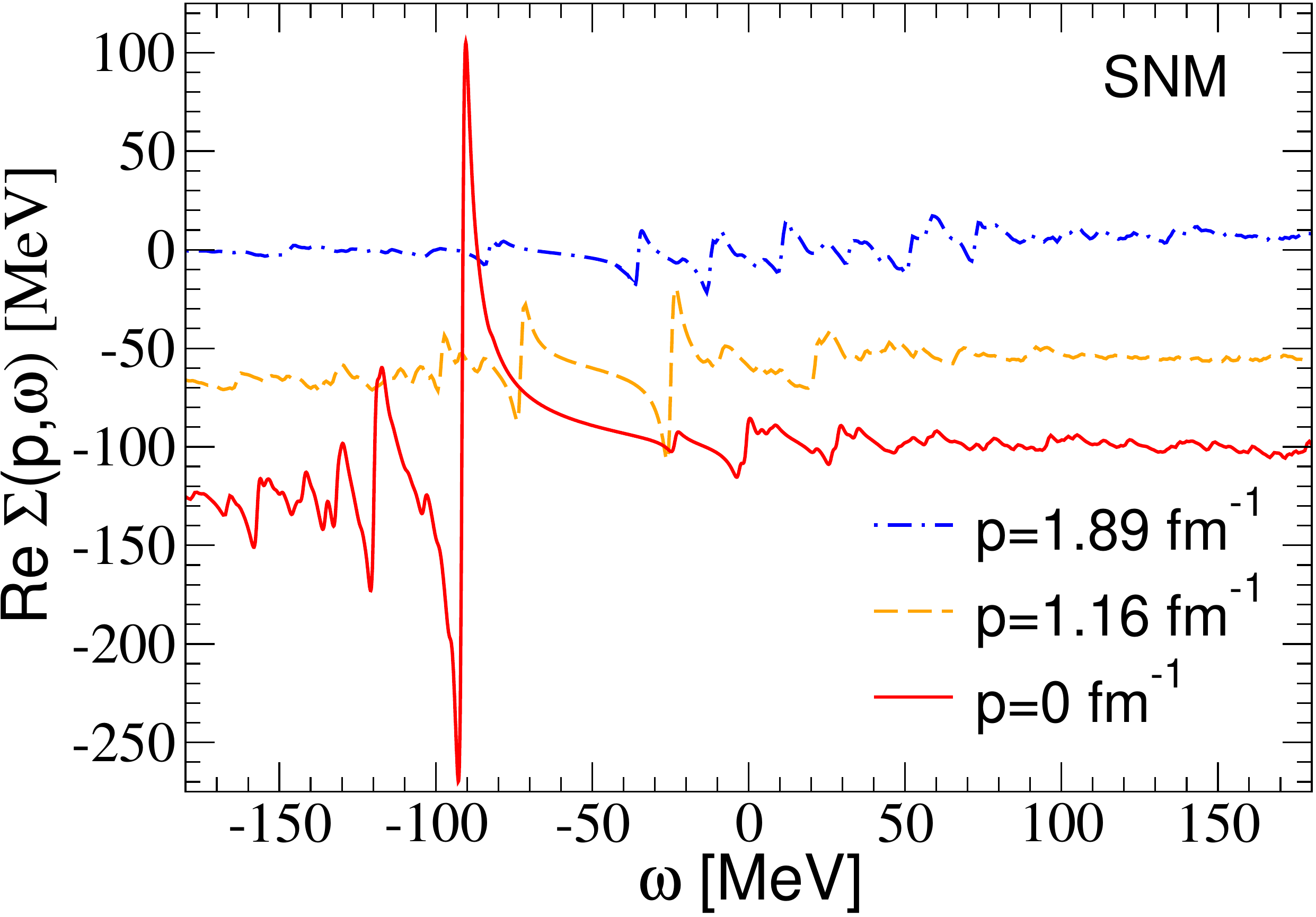}
\caption{Real part of the nuclear self-energy, ${\rm Re} \; \Sigma(p,\omega)$, of PNM (left) and SNM (right) at nominal saturation density ($\rho=0.16$~fm$^{-3}$), obtained from ADC(3).  The Fermi momentum is $k_F$=1.68~fm$^{-1}$ for PNM and $k_F$=1.33~fm$^{-1}$ for SNM.  The plots are shown for fixed momenta at  $p=0$ fm$^{-1}$, at  $p\approx0.87p_F$ (just below $p_F$) and at $p\approx1.42p_F$ (above $p_F$).}
\label{fig:minn_ReSE}
\end{center}
\end{figure}

\begin{figure}[tb]
\begin{center}
\includegraphics[width=0.46\textwidth]{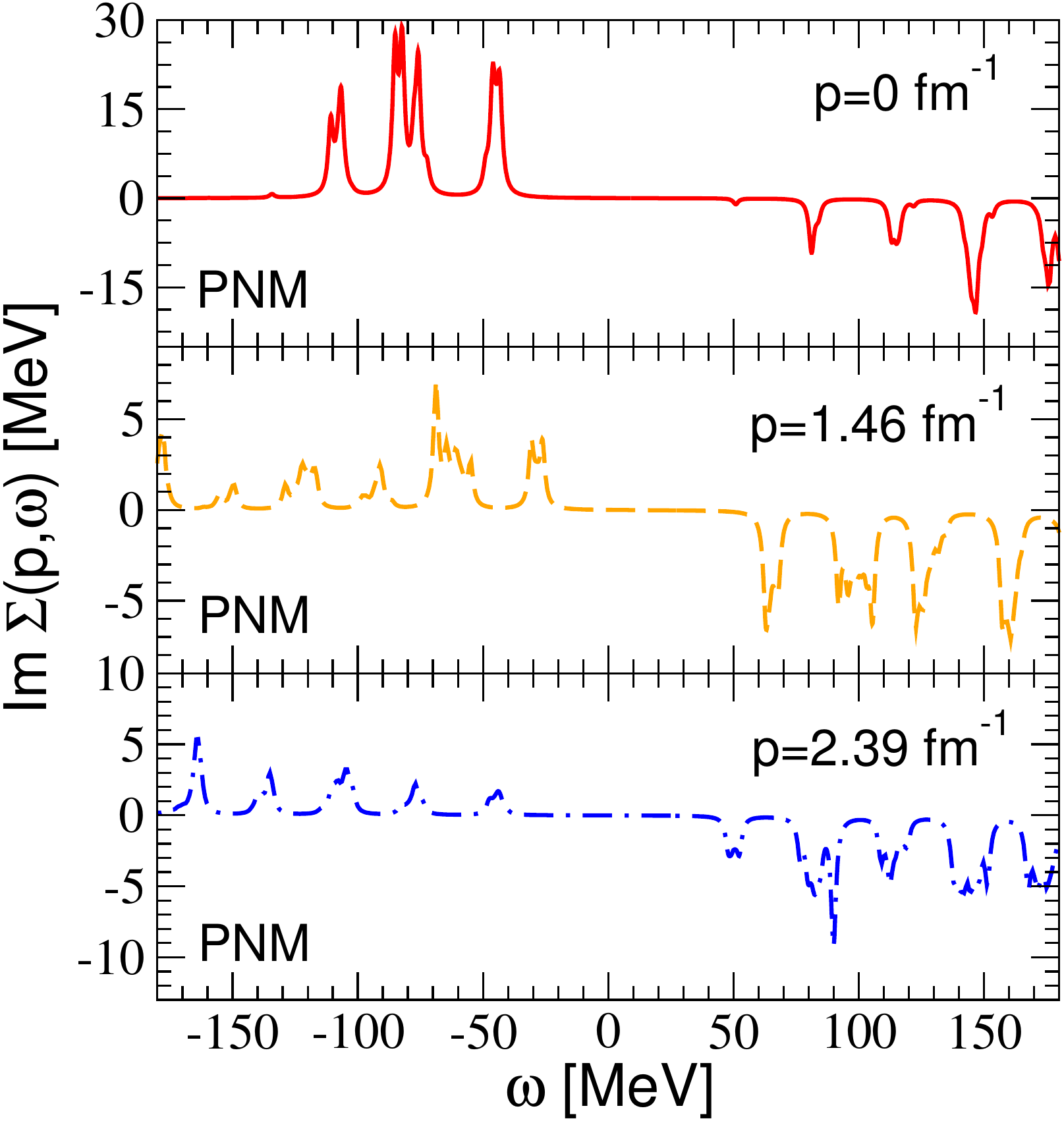}  \hspace{0.06\textwidth}
\includegraphics[width=0.46\textwidth]{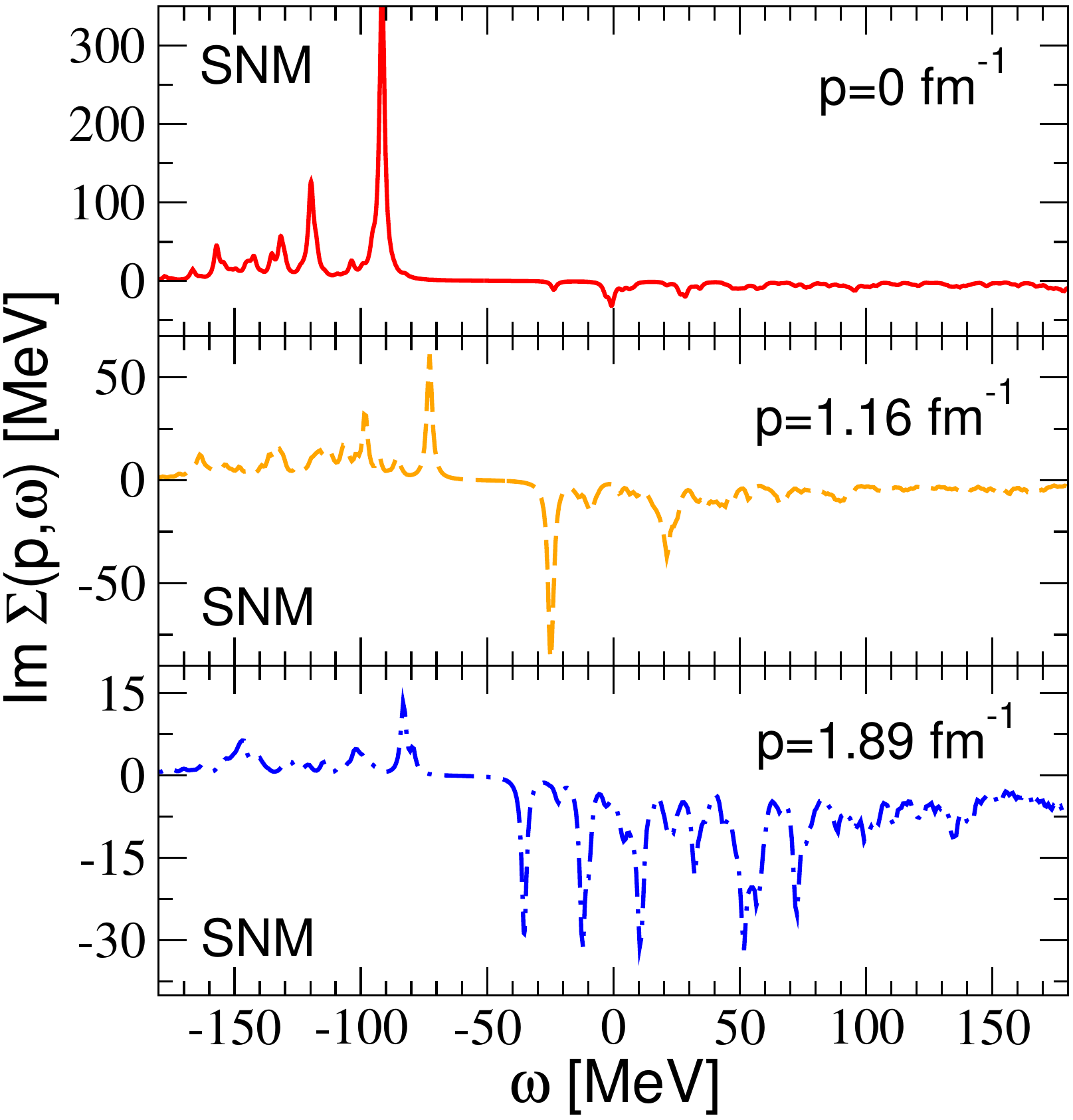}
\caption{Imaginary part of the nuclear self-energy, ${\rm Im} \, \Sigma(p,\omega)$, of PNM (left) and SNM (right) at nominal saturation density ($\rho=0.16$~fm$^{-3}$), as calculated from  ADC(3).  The Fermi momentum is $k_F$=1.68~fm$^{-1}$ for PNM and $k_F$=1.33~fm$^{-1}$ for SNM. Fixed  momenta  of  $p=0$ fm$^{-1}$, at  $p\approx0.87p_F$ and $p\approx1.42p_F$ are shown. }
\label{fig:minn_ImSE}
\end{center}
\end{figure}

The real and imaginary parts of the self-energy, $\Sigma^\star({\bf p}, s_z; \, \omega)$, are shown in Figs.~\ref{fig:minn_ReSE} and ~\ref{fig:minn_ImSE} for values of the 
momentum ${\bf p}_i$ both below and above $p_F$.  Also in this plots, the discrete energy poles are folded by taking a finite value of $\Gamma$ in Eq.~\eqref{eq:MattK_Sig_Leh},
which correspond to using finite width Lorentzians for the imaginary part.
In Fig.~\ref{fig:minn_ReSE}, bot PNM and SNM have a similar dependence on momentum that comes form the kinetic energy term in $\Sigma^{(\infty)}({\bf p})$ but there is more attraction in the second case. This is due to the additional attractive force between protons and neutrons, which makes SNM bound. Superimposed to this trend is the energy dependence  coming form the coupling to ISCs, which fragments and spreads the single particle strength over different energies. 
The imaginary part of the self-energy encodes the strength of the absorption effects that mix single particle degrees of freedom to ISCs ones. Thus, it is also directly connected to the mean free path of nucleons in the system~\cite{ch11_Rios2012MFP}. This term is always positive~(negative) for energies below~(above) the fermi surface. 
 For ${\bf p}_i\approx$~0 the absorption is strongest at low energies.  As one increases $p$, this becomes weak in the energy region of hole states and much more stronger correlations are seen for quasiparticle energies and momenta outside the Fermi sea.  Once again the PNM panel shows weak and more isolated peaks, while SNM is characterized by stronger fragmentation and absorption (hence,  a more collective behavior).

Most of the qualitative features of these self-energies and of the spectral functions just shown are general to extended correlated fermion systems and are  also seen, for example, in the electron gas or liquid $^3$He. It is interesting to compare the plots of Fig.~\ref{fig:minn_adc_sfnct} to the analogous distribution of a finite system, like the one shown in Fig.~\ref{fig:ScptFnctN56}. In the latter case, the spectral function displays orbits form the shell structure rather than peaks  distributed according to kinetic energy. In all cases, correlations alter the simple mean-field view. However, the strength near the Fermi energy tends to remain dominated by single particle structures because of the low density of ISCs (2p1h, 2h1p and beyond) in that region.

\begin{figure}[t]
\begin{center}
\includegraphics[width=0.85\textwidth]{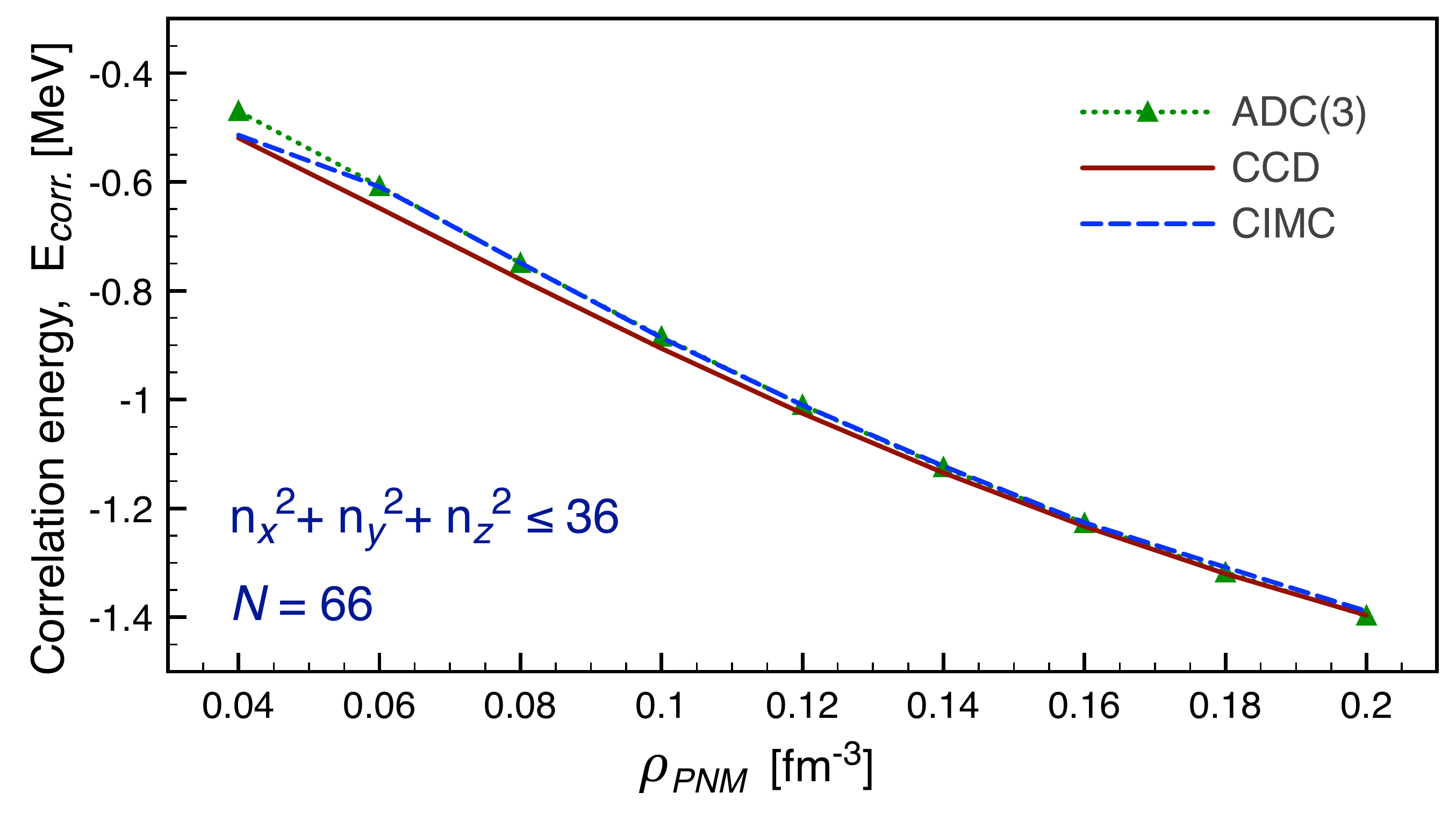}
\caption{Total correlation energy for pure neutron matter obtained from the CCD, the configuration interaction Monte Carlo (CIMC) and the ADC(3)-{\em sc0} methods that are  presented in this book. See also Section 10.3.7 of Ref.~\cite{chapter10} for results based on the IMSRG(2) approach.}
\label{fig:minn_all}
\end{center}
\end{figure}

Figure~\ref{fig:minn_all} compares the results for PNM with  the coupled cluster and Monte Carlo methods introduced in previous chapters.  Note that we show correlation energies, rather than the total energy per particle, to amplify differences among many-body methods. The largest discrepancy is at the lowest density and amounts to $\approx$50~keV/A. This is 10\% of the correlation energy but less than 0.5\% of the total energy.  At larger densities, all methods agree to higher accuracy.  It is interesting to see that ADC(3) initially follows configuration interaction Monte Carlo (CIMC) and then shifts to being closer to CCD as the density increases.

\section{Self-consistent Green's functions at finite temperature in the thermodynamic limit}
\label{sec:scgf_finiteT}

We now concentrate on the study of  infinite systems at finite temperature and will  set ourselves in the thermodynamic limit, that is number of particles $N$ and volume $V$ going to infinity with density $\rho=N/V$ kept constant. The many-body SCGF approach at finite temperature is particularly suited for this kind of study because, for appropriate approximations of the self-energy, it is thermodynamically consistent: a quantity calculated from the microscopic point of view yields the same result as the thermodynamical macroscopic quantity~\cite{ch11_Baym1962}. This consistency is strictly related to the fact that a fully dressed propagator, obtained via iterative solution of Dyson's equation, Eqs.~(\ref{eq:Dyson}), is used in the calculation of the partition function in the Luttinger-Ward formalism~\cite{ch11_Luttinger1960}, from which one extracts the thermodynamical properties of the system. Furthermore, it can be demonstrated that this method fulfills the Hugenholtz van-Hove theorem~\cite{ch11_Hugenholtz1958}, and this once again relates to the fact that the conservation laws of particle number, momentum and energy are preserved in this kind of approximation~\cite{ch11_Baym1961,ch11_Baym1962}.

We will show in this section how to calculate the self-consistent propagator in the {\em ladder approximation}, a specific approximation for the self-energy $\Sigma^\star(\omega)$ where particle-particle and hole-hole intermediate scattering states are resummed to all orders in the so called in-medium $T$-matrix. We will be working with the effective Hamiltonian of Eq.~\eqref{eq:Heff}, considering the two-body averaged three-body force that enters $\widetilde U$ as given in Eq.~\eqref{eq:ueff_3b_first}, and disregarding all irreducible three-body terms. The Koltun sum rule of Eq.~\eqref{eq:Koltun_hW} is then used to obtain the total energy of the many-body system. The great advantage of working at finite temperature is that the appearance of pairing when considering hole-hole intermediate states is washed out by thermal effects~\cite{ch11_Alm1996}. Note that a different possibility is to account for pairing by implementing analogous calculations but in a formalism with both normal and anomalous propagators (as done in Gorkov theory)~\cite{ch11_BozekPair2002,ch11_MutherPair2005}.  Recently, an improved treatment of pairing in the SCGF method when going to zero temperature has been presented in Ref.~\cite{ch11_ding2016}. Within the Luttinger-Ward formalism at finite temperatures, the entropy can then be calculated via the knowledge of the self-consistent propagator, and from the entropy all other thermodynamical quantities are accessible. We will not treat here the calculation of the entropy, for a detailed description we refer the reader to Chapter 3 of Ref.~\cite{ch11_Rios2007PhD}.

In the next section, we will give a few hints on the theoretical formalism and then sketch in the following section the working equations necessary to perform the numerical implementation. The full self-consistent numerical calculation considering the complete off-shell properties of the system and considering fully microscopic potentials was performed by the Gent~\cite{ch11_DewulfSCGF2002},  the T\"ubingen and Barcelona~\cite{ch11_Frick2003,ch11_Frick2004PhD,ch11_Frick2005,ch11_Rios2006C74,ch11_Rios2008,ch11_Rios2009} and the Cracow groups~\cite{ch11_Soma2006,ch11_Soma2008,ch11_Soma2009,ch11_Soma2009phd}.

\subsection{Finite-temperature Green's function formalism}
In a similar way to  Sec.~\ref{sec:scgf_defs}, we start by defining the one-body Green's function, however this time as a statistical average in the grand-canonical ensemble:
\begin{equation}
\label{eq:thermalG}
ig({\bf x}t, {\bf x'}t')= {\rm Tr}\{\widehat{\rho}{\pazocal T}[\widehat\psi({\bf x}t) \widehat\psi^\dagger({\bf x'}t')]\}\,;
\end{equation}
here ${\pazocal T}$ describes the Wick time-ordered product of the quantum field operators for the creation, $\widehat\psi^\dagger({\bf x'}t')$, and destruction, $\widehat\psi({\bf x}t)$, of a single-particle state in the Heisenberg picture. The field operators are related to the operators of creation and destruction, i.e. $a^\dagger_\alpha$ and $a_\alpha$, via $\widehat\psi^\dagger({\bf x'})=\sum_\alpha\psi_\alpha({\bf x})^\dagger a^\dagger_\alpha$ and $\widehat\psi({\bf x})=\sum_\alpha\psi_\alpha({\bf x}) a_\alpha$, where the coefficients are the single-particle wave functions of state $\alpha$ and the sum is over the complete basis set of single-particle quantum numbers. The statistical factor $\widehat \rho$ is defined by:
\begin{equation}
\widehat \rho=\frac{1}{Z}e^{-\beta(\widehat H -\mu\widehat N)}\,,
\end{equation}
where $\beta$=1/T is the inverse temperature, $\mu$ is the chemical potential and $Z$ is the grand-partition function
\begin{equation}
\label{eq:part_fun}
Z={\rm Tr}\,e^{-\beta(\widehat H -\mu\widehat N)}\,,
\end{equation}
with $\widehat H$ the Hamiltonian given in Eq.~(\ref{eq:H}), and $\widehat N$ the particle number operator. The trace in Eq.~(\ref{eq:part_fun}) is to be taken over a full set of energy and particle number eigenstates of the system. The two possible time-ordering products in Eq.~(\ref{eq:thermalG}) are given by:
\begin{equation}
\label{eq:Tproduct}
{\pazocal T}[\widehat\psi({\bf x}t) \widehat\psi^\dagger({\bf x'}t')]=
 \Bigg\{
  \begin{tabular}{c}
  \,\,\,\,$\widehat\psi({\bf x}t) \widehat\psi^\dagger({\bf x'}t'), \quad t>t'$  \\
  $-\widehat\psi^\dagger({\bf x'}t') \widehat\psi({\bf x}t), \quad t'>t$  \; .
  \end{tabular}
\end{equation}
The first time-ordered product in Eq.~(\ref{eq:Tproduct}) describes the creation of a particle state at time $t'$ with position ${\bf x'}$, and the destruction of the propagated particle state at time $t$ with position~${\bf x}$. Analogously, the second time-ordered product describes the destruction of a particle state, or creation of a hole state, at time $t$ with position ${\bf x}$, and the destruction of the propagated hole state at time $t'$ with position~${\bf x'}$. Using Eq.~(\ref{eq:Tproduct}) one can define the correlation functions: 
\begin{eqnarray}
\label{eq:corr_creat}
ig^>({\bf x}t, {\bf x'}t')&=& ~ ~ {\rm Tr}\{\widehat{\rho}[\widehat\psi({\bf x}t) \widehat\psi^\dagger({\bf x'}t')]\} \\
ig^<({\bf x}t, {\bf x'}t')&=& -{\rm Tr}\{\widehat{\rho}[\widehat\psi^\dagger({\bf x'}t')\widehat\psi({\bf x}t)]\}\,.
\label{eq:corr_destr}
\end{eqnarray}
Depending on the specific time ordering, the Green's function defined in Eq.~(\ref{eq:thermalG}) corresponds to one correlation function or the other, i.e. either to Eq.~(\ref{eq:corr_creat}) or to Eq.~(\ref{eq:corr_destr}). It is also useful to define the retarded propagator; this is that part of the one-body Green's function which is related only to the causal  propagation of events, i.e. forward in time:
\begin{equation}
\label{eq:retar_prop}
g^R({\bf x}t, {\bf x'}t')=\theta(t-t')[g^>({\bf x}t, {\bf x'}t')-g^<({\bf x}t, {\bf x'}t')]\,.
\end{equation}

In the following we will be dealing with the imaginary time domain, also known as Matsubara formalism to solve for the Green's function. One could equivalently well work in the real-time domain and reach the same result~\cite{ch11_Soma2009phd}. The quantum field operators of creation and destruction  in  Heisenberg picture
\begin{equation}
\label{eq:heis_field}
\widehat\psi^{(\dagger)}({\bf x}t)=e^{i\widehat Ht}\widehat\psi^{(\dagger)}({\bf x}0)e^{-i\widehat Ht}\,
\end{equation}
carry a resemblance between the thermal weight factor $e^{\beta\widehat H}$ and the time evolution operator $e^{i\widehat Ht}$ when considering the imaginary time domain $t=-i\beta$. If one includes the expression (\ref{eq:heis_field}) in the definition of the correlation functions, Eqs.~\eqref{eq:corr_creat} and~\eqref{eq:corr_destr}, it can  be proved that for a certain imaginary time domain there is absolute convergence of the two expressions, specifically in the intervals $-i\beta<t-t'<0$ for $g^>$ and $0<t-t'<i\beta$ for $g^<$. Furthermore, it can be shown that the two correlation functions are related to one another at one of their imaginary time boundaries, providing the important relation: 
\begin{equation}
\label{eq:qprelation}
g^<({\bf x},t=0;{\bf x'}, t')=e^{\beta\mu}g^>({\bf x},t=-i\beta;{\bf x'},t')\,.
\end{equation}
Thanks to the invariance under space translation of an infinite system and to time translational invariance, the Green's function only depends on the differences \hbox{${\bf r}={\bf x}-{\bf x'}$} and~\hbox{$\tau=t-t'$}. Consequently, by exploiting the quasi-periodicity relation of the Green's function along the imaginary time axis given in Eq.~(\ref{eq:qprelation}), one can write a discrete Fourier representation for the one-body Green's function in the frequency domain:
\begin{equation}
g({\bf r},\tau)=\int \frac{{\rm d}^3p}{(2\pi)^3}e^{i{\bf p}{\bf r}}\frac{1}{-i\beta}\sum_\nu e^{-iz_\nu\tau}g({\bf p},z_\nu)\,,
\end{equation}
where $z_\nu=\frac{\pi\nu}{-i\beta}+\mu$ are the Matsubara frequencies for odd integers $\nu= \pm1, \pm3, \pm5, ...$ The Fourier coefficients are then given by the inverse transformation:
\begin{equation}
\label{eq:Fouriercoeff}
g({\bf p},z_\nu)=\int {\rm d}^3r\int_0^{-i\beta}{\rm d}\tau\,e^{-i{\bf p}{\bf r}+iz_\nu\tau}g({\bf r},\tau)\,.
\end{equation}
These coefficients are evaluated for an infinite set of complex frequencies $z_\nu$, corresponding to the imaginary time domain, however one would like to understand the properties of the physical propagator, i.e. in the real time and frequencies domain. To do so let's go back to the expressions of the correlation functions, Eqs.~(\ref{eq:corr_creat}) and (\ref{eq:corr_destr}), and write down their Fourier transform:
\begin{eqnarray}
\label{eq:FTpart}
g^>({\bf p},\omega) &=& i\int{\rm d}^3r\int_{-\infty}^{+\infty}{\rm d}\tau\,e^{-i{\bf pr}+i\omega \tau}g^>({\bf r},\tau)\,,\\
\label{eq:FThole}
g^<({\bf p},\omega) &=& -i\int{\rm d}^3r\int_{-\infty}^{+\infty}{\rm d}\tau\,e^{-i{\bf pr}+i\omega \tau}g^<({\bf r},\tau)\,.
\end{eqnarray}
These two quantities now define the spectral probability to attach or remove a particle with an energy $\omega$ and momentum  {\bf p} to or from the many-body system (we omit for simplicity the spin and isospin quantum numbers). The sum of these two functions is a positive quantity and yields the spectral function at finite temperatures:
\begin{equation}
\label{eq:spec_fun}
A({\bf p},\omega)=g^>({\bf p},\omega)+g^<({\bf p},\omega)\,.
\end{equation}
An important feature of the spectral function is that it fulfills the sum rule
\begin{equation}
\int_{-\infty}^{+\infty}\frac{{\rm d}\omega}{2\pi}A({\bf p},\omega)=1\,,
\end{equation} 
which is consistent with the interpretation of $A({\bf p},\omega)$ as a probability of leaving the system in a state of energy $\omega$ by either adding or removing a nucleon of momenutm ${\bf p}$. Below, we show how $A({\bf p},\omega)$ relates to its zero temperature counterpart, Eqs.~\eqref{eq:SpSh}.

Using Eq.~(\ref{eq:qprelation}) in  Eqs.(\ref{eq:FTpart}) and (\ref{eq:FThole}), we can write the Fourier transform of the periodicity condition
\begin{equation}
g^>({\bf p},\omega)=e^{\beta(\omega-\mu)}g^<({\bf p},\omega)\,,
\end{equation}
and considering the definition of the spectral function, we can write the correlation functions in momentum and frequency as:
\begin{eqnarray}
\label{eq:FTpart_spec}
g^<({\bf p},\omega) &=& f(\omega)A({\bf p},\omega)\,,\\
\label{eq:FThole_spec}
g^>({\bf p},\omega) &=&[1-f(\omega)]A({\bf p},\omega)\,,
\end{eqnarray}
where $f(\omega)=\frac{1}{e^{\beta(\omega-\mu)}+1}$ is the Fermi-Dirac distribution function. These expressions show that, once the spectral function is known, it is easy to access the correlation functions. A similar relation can be found between the spectral function and the Fourier coefficients of Eq.~\eqref{eq:Fouriercoeff}:
 \begin{equation}
\label{eq:FT_fullprop}
g({\bf p},z_\nu)=\int^{+\infty}_{-\infty} \frac{{\rm d}\omega'}{2\pi} \frac{A({\bf p},\omega')}{z_\nu-\omega'}\,.
\end{equation}
The previous expression is performed for a given infinite set of Matsubara frequencies in the complex plane. However we would like to extend this to the entire complex plane, especially close to the real axis, which corresponds to physical frequencies. It can be demonstrated that this analytical continuation is possible and one can safely replace $z_\nu\rightarrow z$, where $z$ is a continuous energy variable in the complex  plane~\cite{ch11_Baym1961}. Eq.~\eqref{eq:FT_fullprop} then relates the Green's function $g({\bf p},z)$ in the complex plane to the spectral function $A({\bf p},\omega)$ and is referred to as the {\em spectral decomposition} of the single-particle propagator. Similarly, one could write the real-time Fourier transform for the retarded propagator defined in Eq.~(\ref{eq:retar_prop}):
 \begin{equation}
\label{eq:FT_retarprop}
g^R({\bf p},\omega)=\int^{+\infty}_{-\infty} \frac{{\rm d}\omega'}{2\pi} \frac{A({\bf p},\omega')}{\omega_+-\omega'}\,,
\end{equation}
with $\omega_+=\omega+i\eta$. This quantity is equal to evaluating the Green's function slightly above the real axis, i.e. $g^R({\bf p},\omega)=g({\bf p},\omega_+)$. This equality is of fundamental importance. In fact, it tells us that, by knowledge of the spectral function, there exists a Green's function $g({\bf p},z)$ which corresponds both to the Green's function at the Matsubara frequencies, $z=z_\nu$, and also to the retarded propagator for frequencies slightly above the real axis, $z=\omega+i\eta$. This means that the information carried by the coefficients in Eq.~\eqref{eq:FT_fullprop} can be analytically continued to the real axis, and so to a physical propagator. Furthermore, exploiting the Plemelj identity,
\begin{equation}
\frac{1}{\omega\pm i\eta}=\frac{\pazocal P}{\omega}\mp i\pi\delta(\omega)\,,
\end{equation}
one can separate the real and imaginary part of the propagator in Eq.~\eqref{eq:FT_retarprop}, and it can be checked that the imaginary part of the retarded propagator is proportional to the spectral function:
\begin{equation}
\label{eq:spec_img}
A({\bf p},\omega)=-2 \, {\rm Im} \, g({\bf p},\omega_+)\,.
\end{equation}
Furthermore, the Dyson equation given in Eq.~\eqref{eq:Dyson} can be rewritten in an algebraic form as follows:
\begin{equation}
g({\bf p},\omega_+)=\frac{1}{[g^{(0)}({\bf p},\omega_+)]^{-1}-\Sigma^\star({\bf p},\omega_+)}\,,
\label{eq:G_algebraic}
\end{equation}
and combining Eq.~(\ref{eq:G_algebraic}) with Eq.~(\ref{eq:spec_img}),
one can express the spectral function as:
\begin{equation}
A({\bf p},\omega)=\frac{-2 \, \mathrm{Im} \, \Sigma^\star({\bf p},\omega_+)}
{[\omega-\frac{p^2}{2m}-\mathrm{Re}\Sigma^\star({\bf p},\omega)]^2+[\rm{Im} \, \Sigma^\star({\bf p},\omega_+)]^2} \,.
\label{eq:spectral_fun}
\end{equation}
The numerical calculation that one has to perform requires self-consistency between Eq.(\ref{eq:spectral_fun}) and  an  appropriate approximation for~$\Sigma^\star({\bf p},\omega)$. Self-consistency is achieved once the spectral function inserted in the calculation of the irreducible self-energy is equal to the one obtained by solving Eq.~(\ref{eq:spectral_fun}).

Before going on, it is interesting to point out that in the limit of zero temperature, the spectral decomposition of the one-body propagator given in Eq.~(\ref{eq:FT_fullprop}) can be separated into two pieces:
\begin{equation}
g({\bf p},\omega)=\int_{\varepsilon_\textrm F}^\infty\mathrm d\omega'\frac{S^p({\bf p},\omega')}{\omega-\omega'+i\eta}
+\int_{-\infty}^{\varepsilon_\textrm F}\mathrm d\omega'\frac{S^h({\bf p},\omega')}{\omega-\omega'-i\eta}\,,
\label{eq:Lehm_infty}
\end{equation}
The $S^p({\bf p},\omega)$ and $S^h({\bf p},\omega)$ correspond to the particle and hole spectral functions, which were already introduced in Eqs.~(\ref{eq:SpSh}). Notice however that, unlike in Eqs.~(\ref{eq:SpSh}), we have one single Fermi energy $\varepsilon_\textrm F$ ($\varepsilon_\textrm F$ = $\varepsilon_0^+$ = $\varepsilon_0^-$) in the integrals domain because the gap disappears in an infinite gas or a normal Fermi liquid. In an uncorrelated system, this energy defines the last filled level and hence corresponds to the energy needed to remove a particle from the many-body ground state. In the case of an interacting system, not in the superfluid nor in the superconducting phase, $\varepsilon_\textrm F$ equals the chemical potential $\mu$, and corresponds to the minimum energy necessary to add or remove a particle to/from the many-body system. Consequently, the expression for the spectral function given in Eq.~(\ref{eq:spectral_fun}) can be divided into two parts:
\begin{subequations}
\label{eq:S_self_all}
\begin{align}
S^p({\bf p},\omega)={}&-\frac{1}{\pi}\frac{\mathrm{Im} \, \Sigma^\star({\bf p},\omega)}
{(\omega-\frac{p^2}{2m}-\mathrm{Re} \, \Sigma^\star({\bf p},\omega))^2+(\rm{Im} \, \Sigma^\star({\bf p},\omega))^2} \quad \omega>\varepsilon_\textrm F\,,\qquad\,\,\,
\label{eq:Sp_self}
\\ 
S^h({\bf p},\omega)={}&\frac{1}{\pi}\frac{\mathrm{Im} \, \Sigma^\star({\bf p},\omega)}
{(\omega-\frac{p^2}{2m}-\mathrm{Re} \, \Sigma^\star({\bf p},\omega))^2+(\rm{Im} \, \Sigma^\star({\bf p},\omega))^2} \quad\,\,\,\, \omega<\varepsilon_\textrm F\,,\quad
\label{eq:Sh_self}
\end{align}
\end{subequations}
resembling the structure of Eqs.~(\ref{eq:SpSh}).

\vskip 0.2 cm
With all the basic formalism in place, we still need to devise a proper conserving approximation to the self-energy~$\Sigma^\star({\bf p},\omega)$.  For applications to infinite nucleonic matter this is done by summing infinite ladders of two-particle and two-hole configurations inside the medium. Hence, the first two diagram of this expansion are those of Figs.~\ref{fig:2ndOrd}a) and~\ref{fig:3rdOrd}a). This approximation is analogous to the series generated in Eq.~\eqref{eq:pp_ladder} and {\bf Exercise 11.3} except that it is resummed in the RPA way\footnote{Note that particle-hole summations, corresponding to the {\em ring} diagram of Fig.~\ref{fig:3rdOrd}b), represent a formidable task in nucleonic matter and have been almost always disregarded in SCGF studies of infinite matter. This in contrast to Green's function studies of the electron gas and solid state materials, where rings are necessary to screen the Coulomb interaction while ladders can often be neglected.}.
In the next subsection we will  sketch the main steps that have to be taken to perform the numerical implementation of SCGF calculations at finite temperature and introduce the working equations of the ladder expansion of the self-energy along the way.

\subsection{Numerical implementation of the ladder approximation}
Figure~\ref{fig:num_impl} shows the iterative scheme  that needs to be implemented numerically for self-consistent calculations. This is for the case of both two-body and three-body forces, when working with the Hamiltonian $\widetilde H_1$ of Eq.~\eqref{eq:Heff} and disregarding irreducible three-body terms.
The fundamental quantities that one has to compute are the non-interacting 
 two-body Green's function, the in-medium $T$-matrix and the irreducible self-energy, which are depicted in the three light blue boxes  
 with their respective Feynman diagrams.  The diagrams are a direct way to write down the complicated mathematical expressions that one has to solve numerically (see Appendix~1 for the T=0 case). The one-body and two-body effective nuclear potentials are depicted in the central orange boxes. These are similar to the contributions given in Fig.~\ref{fig:EffOps}, except for the one-body effective potential in which the first term is zero for infinite matter and the last term where we approximate the contribution of the three-nucleon force by averaging only with two independent correlated density matrices. Fig.~\ref{fig:num_impl} shows the correct multiplying factor, as in Eq.~\eqref{eq:ueff_3b_first}~\cite{ch11_Carbone2013NM3nf,ch11_Carbone2014}. As can be seen from this scheme, all quantities in blue or orange boxes are fed with the spectral function, the left red box, which is then  computed iteratively by solving Dyson's equation, in the form of Eq.~(\ref{eq:spectral_fun}), until convergence. The criteria for reaching self-consistency is usually to compare the chemical potentials of two consecutive iterations, which is computed using $A({\bf p},\omega)$.

For clarity, we will distinguish between the wording \emph{calculation} and \emph{iteration}: we will refer to {calculation} as the whole set of several {iterations} necessary to get to a converged result for the spectral function, so Fig.~\ref{fig:num_impl} depicts exactly one iteration. For a more in depth explanation of the numerical details the reader can refer to Refs.~\cite{ch11_Frick2004PhD,ch11_Rios2007PhD}.

\vskip 0.2 cm

Each calculation is performed at a specific density $\rho$ and temperature T of the system. One starts the first iteration with a guess of the spectral function, which is given in terms of the imaginary, $\mathrm{Im}\,\Sigma^\star({\bf p},\omega)$, and real, $\mathrm{Re}\,\Sigma^\star({\bf p},\omega)$, parts of the irreducible self-energy. When possible, it is convenient to start with a converged solution   for these quantities at different values of~$\rho$ and~T. 

\begin{figure}[t]
\begin{center}
\includegraphics[width=1.0\textwidth]{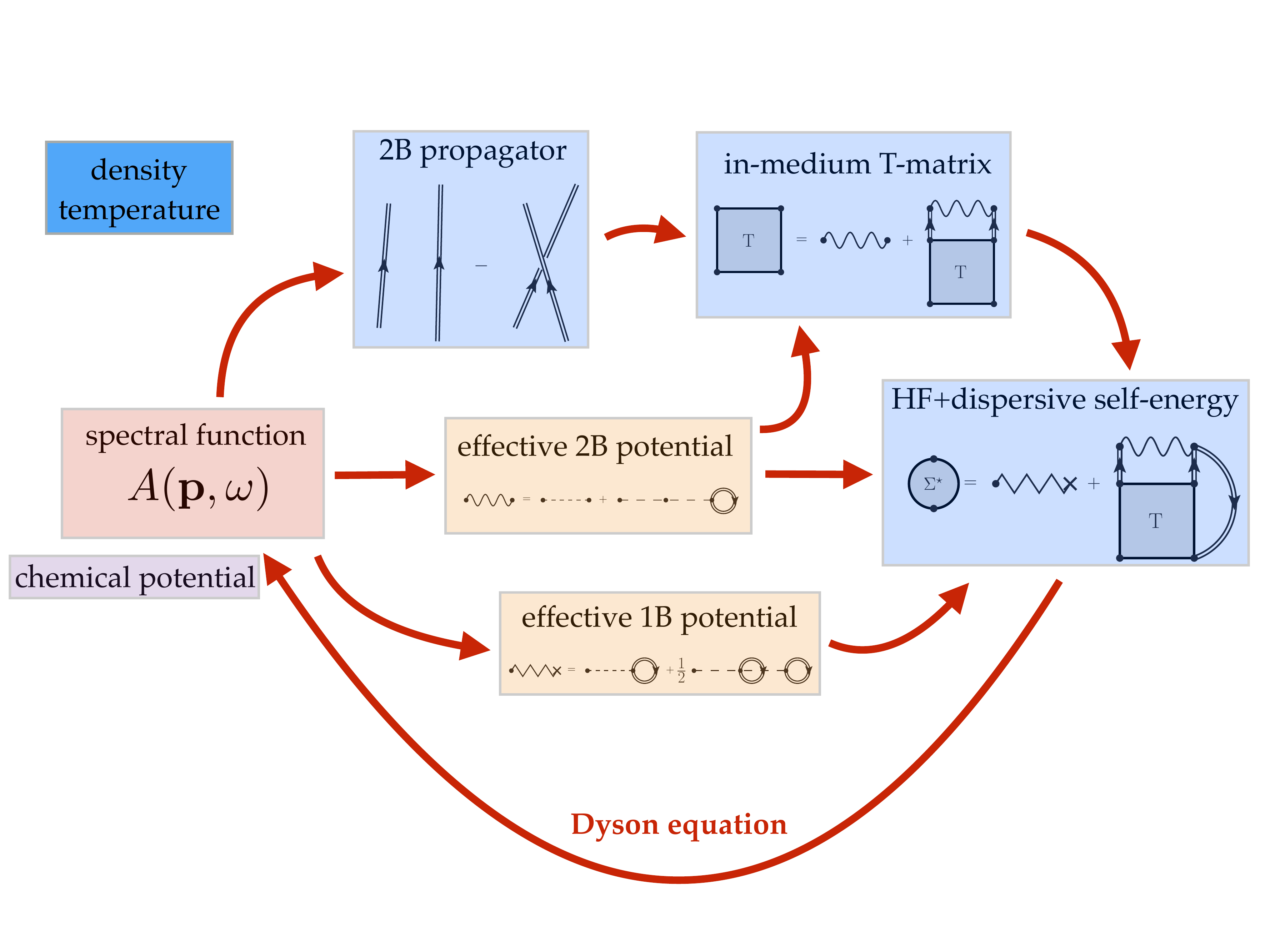}
\caption{The structure of a ladder SCGF calculation including both two-body and three-body forces through the definition of effective interactions (see text for details). Each quantity is also represented via the corresponding Feynman diagram.}
\label{fig:num_impl}
\end{center}
\end{figure}

\begin{itemize}
\item {\bf Numerical tips for the $({\bf p},\omega)$ meshes:}
The mesh of the single-particle momentum ${\bf p}$ for the self-energy is adjusted during the first iteration to be more dense around the Fermi momentum $p_{\rm F}$ corresponding to the specific density considered: 
$N_{\bf p}=70$ mesh points are enough, considering linear meshes at low momentum and around the Fermi momentum, and a logarithmic mesh for the tail all the way up to a value $\sim10p_{\rm F}$. The mesh in the single-particle energy $\omega$ has to be very dense because of the complicated features of the spectral function, especially near the quasiparticle peak. Storing a dense mesh at each iteration is demanding in terms of memory; for this reason one saves separately the imaginary and real part in a dilute linear mesh, typically of $N_{\omega}\approx6000$ points in the interval \hbox{[-2000:15000] MeV}. This is interpolated during the iterations to denser meshes of $N_{\omega}\approx30000$ points, in order to have a good description of the spectral function in the energy domain. However, the mesh in energy is adjusted in different ways during the iteration according to the specific quantities that one has to calculate (two-body propagator, $T$-matrix, etc.) as it will be explained later on.
\end{itemize}

We will now enumerate the steps to perform a complete iteration.
\vskip 0.2 cm
\noindent {\bf 1.}  Given a previously computed self-energy, the first step is to extract the corresponding single particle spectrum, which describes the centroid position of quasiparticle peaks for each momentum:
\begin{equation}
\varepsilon({\bf p})=\frac{p^2}{2m}+{\rm Re} \, \Sigma({\bf p},\varepsilon({\bf p}))\, ,
\label{eq:e_of_p}
\end{equation}
which will be used throughout the new iteration. 

\vskip 0.2 cm
\noindent {\bf 2.} The density $\rho$, temperature T and spectral function $A({\bf p},\omega)$ are the inputs to calculate the next fundamental quantity: the chemical potential $\mu$. This is obtained from the sum rule for the density:
\begin{equation}
\rho=\nu_d \int\frac{{\rm d}{\bf p}}{2\pi^3}\int_{-\infty}^{+\infty}\frac{{\rm d}{\omega}}{2\pi}A({\bf p},\omega)f(\omega,\mu)\,, 
\label{eq:chempot_micro}
\end{equation}
where $\nu_d$ is the degeneracy of the system ($\nu_d$=2 for pure neutron and $\nu_d$=4 for isospin symmetric matter), the temperature  enters through the Fermi-Dirac function, $f(\omega,\mu)$, and we have also made explicit its dependence on~$\mu$.
\begin{itemize}
\item {\bf Numerical tips for the $\mu$ mesh:} One chooses a sample mesh of chemical potentials $\mu$ to insert in $f(\omega,\mu)$ and then solves Eq.~(\ref{eq:chempot_micro}). For each point $\mu$ one gets a value of density $\rho$. Parametrizing $\rho$ as a function of $\mu$, one can then find the value of $\mu$ which corresponds to the correct  density of the system. The mesh of $\mu$ can be initially distributed around the value of the single-particle spectrum calculated at $p_{\rm F}$ (in the case of a zero temperature calculation the relation $\varepsilon(p_{\rm F})=\mu$ holds), and then adjust the mesh testing if the limits include the value of the external density. 
\end{itemize}
It must be noted that both the single-particle spectrum $\varepsilon({\bf p})$ and the spectral function $A({\bf p},\omega)$ that enter Eq.~\eqref{eq:chempot_micro} come from a previous iteration that was based on a different value of $\mu$. In this sense, the old value of the chemical potential is implicitly carried over throughout the new iteration.
However, these will end up coinciding when self-consistency is reached at the end of the calculation.
%

\vskip 0.2 cm
\noindent{\bf 3.} At this point the imaginary part of the non-interacting two-body Green's function can be computed. The lowest order approximation of the two-body propagator corresponds to the independent propagation of two fully dressed particles and was discussed in {\bf Example 11.1} for the case of zero temperature. This includes two terms, a direct and an exchange one, as depicted diagrammatically in Fig.~\ref{fig:num_impl}. 
Since we are working with dressed propagators we should refer to Eqs.~\eqref{eq:GIIf_int} and~\eqref{eq:GIIf}, however both the  direct and exchange terms must be included in our $G^{II,f}_{pphh}$. 
The imaginary part of this quantity extended to finite temperatures reads:
\begin{equation}
\label{eq:gII_imag}
{\rm Im} \, G^{II,f}_{pphh}(\Omega_+;{\bf p},{\bf p'}) = -\frac{1}{2}\int_{-\infty}^{+\infty}\frac{{\rm d}{\omega}}{2\pi}A({\bf p},\omega)A({\bf p'},\Omega-\omega)[1-f(\omega)-f(\Omega-\omega)]\,.
\end{equation}
where $\Omega_+$ is the sum of the energies of the two particles close to the real axis. This expression is derived from a sum over Matsubara frequencies of a function with a double pole on the real-energy axis via use of the Cauchy theorem~\cite{ch11_Rios2007PhD}.
\begin{itemize}
\item {\bf Numerical tips for the $\omega$ mesh:} The integrand of Eq.~(\ref{eq:gII_imag}) will be particularly hard to resolve in  regions where the two spectral functions are peaked, at energies where $\omega\sim\varepsilon({\bf p})$ and $\omega\sim\Omega-\varepsilon({\bf p'})$. It can be shown that a convenient variable change  makes these energies independent of the momenta ${\bf p}$ and ${\bf p'}$, so that one is safe with defining an energy mesh accurately distributed around only two specific points~(see Ref.~\cite{ch11_Rios2007PhD} for details). To obtain the spectral function in this specific mesh one interpolates the imaginary and real self-energies to this mesh and then uses Eq.~(\ref{eq:spectral_fun}).
\end{itemize}

\noindent{\bf 4.} From the imaginary part it is then possible to obtain the real-part of the non-interacting two-body Green's function via a dispersion relation:
\begin{equation}
{\rm Re} \, G^{II,f}_{pphh}(\Omega;{\bf p},{\bf p'}) = -{\pazocal P}\int_{-\infty}^{+\infty}\frac{{\rm d}{\Omega'}}{\pi}\frac{{\rm Im} \, G^{II,f}_{pphh}(\Omega'_+;{\bf p},{\bf p'})}{\Omega-\Omega'}\,.
\end{equation}

\noindent{\bf 5.} In practice, $G^{II,f}_{pphh}$ has to be averaged over angles. This is necessary to circumvent the coupling of partial waves with different total angular momentum $J$ which appear in $G^{II,f}_{pphh}$. The average is performed over the angle formed by the center of mass momentum ${\bf P}={\bf p}+{\bf p'}$ and the relative momentum of the two nucleons ${\bf k}=({\bf p}-{\bf p'})/2$. This strategy will facilitate  solving the in-medium $T$-matrix equations to evaluate the effective interaction in the medium. The average reads:
\begin{equation}
\overline{G^{II,f}_{pp,hh}}(\Omega_+;{P},{ k})=\frac{1}{2}\int_{-1}^{+1}{\rm d\,cos}\theta~G^{II,f}_{pp,hh}(\Omega_+;|{\bf P}/2+{\bf k}|,|{\bf P}/2-{\bf k}|)\,.
\label{eq:GIIavrgd}
\end{equation}

\noindent{\bf 6.} The two-body propagator together with the nuclear potential are then used to obtain the in-medium $T$-matrix. The exact equation for this is of Lippmann-Schwinger type:
\begin{equation}
\label{eq:t-matrix}
\langle{\bf k'}|T(\Omega_+,{\bf P})|{\bf k}\rangle=\langle{\bf k'}|\widetilde V|{\bf k}\rangle+\int {\rm d}{\bf k_1}\langle{\bf k'}|\widetilde V|{\bf k_1}\rangle{G^{II,f}_{pp,hh}}(\Omega_+;{\bf P},{\bf k_1})\langle{\bf k_1}|T(\Omega_+;{\bf P})|{\bf k}\rangle\,.
\end{equation}
As explained previously, this is a ladder resummation of particle-particle and hole-hole diagrams, this differs with respect to the Brueckner G-matrix presented in Chapter 8 because it includes hole-hole diagrams and considers the full off-shell description of the spectral function (that is, the dressed propagator). As seen from Fig.~\ref{fig:num_impl}, the potential to be included is the sum of a bare two-body potential and an averaged three-body one. Details on the numerical solution for the averaged three-body force are given in the next section, while working equations for three-nucleon chiral forces are reported in Appendix 2.

Here, we make an approximation and substitute the  two-body propagator with its  angle-averaged version~\eqref{eq:GIIavrgd}. Since the latter depends only on the magnitudes of momenta ${ P}$ and  ${ k}$, our Eq.~\eqref{eq:t-matrix} reduces to a one dimensional integral and decouples in total angular momentum, spin and isospin:
\begin{equation}
\label{eq:t-matrix_av}
\langle{k'}|T^{J \,S \,T}(\Omega_+,{ P})|{ k}\rangle=\langle{ k'}|{\widetilde V}^{J \,S \,T}|{ k}\rangle+\int_0^\infty {\rm d}{ k_1} \; { k_1}^2 \, 
\langle{ k'}|{\widetilde V}^{J \,S \,T}|{ k_1}\rangle \, \overline{G^{II,f}_{pp,hh}}(\Omega_+;{ P},{ k_1})\langle{ k_1}|T^{J \,S \,T}(\Omega_+;{ P})|{ k}\rangle\,.
\end{equation}
Going beyond Eq.~\ref{eq:GIIavrgd} with fully dressed (off shell) propagators is extremely diffcult and, to our knowledge, there is no available implementation of SCGF that can treat Eq.~\eqref{eq:t-matrix} exactly. However, estimates in  Brueckner-type calculations suggests that the error introduced by the angle averaging is small~\cite{ch11_Sartor1996BGmtx,ch11_Suzuki2000Qop}.
Eq.~\ref{eq:t-matrix_av} is a one dimensional integral equation for each allowed combination of $J,\,S,\,T$ and has at most two coupled values of $L$, due to the tensor component of the nuclear interaction. It must be noted that the nuclear interaction $\widetilde V$ considered in Eq.~(\ref{eq:t-matrix}) is the effective two-body operator given in Eq.~(\ref{eq:V_eff}). By means of a discretization procedure, the equation for the $T$-matrix is converted into a complex matrix equation which can be solved via standard numerical techniques~\cite{ch11_Rios2007PhD}. A matrix inversion has to be performed to solve this equation. This can be quite demanding if the dimension of the matrix is large. 
\begin{itemize}
\item {\bf Numerical tips for the ${\bf k_1}$ and $\Omega$ mesh:} It is important to sample in a correct manner the number of integration mesh points without loosing physical information. This is achieved by sampling conveniently the region where $G^{II,f}_{pp,hh}$ is maximum in the relative momentum ${\bf k_1}$ (for $\Omega>0$ this is close to the pole $k_1=\sqrt{m\Omega}$) and the high relative momentum region, where $G^{II,f}_{pp,hh}$ might not be negligible due to correlations. Furthermore  the $T$-matrix has a node for $\Omega=2\mu$, so an accurate mesh for the bosonic energies around this value is needed for the computation of the self-energy at the next steps.
\end{itemize}
At low temperatures, the appearance of bound states signals the onset of the pairing instability. This would directly appear as a pole in the matrix which has to be inverted to solve the Lippmann-Schwinger equation, for ${\bf P}=0$ and $\Omega=2\mu$~\cite{ch11_Thouless1960}. However, this should be seen only below a critical temperature which is around T$_c\sim4$ MeV. For this reason, calculations should not go below this border line in temperature. Especially in the case of symmetric nuclear matter, convergence starts to become slow and difficult to control when approaching this temperature and for increasing densities. This is due to the neutron-proton pairing in the coupled $^3S_1-^3D_1$ channel. In pure neutron matter, where this channel is not available, convergence is good for higher densities, and even for lower temperatures. 

\vskip 0.2 cm
\noindent{\bf 7.} The remaining step in the SCGF method is the computation of the self-energy from the $T$-matrix. The first quantity to be obtained is the imaginary part of the self-energy:  
\begin{equation}
{\rm Im} \, \Sigma^\star({\bf p},\omega_+)=\int\frac{{\rm d}{\bf p'}}{(2\pi)^3}\int_{-\infty}^{+\infty}\frac{{\rm d}{\omega'}}{2\pi}\langle{\bf pp'}|{\rm Im} \, T(\omega_++\omega',{\bf P})|{\bf pp'}\rangle A({\bf p'},\omega')[f(\omega')+b(\omega+\omega')] \,,
\label{eq:Im_self_en}
\end{equation}
where $b(\Omega)=\frac{1}{e^{\beta(\Omega-2\mu)}-1}$ is the Bose function.
We recall that the expression~\eqref{eq:Im_self_en} is also obtained from a summation over Matsubara frequencies of a function with two poles on the real energy axis~\cite{ch11_Rios2007PhD}. 
\begin{itemize}
\item {\bf Numerical tips for the ${\bf p'}$  and $\omega'$ meshes:} A momenta and energy integrals have to be performed, taking special care for the pole in energy of the Bose function $b(\Omega)$. This pole is canceled by the node we had previously mentioned in the $T$-matrix, for this reason it comes in hand that we had already defined a convenient mesh for $\Omega$ around the node. 
\end{itemize}
 
\noindent{\bf 8}. The real part of the self-energy is then obtained from its imaginary part by means of the dispersion relation: 
\begin{equation}
{\rm Re} \, \Sigma^\star({\bf p},\omega)=\Sigma^{(\infty)}({\bf p})-{\pazocal P}\int_{-\infty}^{+\infty}\frac{{\rm d}{\omega'}}{\pi}\frac{{\rm Im} \, \Sigma^\star({\bf p},\omega'_+)}{\omega-\omega'}\,.
\label{eq:self_en}
\end{equation}
The $\Sigma^{(\infty)}$ is the correlated Hartree-Fock part of the single-particle self-energy that is defined by Eqs.~\eqref{eq:U_eff} and~\eqref{eq:UeffSig}. We now  approximate this according to Eq.~\eqref{eq:ueff_3b_first}, where the three-body interaction is averaged over two non interacting particles. This can be explicitly written as:  
\begin{equation}
\label{eq:HF_self}
\Sigma^{(\infty)}({\bf p})=\int\frac{{\rm d}{\bf p'}}{(2\pi)^3}n({\bf p'})\Big[\langle{\bf pp'}|V^{\rm 2NF}|{\bf pp'}\rangle +\frac{1}{2}\langle{\bf pp'}|\widetilde V^{\rm 3NF}|{\bf pp'}\rangle\Big] \,,
\end{equation}
where $V^{\rm 2NF}$ and $\widetilde V^{\rm 3NF}$ correspond respectively to the first and second terms in Eq.~(\ref{eq:V_eff}). $\widetilde V^{\rm 3NF}$ is a one-body averaged three-nucleon force, detailed description on how to calculate this quantity and the momentum distribution $n({\bf p})$, together with an additional numerical sample code, are given in Sec.~\ref{subsec:average_3bf}.
%

Finally, the spectral function can be obtained via Eq.~(\ref{eq:spectral_fun})  and the procedure starts again from step ${\bf 1.}$ until a consistent result is achieved for the chemical potential. It must be kept in mind that, according to the mesh points in which the spectral function is needed, the interpolation is done on the imaginary and real part of the self-energy, and not directly on the spectral function. This is done in order to avoid incorrect samplings of the structure of the spectral function which could induce numerical inaccuracies. We must point out that the energy mesh for the evaluation of the spectral function must be accurate enough to reproduce not only the quasiparticle peak region but also the low and high-energy tails that characterize the spectral function (especially for large momenta in the case of hard interactions).

To calculate the total energy of the system, we make use of the modified Koltun sum rule given in Eq.~(\ref{eq:Koltun_hW}). Consequently we need to evaluate the expectation value of the three-body operator $\langle \widehat W\rangle$. As already stated in Sec.~\ref{sec:scgf_obs}, we approximate this expectation value to its first-order term, which in infinite matter corresponds to the integral over three independent but fully correlated momentum distributions $n({\bf p})$. The integral to be evaluated is given by the expression:
\begin{equation}
\langle \widehat W\rangle\simeq\frac{\nu_d}{\rho}\frac{1}{6}\int \frac{{\rm d}{\bf p}}{(2\pi)^3}\int\frac{{\rm d}{\bf p'}}{(2\pi)^3}n({\bf p})n({\bf p'})\langle{\bf pp'}|\widetilde V^{\rm 3NF}|{\bf pp'}\rangle\,,
\label{3B_exp}
\end{equation}
with $\nu_d$ the degeneracy of the system and the averaged three-body force, $\widetilde V^{\rm 3NF}$, is discussed in the next section. Once $\widetilde V^{\rm 3NF}$ is  known, the total energy per nucleon of the system can be calculated via the modified Koltun sum rule:
\begin{equation}
\frac{E}{A}=\frac{\nu_d}{\rho}\int\frac{{\rm d}{\bf p}}{(2\pi)^3}\int\frac{{\rm d}\omega}{2\pi}\frac{1}{2}\Big\{\frac{p^2}{2m}+\omega\Big\}A({\bf p},\omega)f(\omega)-\frac{1}{2}\langle \widehat W\rangle\,,
\end{equation}
which is equivalent to Eq.~(\ref{eq:Koltun_hW}).

\subsection{Averaged three-body forces: numerical details.}
\label{subsec:average_3bf}
The inclusion of one-body averaged three-nucleon forces $\widetilde V^\mathrm{3NF}$ enters the calculations presented in the previous section through Eqs.~(\ref{eq:t-matrix}), (\ref{eq:HF_self}) and \eqref{3B_exp}. Its computation requires traces over the spin and isospin quantum numbers of the averaged particle, in this case the third particle, and an integration over its momentum ${\bf p}_3$:
\begin{equation}
\langle {\bf p}'_1 {\bf p}'_2|\widetilde V^\mathrm{3NF}|{\bf p}_1 {\bf p}_2\rangle_A =
\mathrm{Tr}_{\sigma_3}\mathrm{Tr}_{\tau_3}
\int \frac{{\mathrm d}{\bf p}_3}{(2\pi)^3}n({\bf p}_3)
\langle {\bf p}_1' {\bf p}_2' {\bf p}_3 | W^\mathrm{3NF}
(1-P_{13}-P_{23})
| {\bf p}_1 {\bf p}_2 {\bf p}_3 \rangle_{A_{12}}\ \,,
\label{eq:dd3bf_new}
\end{equation}
where ${\bf p}_i$ are single-particle momenta of particles $i=1,2,3$ and $W^\mathrm{3NF}$ is the third term on the right hand side of Eq.~\eqref{eq:H}; we have omitted the spin/isospin indices in the potential matrix elements for simplicity. The ket on the right hand side of Eq.~\eqref{eq:dd3bf_new} is antisymmetrized only with respect to particles 1 and 2, i.e. ${\rm A}_{12}=(1-{\rm P}_{12})/2$; this part  is not affected by the averaging procedure over the third particle.  ${\rm P}_{ij}=(1+\boldsymbol \sigma_i\cdot\boldsymbol \sigma_j)(1+\boldsymbol \tau_i\cdot\boldsymbol \tau_j)/4$ is the permutation operator of momentum and spin/isospin quantum numbers of particles $i$ and $j$. The momentum distribution that appears in Eq.~(\ref{eq:dd3bf_new}) can be obtained directly from the spectral function, via the relation:
\begin{equation}
\label{eq:mom_dist}
n({\bf p})=\int_{-\infty}^{+\infty}\frac{{\rm d}\omega}{2\pi}A({\bf p},\omega)f(\omega)
\end{equation}

Let us give some details on the numerical implementation of Eq.~\eqref{eq:dd3bf_new} with regards to the mesh for the internal momentum ${\bf p}_3$ and the calculation of the distribution $n({\bf p}_3)$ via Eq.~\eqref{eq:mom_dist}:
\begin{itemize}
\item We start with the definition of the mesh necessary to calculate the integral over the internal momenta ${\bf p}_3$. Considering that in the integral we deal with a dressed distribution function $n({\bf p}_3)$, which may have populated states at high momentum, we need to cover momenta up to a certain high value in which it is sure that the $n({\bf p}_3)$ has reached zero. One may choose to compose this of an arbitrary number \verb|imesh| of Gauss-Legendre meshes (in the example shown below, \verb|imesh=4|), with each mesh spanning a region of width $2/3 p_{\rm F}$. This width is chosen to cover accurately the behavior of the distribution function below, across and above the Fermi momentum $p_\textrm F$. Finally, high-momentum points are included through a tangential mesh. We have 100  points in the Gauss-Legendre meshes, and 50 in the tangential one. 
\item One then needs to calculate the momentum distribution function by means of Eq.~(\ref{eq:mom_dist}). To do so, we extract the spectral function on the fly, from the self-energy of the previous iterative step. The values of the imaginary and real part of the self-energy are stored at each iteration for different points in the momentum and energy space: for the momentum we typically have $N_{\bf p}=70$ mesh points with values ranging from 0 to 3000~MeV/c; for the energy, it is sufficient to cover a smaller range of values than the one actually stored, but a much finer mesh is useful to simplify the integrations. We perform a spline interpolation of the stored energy values of the imaginary and real parts of the self-energy to a fine linear energy mesh of \hbox{$N_{\omega,\textrm{spline}}=30000$} in the interval of  \hbox{$\approx$[-2000:5000]~MeV}.
 These values are used to calculate the spectral function (see Eq.~(\ref{eq:spectral_fun})) necessary to evaluate  Eq.~(\ref{eq:mom_dist}) correctly: with the fine energy mesh this integration is easily performed across the quasiparticle peak via the trapezoidal rule. Finally, one can linearly interpolate the values obtained for $n({\bf p})$ to the mesh of ${\bf p}_3$ in order to perform the integration of the  averaged force, Eq.~\eqref{eq:dd3bf_new}. 
In doing this, the values of $n({\bf p}_3)$ outside the range of the original $N_{\bf p}=70$ mesh are set to zero.
\end{itemize}

\noindent
Here we show a simple Fortran code to perform the previous two steps (\verb|gauss()| is a standard routine to generate a gauss-legendre mesh; \verb|splin()| and \verb|splin2()| are used to perform spline interpolations; \verb|linint()|  performs linear interpolations):

\vspace*{0.3cm}
\lstset{alsolanguage=[90]Fortran}
\begin{lstlisting}
	! ... MOMENTA MESH FOR INTEGRALS OVER MOMENTUM DISTRIBUTION 

	write(*,*) "Correlated distribution function for averaged 3BF integration"

	! choose number of mesh regions for momenta p3, (imesh-1) gauss set + 1 tangent set for farther points
	imesh = 4 
   
	! choose number of points for gauss and tangent sets
	Np1=100   ! gauss
	Np2=50      !tangent
	Np3=(imesh-1)*Np1+Np2     ! total number of mesh points 
    
	itmp = MAX(Np1,Np2)  ! for the auxiliary arrays always allocate the largest between Np1 and Np2
	ALLOCATE(xp3(Np3),wp3(Np3))
	ALLOCATE(xaux( itmp ),waux( itmp ))  
    
	! initialize variables
	xp3=0d0
	wp3=0d0
    
	! first mesh point
	pin = 0d0
     
	do im = 1, imesh-1  ! loop over linear regions
    
			! reset auxiliary variables at each region
			xaux=0d0 
			waux=0d0
    
			pfin = im*(2d0/3d0)*pF  ! set final point of mesh region according to Fermi momentum pF
        
			! ... gaussian set of points for momenta p3 from pin to pfin
			call gauss(pin,pfin,Np1,xaux,waux)
		
			! copy points to final vector for mesh p3
			do ip3=1,Np1
					xp3(ip3+(im-1)*Np1)= xaux(ip3)
					wp3(ip3+(im-1)*Np1)= waux(ip3)
			enddo

			pin = pfin ! set last point of previous region to first point of next region
	enddo

	! ... create the tangent set for higher momenta
	call gauss(0d0,1d0,Np2,xaux,waux)  ! gauss set [0,1] to be mapped to the interval [pin,+infinity]

	c=10d0*pF/tan(pi/2.d0*xaux(Np2))
	do ip3=1,Np2
			xp3(ip3+(im-1)*Np1)=c*tan(pi/2.d0*xaux(ip3))+pin
			xxw=cos(pi/2.d0*xaux(ip3))
			xxw=xxw*xxw
			wp3(ip3+(im-1)*Np1)=pi/2.d0*c/xxw*waux(ip3)
	enddo


	! ... obtaining correlated momentum distribution

	! ... FINE ENERGY MESH WHERE CALCULATIONS ARE DONE
	! ... allocate energy mesh for calculation of momentum distribution
	N_fine=30000
	ALLOCATE(xmom(N_fine))
	ALLOCATE(xmp(Np))
     
	wi=-2000.d0 !MeV  initial energy for spectral function
	wf=5000.d0  !MeV  final energy for spectral function
	dw=(wf-wi)/dble(N_fine-1)
        
	! ... LOOP OVER PMESH        
	do ip=1,Np   ! this is the mesh of stored momenta (usually Np ~ 70)
           
			edp=xpmesh(ip)**2/(2.d0*xmass)  ! kinetic spectrum

			do iw=1,Nwac
					auxre(iw)=xreal_sigma(ip,iw)    ! real part of self-energy
					auxim(iw)=ximag_sigma(ip,iw)  !imaginary part of self-energy
			enddo
	   
			! obtain derivatives of the self-energy for later splines
			call spline(w_actual,auxim,Nwac,yspl,yspl,d2im)  
			call spline(w_actual,auxre,Nwac,yspl,yspl,d2re)   
	  
			! ... LOOP OVER WFINE
			do iif=1,N_fine
					w_fine = wi + dble(iif-1)*dw  
					wfine(iif)=w_fine
					fdfine=fermi(t,xmu,w_fine)  !Fermi-Dirac distribution
	      
					! .. Spline interpolation in fine energy mesh
					call splin2(w_actual,auxim,d2im,Nwac,w_fine,ximsig)
					call splin2(w_actual,auxre,d2re,Nwac,w_fine,xresig)
	      
					! ... Spectral function
					sf=-ximsig/( (w_fine - edp - xresig)**2 + ximsig**2 )/pi
	      
					! ... momentum distribution
					xmom(iif)=sf*fdfine
	   
			enddo ! END LOOP OVER WFINE

			! performs the integration over energy (trapezoidal rule)
			call trapz(w_fine,xmom,N_fine,mom)
			xmp(ip)=mom

	enddo ! END LOOP OVER MOMENTA
	
	! ... interpolation of momentum distribution to mesh xp3 for integrals
	call linint(xpmesh,xmp,Np,xp3,xnp3,Np3)

	! ... set the extrapolated values of n(p) to zero, mesh points xp3 beyond initial mesh xpmesh 
	do ip3=1,Np3
			xnp0=xp3(ip3)
			if(xnp0.gt.xpmesh(Np)) xnp3(ip3)=0d0
			if(xnp0.lt.0d0) xnp3(ip3)=0d0
	enddo

	DEALLOCATE(xaux,waux,xmom,xmp)
  
 \end{lstlisting}
 \vspace*{0.3cm}

Note that the chemical potential $\mu$ enters the calculation of the averaged three-body force, via the Fermi-Dirac function in the expression for momentum distribution, Eq.~\eqref{eq:mom_dist}. For this reason it is best to compute Eq.~\eqref{eq:dd3bf_new} after step ${\bf 2}$ of the iterative procedure presented in the previous section. For further details on including three-body forces in a SCGF infinite matter calculation we refer the reader to Ref.~\cite{ch11_Carbone2014PhD}.

\section{Concluding remarks}
This chapter concludes an overview of the major methods based on Fock space, which are covered in chapters 8, 10 and 11 of this book. All these approaches have the common feature that their computing requirements scale only polynomially with the increase of particle number. This feature has permitted to push {\em ab initio} studies of atomic nuclei up to medium-mass isotopes: a progress that would have seemed unthinkable until just a decade ago. 

Here, we have focused on many-body Green's function theory, which is arguably the most complex of these formalisms but it has the advantage of providing a unique and global view of the many-particle structure and dynamics.   The spectral function is extracted directly from the physics information contained in the one-body Green's function and gives an intuitive understanding of correlations (that is, features that go beyond a simple mean-field description) of the system. Besides, expectation values of observables can be calculated easily, including binding energies.
The formalism of SCGF is so vast that even a dedicated monograph would not be able to cover it in full. In this chapter, we have focused on presenting the two most important techniques that are currently used in modern {\em ab initio} nuclear theory.
 In the first case, the Algebraic Diagrammatic Construction method proves to be particularly suited for the study of finite nuclei, but can as well be applied to infinite matter, as was demonstrated in this chapter. In the second case, we looked at how one can solve the  Dyson equation directly in momentum space for extended systems.  The latter is an important aspect since the formalism allows to construct a fully-dressed propagator at finite temperature, which grants the method to be thermodynamically consistent, preserving all the fundamental laws of conservation. 
For these cases we also discussed the most relevant steps and knowhow necessary for implementing SCGF calculations. Furthermore, we provided working numerical codes that can solve the same toy models used as examples throughout this book: a four-level pairing Hamiltonian and neutron matter with a Minnesota force.   While these applications are simple, the codes we provide already contain the most crucial elements and could be easily extended to real applications (in nuclear physics and other fields too!). We hope this chapter can be the starting point for readers interested in working with many-body Green's functions, starting from the sample codes presented and making use of numerous tips provided for the numerical solutions.

What we did not touch upon, due to lack of space, are the most advanced techniques that have been introduced in recent years or that are still under development. Improving accuracy in calculating open shell isotopes, describing excited spectra,  accessing deformed nuclei and describing pairing and superfluidity at finite temperatures are some among the compelling challenges that are to be addressed and that will be crucial to the study of exotic nuclei at future radioactive beam facilities.
Likewise, the methods described in this chapter can be extended to novel applications in nuclear physics,  besides the structure and reactions with unstable nuclei. Examples are: understanding the response to electroweak probes and the interaction of high energy neutrinos with matter; the spectral function (and hence the individual behavior) of hyperons in finite nuclei and neutron star matter; how thermodynamic properties of nuclear matter impact stellar evolution.
With still much room for further development, Fock space methods, and the SCGF approach in particular, are possibly the most promising frontier for advancing first principle computations on large and complex nuclei. All in all, this is an exciting time not only for computational nuclear physics itself  but also for the quest of an accurate understanding of nuclear structure and related topics.

\begin{acknowledgement}
We thank O. Benhar, A. Cipollone, W. H. Dickhoff, C. Drischler, T. Duguet, K. Hebeler, M. Hjorth-Jensen, J. W. Holt, A. Lovato, G. Mart\'inez-Pinedo, H. M\"uther, P. Navr\'atil, A. Polls, A. Rios, J. Schirmer, A. Schwenk, V. Som\`a and D. Van Neck  for several fruitful collaborations and  enlightening discussions over the years.
This work was supported by the UK Science and Technology Facilities Council (STFC) under Grant No. ST/L005743/1, %
the  Deutsche Forschungsgemeinschaft through Grant SFB 1245
and by the Alexander von Humboldt Foundation through a Humboldt Research Fellowship for Postdoctoral Researchers.
\end{acknowledgement}

\section*{Appendix 1: Feynman rules for the one-body propagator and the self-energy}
\addcontentsline{toc}{section}{Appendix 1: Feynman rules for the one-body propagator and self-energy}
\label{app:Feyn_rules}

We present the Feynman rules associated with the diagrams arising
in the perturbative expansion of Eq.~(\ref{gpert}) at zero temperature. The rules are given both in time and energy formulation and a specific example is given at the end.
We provide the general rules for $p$-body propagators. These arise from a trivial generalization of the perturbative 
expansion of the one-body propagator in Eq.~(\ref{gpert})~\cite{ch11_Carbone2013Nov}. 
At  $k$th order in perturbation theory, any contribution from the time-ordered product in 
Eq.~(\ref{gpert}), or its generalization, is represented by a diagram with $2p$ external 
lines and $k$ interaction lines (called {\em vertices}), 
all connected by means of oriented fermion lines. 
These fermion lines arise from contractions between annihilation and creation operators.
In the following we will explicitly include the $\hbar$ factors.
Applying the Wick theorem to the terms at each order of the above expansion results in the following Feynman rules. At order $k$ in the perturbation series:
\begin{description}
\item[\underline{Rule 1}:] Draw all, topologically distinct and connected diagrams with $k$ vertices,  $p$ incoming and $p$ outgoing external lines, using directed arrows. 
Each vertex representing a \hbox{$n$-body} interaction must have $n$ incoming and $n$ outgoing lines.
For diagrams describing interaction kernels the external lines are not present.
\item[\underline{Rule 2}:] Each oriented fermion line represents a Wick contraction, leading to the unperturbed propagator  
${\rm i}\hbar g_{\alpha\beta}^{(0)}(t_\alpha-t_\beta)$ [or ${\rm i}\hbar g_{\alpha\beta}^{(0)}(\omega_i)$]. 
In time formulation, the $t_\alpha$ and $t_\beta$ label the times of the vertices respectively  at the end and at the beginning of the line. 
In energy formulation, $\omega_i$ denotes the energy carried by the propagator along its oriented line.
\item[\underline{Rule 3}:] Each fermion line starting from and ending at the \emph{same} vertex is an 
equal-time propagator and contributes:  $-{\rm i}\hbar g_{\alpha\beta}^{(0)}(0^-)=\rho_{\alpha\beta}^{(0)}$.
\item[\underline{Rule 4}:] For each one-body, two-body or three-body vertex, write down a factor $\frac{\rm i}{\hbar} U_{\alpha \beta}$, \, $-\frac{\rm i}{\hbar} V_{\alpha\gamma,\beta\delta}$  or  $-\frac{\rm i}{\hbar} W_{\alpha\gamma\xi,\beta\delta\theta}$, respectively. For effective interactions, the factors are $-\frac{\rm i}{\hbar} \widetilde{U}_{\alpha \beta}$, \, $-\frac{\rm i}{\hbar} \widetilde{V}_{\alpha\gamma,\beta\delta}$.
\end{description}
When propagator renormalization is considered, only skeleton diagrams are used in the 
expansion. In that case, the previous rules apply with the substitution 
${\rm i} \hbar g_{\alpha\beta}^{(0)} \to 
{\rm i} \hbar g_{\alpha\beta}$.
Furthermore, note that  Rule 3 generates interaction-reducible diagrams and therefore it is not encountered when working with the
effective Hamiltonian~\eqref{eq:Heff}. However, the correlated density matrix $\rho_{\alpha\beta}$ enters the calculations of $\widetilde{U}$ and $\widetilde{V}$ through  Eqs.~\eqref{eq:UV_eff}.
\begin{description}
\item[\underline{Rule 5}:] Include a factor $(-1)^{L}$ where $L$ is the number of closed fermion loops. This sign comes from the odd permutation of  operators needed to create a loop.
The loops of a single propagator are already accounted for by Rule 3 and must {\em not} be included in the count for $L$.
\item[\underline{Rule 6}:] For a diagram representing a $2p$-point Green's function, add a factor $(-{\rm i}/\hbar)$, whereas for a $2p$-point interaction kernel without external lines (such as $\Sigma^\star(\omega)$) add a factor ${\rm i}\hbar$.
\end{description}
The next two rules require a distinction between the time and the energy representations. 
In the time representation:
\begin{description}
\item[\underline{Rule 7}:] Assign a time to each interaction vertex. All the fermion lines connected to the same vertex $i$ share the same time,~$t_i$. 
\item[\underline{Rule 8}:] Sum over all the internal quantum numbers and integrate over all internal times from $-\infty$ to $+\infty$. 
\end{description}
Alternatively, in energy representation:
\begin{description}
\item[\underline{Rule 7'}:]  Label each fermion line with an energy $\omega_i$, 
under the \emph{constraint} that the total incoming energy equals the total outgoing energy at 
each interaction vertex, \hbox{$\sum_i\omega_i^{in}=\sum_i\omega_i^{out}$}.
\item[\underline{Rule 8'}:] Sum over all the internal quantum numbers and integrate over each independent internal energy, with an extra factor $\frac{1}{2\pi}$, i.e. $\int^{+\infty}_{-\infty} \frac{{\rm d}\omega_i}{2\pi}$.
\end{description}
Each diagram is then multiplied by a combinatorial factor S that  originates from the number of 
equivalent Wick contractions that lead to it. This symmetry factor 
represents the order of the symmetry group for one specific diagram or, in other words, 
the order of the permutation group of both open and closed lines, once the vertices are fixed. 
Its structure, assuming only 2BFs and 3BFs, is the following :
\begin{equation}
S=\frac{1}{k!}\frac{1} {[(2!)^2]^{q} [(3!)^2]^{k-q} }\binom{k}{q} \; C
= \prod_i S_i \; .
\label{diagsymfac}
\end{equation}
Here, $k$ represents the order of expansion. 
$q$ ($k-q$) denotes the number of two-body (three-body) vertices in the diagram.
The binomial factor counts the number of terms in the expansion $(\widehat{V}+\widehat{W})^k$ 
that have $q$ factors of $\widehat{V}$ and $k-q$ factors of $\widehat{W}$.
Finally, $C$ is  the overall number of \emph{distinct} contractions and reflects 
the symmetries of the diagram. Stating general rules to find $C$ is not simple. 
For example, explicit simple rules valid for the well-known $\lambda \phi^4$ scalar  theory are still 
an object of debate~\cite{ch11_Feyn_rules}. 
An explicit calculation for $C$ has to be carried out diagram by diagram 
\cite{ch11_Feyn_rules}. Eq.~(\ref{diagsymfac}) can normally be factorized in a product factors $S_i$,
each due to a particular symmetry of the diagram. In the following, we list a series of specific examples which is,
by all  means, not exhaustive.
\begin{description}
 \item[\underline{Rule 9}:]  For each group of $n$ symmetric lines, or symmetric groups-of-lines as defined below, multiply by a symmetry factor $S_i$=$\frac{1}{n!}$. The overall symmetry factor of the diagram will be $S=\prod_i S_i$.
Possible cases include:
  \end{description}
\begin{itemize}
 \item[(i)]\quad {\em Equivalent lines}.  
 $n$ equally-oriented fermion lines are said to be equivalent if they start from the same initial vertex and end on the same final vertex.
 \item[(ii)]\quad {\em Symmetric and interacting lines}.  
 $n$ equally-oriented fermion lines that start from the same initial vertex and end on the same final 
 vertex, but are linked via an interaction vertex to one or more close fermion line blocks. 
 The factor arises as long as the diagram is {\em invariant} under the permutation of the two blocks.
 \item[(iii)]\quad {\em Equivalent groups of lines}. 
 These are blocks of interacting lines (e.g. series of bubbles) that are equal to each other: 
           they all start from the same initial vertex and end on the same final vertex.
 \end{itemize} 

 Rule 9(i)  is the most well-known case and applies, for instance, to the two second-order diagrams 
 of Fig.~\ref{fig:2ndOrd}. Diagram a) in Fig.~\ref{fig:2ndOrd} has 2 upward-going equivalent lines and requires a symmetry factor $S_e$=$\frac1{2!}$. In contrast, diagram b) in Fig.~\ref{fig:2ndOrd} has 3 upward-going equivalent lines and 2 downward-going equivalent lines, that give a factor $S_e$=$\frac1{2! \, 3!}$=$\frac1{12}$. For an extended explanation on how to calculate this combinatorial factor and examples for rules 9(ii) and 9(iii) we refer to Ref.~\cite{ch11_Carbone2013Nov}.
 
 As an example of the application of the above Feynman rules, we give here the formulae for diagram c) in Fig.~\ref{fig:3rdOrd}. There are two sets of upward-going equivalent lines, which contribute to a
symmetry factor $S_e=\frac{1}{2^2}$. Considering the overall factor of Eq.~(\ref{diagsymfac}) and the
presence of one closed fermion loop, one finds:

\begin{eqnarray}
\Sigma^{(c)}_{\alpha \beta}(\omega)=
- \frac{(i \hbar)^{4} }{4}
\int\frac{{\rm d}\omega_1}{2\pi} \cdots \int\frac{{\rm d}\omega_4}{2\pi}
&&\sum_{\substack{ \gamma\delta\nu \mu\epsilon\lambda \\ \xi\eta\theta \sigma\tau\chi}} 
\widetilde{V}_{\alpha\gamma,\delta\nu} \; g^{(0)}_{\delta\mu}(\omega_1) \, g^{(0)}_{\nu\epsilon}(\omega_2)  \;
W_{\mu\epsilon\lambda,\xi\eta\theta} \; g^{(0)}_{\xi\sigma}(\omega_3) \, g^{(0)}_{\eta\tau}(\omega_4)  
\nonumber\\ &&
\qquad \times \, g^{(0)}_{\theta\gamma}(\omega_1+\omega_2-\omega)
 \; \widetilde{V}_{\sigma\tau,\beta\chi} \;
g^{(0)}_{\chi\lambda}(\omega_3+\omega_4-\omega)  \, .  \quad
\end{eqnarray}

\section*{Appendix 2: Chiral next-to-next-to-leading order three-nucleon forces}
\addcontentsline{toc}{section}{Appendix 2: Chiral next-to-next-to-leading order three-nucleon forces}
\label{app:scgf_3NF}

We report the working equations that result from performing analytically the average of Eq.~(\ref{eq:dd3bf_new}) in the specific case of 
 leading order three-nucleon forces, i.e. next-to-next-to-leading order (NNLO), in the chiral effective filed theory expansion~\cite{ch11_vKol1994,ch11_Epelbaum2002Dec2}. 
At NNLO we have a two-pion exchange (TPE), one-pion exchange (OPE) and a contact three-nucleon forces (3NF), given respectively by the following expressions:

\begin{align}
W^\mathrm{3NF}_\mathrm{TPE} ={}&\quad  \sum_{i\neq j\neq k}  \frac{g_A^2}{8F_\pi^4}
\frac{(\boldsymbol\sigma_i\cdot{\bf q}_i)(\boldsymbol\sigma_j\cdot{\bf q}_j)}{({\bf q}_i^2 + M_\pi^2)
({\bf q}_j^2 + M_\pi^2)}
F_{ijk}^{\alpha\beta}\tau_i^{\alpha}\tau_j^{\beta} \; ,
\label{eq:tpe} \\
W^\mathrm{3NF}_\mathrm{OPE} ={}&  -\sum_{i\neq j\neq k} \frac{c_D g_A}{8F_\pi^4\Lambda_\chi}
\frac{\boldsymbol\sigma_j\cdot{\bf q}_j}{{\bf q}_j^2 + M_\pi^2}(\boldsymbol\tau_i\cdot\boldsymbol\tau_j)
(\boldsymbol\sigma_i\cdot{\bf q}_j)  \; ,
\label{eq:ope} \\
W^\mathrm{3NF}_\mathrm{cont} ={}& \quad  \sum_{j\neq k} \frac{c_E}{2F_\pi^4\Lambda_\chi}
\boldsymbol\tau_j \cdot \boldsymbol\tau_k \; ,
\label{eq:cont}
\end{align}
where the ${\bf p}_i$ are the initial and ${\bf p'}_i$ are the final single-particle momenta of the $i$th nucleon ($i=1,2,3$), the ${\bf q}_i = {\bf p'}_i -{\bf p}_i$ are the transferred momenta and ${\bf\sigma}_i$ and ${\bf\tau}_i$ are the sipn and isospin matrices.  
The physical constants appearing in these expressions are the  axial-vector coupling constant $g_A$, the average pion mass $M_\pi$, the weak pion decay constant $F_\pi$ and the chiral symmetry breaking constant $\Lambda_\chi \sim$700~MeV.
The quantity $F_{ijk}^{\alpha\beta}$ in the TPE contribution~(\ref{eq:tpe}) is
\begin{align}
F_{ijk}^{\alpha\beta}=\delta^{\alpha\beta} [-4M_\pi^2c_1+2 c_3{\bf q}_i\cdot{\bf q}_j]
+\sum_\gamma c_4\epsilon^{\alpha\beta\gamma}\tau^\gamma_k\boldsymbol\sigma_k\cdot[{\bf q}_i\times{\bf q}_j]\,.
\label{eq:tpe_tensor}
\end{align}

The force is regularized with a function that in Jacobi momenta reads:
\begin{equation}
\label{eq:regulator}
f({\bf p}_1,{\bf p}_2,{\bf p}_3)=f(p,q)=\exp{\left[-\frac{(p^2+3q^2/4)}{\Lambda^2_\textrm{3NF}}\right]^n}\,,
\end{equation}
where ${\bf p}=({\bf p}_1-{\bf p}_2)/2$ and ${\bf q}=2/3({\bf p}_3-({\bf p}_1+{\bf p}_2)/2)$ are identified only in this expression as the Jacobi momenta. $\Lambda_\textrm{3NF}$ defines the cutoff value applied to the 3NF in order to obtain a three-body contribution which dies down similarly to the two-body part one. The regulator function is applied both on incoming $({\bf p},{\bf q})$ and outgoing $({\bf p'},{\bf q'})$ Jacobi momenta. In present numerical calculations, the approximation of ${\bf P}\equiv {\bf p}_1+{\bf p}_2=0$ is used to facilitate the solution of equations. The averaged terms presented in the following are calculated only for equal relative incoming and outgoing momentum, i.e. ${\bf k}={\bf k'}$ with ${\bf k}=|{\bf p}_1-{\bf p}_2 |/2$ and ${\bf k'}=|{\bf p}'_1-{\bf p}'_2|/2$; an extrapolation is then applied to obtain the off-diagonal potential matrix elements~\cite{ch11_Carbone2014}. Given these conditions, the regulator on incoming and outgoing momenta can be defined as a function of $f(k,p_3)$.\\

{\bf Symmetric Nuclear Matter.} Let's start with the isospin-symmetric case of nuclear matter. Evaluating Eq.~(\ref{eq:dd3bf_new}) for the TPE term of Eq.~(\ref{eq:tpe}) leads to three contracted in-medium two-body interactions. 

\vskip 0.2 cm
\noindent
{\bf TPE-1:} The first term is an isovector tensor term, this corresponds to a 1$\pi$ exchange contribution with an in-medium pion propagator:
\begin{equation}
\widetilde V_\mathrm{TPE-1}^\mathrm{3NF}=\frac{g_A\,\rho_f}{2 F_\pi^4}
\frac{(\boldsymbol\sigma_1\cdot{\bf q})(\boldsymbol\sigma_2\cdot{\bf q})}{[q^2 + M_\pi^2]^2}
\boldsymbol\tau_1\cdot\boldsymbol\tau_2[2 c_1M_\pi^2+ c_3\,q^2]\,.
\label{eq:tpe_dd_1}
\end{equation}
$\rho_f$ defines the integral of the correlated momentum distribution function weighed by the regulator function $f(k,p_3)$
\begin{equation}
\frac{\rho_f}{\nu_d}=\int \frac{{\mathrm d}{\bf p}_3}{(2\pi)^3}n({\bf p}_3)f(k,p_3)\,,
\label{eq:rho_f}
\end{equation}
where $\nu_d$ is the degeneracy of the system, $\nu_d=2$ for pure neutron matter and $\nu_d=4$ in the isospin symmetric case. If the regulator function included in Eq.~(\ref{eq:rho_f}) were not dependent on the internal integrated momentum $p_3$, the integral would reduce to the value of the total density of the system, $\rho$, divided by the degeneracy and multiplied by an external regulator function.

\vskip 0.2 cm
\noindent 
{\bf TPE-2:} The second term is also a tensor contribution to the in-medium nucleon-nucleon interaction. It adds up to the previous term. 
This term includes vertex corrections to the 1$\pi$ exchange due to the presence of the nuclear medium:
\begin{align}
\widetilde V_\mathrm{TPE-2}^\mathrm{3NF} ={}& \frac{g_A^2}{8\pi^2F_\pi^4}
\frac{(\boldsymbol\sigma_1\cdot{\bf q})(\boldsymbol\sigma_2\cdot{\bf q})}{q^2 + M_\pi^2} \boldsymbol\tau_1\cdot\boldsymbol\tau_2 \nonumber  \\
 & ~\times
\Big\{-4c_1M_\pi^2\left[\Gamma_1(k)+\Gamma_0(k)\right]
-(c_3+c_4)\left[q^2(\Gamma_0(k)+2\Gamma_1(k)+\Gamma_3(k))+4\Gamma_2(k)\right]
+4c_4{\pazocal I}(k)\Big\}\,.
\label{eq:tpe_dd_2}
\end{align}
We have introduced the functions $\Gamma_i(k)$~($i$=0-3)  and ${\pazocal I}(k)$, which are integrals over a single pion propagator:
\begin{eqnarray}
\label{eq:gamma0}
\Gamma_0(k) &=& \qquad \int\frac{{\mathrm d}{\bf p}_3}{2\pi}n({\bf p}_3)
\frac{1}{[{\bf k}+{\bf p}_3]^2 + M_\pi^2}f(k,p_3)\, ,
\\ \label{eq:gamma1}
\Gamma_1(k) &=&\frac{1}{k^2}\int\frac{{\mathrm d}{\bf p}_3}{2\pi}n({\bf p}_3)
\frac{ {\bf k}\cdot{\bf p}_3}{[{\bf k}+{\bf p}_3]^2 + M_\pi^2}f(k,p_3)\,, 
\\ \label{eq:gamma2}
\Gamma_2(k) &=&\frac{1}{2k^2}\int\frac{{\mathrm d}{\bf p}_3}{2\pi}n({\bf p}_3)
\frac{p_3^2k^2-({\bf k}\cdot{\bf p}_3)^2}{[{\bf k}+{\bf p}_3]^2 + M_\pi^2}f(k,p_3)\,, 
\\ \label{eq:gamma3}
\Gamma_3(k) &=&\frac{1}{2k^4}\int\frac{{\mathrm d}{\bf p}_3}{2\pi}n({\bf p}_3)
\frac{3({\bf k}\cdot{\bf p}_3)^2-p_3^2k^2}{[{\bf k}+{\bf p}_3]^2 + M_\pi^2}f(k,p_3)\,, 
\\ \label{eq:i_integral}
{\pazocal I}(k) &=& \qquad  \int \frac{{\mathrm d}{\bf p}_3}{2\pi}n({\bf p}_3)
\frac{[{\bf p}_3\pm{\bf k}]^2}{[{\bf p}_3+{\bf k}]^2 + M_\pi^2}f(k,p_3)\,.
\end{eqnarray}

\vskip 0.2 cm
\noindent
{\bf TPE-3:} The last TPE contracted term includes in-medium effects for a 2$\pi$ exchange two-body term:
\begin{eqnarray}
\nonumber 
\widetilde V_\mathrm{TPE-3}^\mathrm{3NF} = \frac{g_A^2}{16\pi^2F_\pi^4}
  &\Big\{ & -12c_1M_\pi^2\big[2\Gamma_0(k)-G_0(k,q)(2M_\pi^2+q^2)\big]
\\\nonumber &&   
- \, c_3\big[12\pi^2\rho_f-12(2M_\pi^2+q^2)\Gamma_0(k)
- \, 6q^2\Gamma_1(k)+3(2M_\pi^2+q^2)^2G_0(k,q)\big] 
\\\nonumber &&  
+  \, 4c_4 \boldsymbol\tau_1\cdot\boldsymbol\tau_2 \big[ (\boldsymbol\sigma_1\cdot\boldsymbol\sigma_2)\, q^2 
   -(\boldsymbol\sigma_1\cdot{\bf q})(\boldsymbol\sigma_2\cdot{\bf q})\big]
G_2(k,q)
\\\nonumber &&  
-\, (3c_3+c_4\boldsymbol\tau_1\cdot\boldsymbol\tau_2)\,i(\boldsymbol\sigma_1+\boldsymbol\sigma_2)\cdot({\bf q}\times{\bf k})
\\\nonumber &&  \qquad \qquad \qquad \qquad \times
\big[2\Gamma_0(k)+2\Gamma_1(k)-(2M_\pi^2+q^2)G_0(k,q)+2G_1(k,q)\big]
\\\nonumber &&  
- \, 12c_1M_\pi^2\,  i(\boldsymbol\sigma_1+\boldsymbol\sigma_2)\cdot({\bf q}\times{\bf k})
\big[G_0(k,q)+2G_1(k,q)\big]
\\ &&  
+ \, 4c_4\boldsymbol\tau_1\cdot\boldsymbol\tau_2\boldsymbol\sigma_1\cdot({\bf q}\times{\bf k})\boldsymbol\sigma_2\cdot({\bf q}\times{\bf k})
\big[G_0(k,q)+4G_1(k,q)+4G_3(k,q)\big]\Big\} \,. \qquad \qquad 
\label{eq:tpe_dd_3}
\end{eqnarray}
Here we have introduced the function $G_0(k,q)$, which is an integral over the product of two different pion propagators and defined as follows:
\begin{equation}
G_{0,\star,\star\star}(k,q) =
\int \frac{{\mathrm d}{\bf p}_3}{2\pi}n({\bf p}_3)
\frac{\{p_3^0,p_3^2,p_3^4\}}{\big[[{\bf k}+{\bf q}+{\bf p}_3]^2+M_\pi^2\big]\big[[{\bf p}_3+{\bf k}]^2+M_\pi^2\big]}f(k,p_3) \, ,
\label{eq:G_0} 
\end{equation}
where the subscripts $0$, $\star$ and $\star\star$ refer respectively to the powers $p_3^0$, $p_3^2$ and $p_3^4$ in the numerator. 
The functions $G_{\star}(k,q)$ and $ G_{\star\star}(k,q)$ have been introduced to define the remaining functions, $G_1(k,q)$, $G_2(k,q)$ and $G_3(k,q)$:
\begin{align}
\label{eq:G_1}
G_1(k,q)={}& \frac{\Gamma_0(k)-(M_\pi^2+k^2)G_0(k,q)-G_\star(k,q)}{4k^2-q^2}\,,
\\
\label{eq:G_1star}
G_{1\star}(k,q)={}& \frac{3\Gamma_2(k)+k^2\Gamma_3(k)-(M_\pi^2+k^2)G_\star(k,q)-G_{\star\star}(k,q)}{4k^2-q^2}\,,
\\
\label{eq:G_2}
G_2(k,q)={}& (M_\pi^2+k^2)G_1(k,q)+G_\star(k,q)+G_{1\star}(k,q)\,,
\\
\label{eq:G_3}
G_3(k,q)={}& \frac{\Gamma_1(k)/2-2(M_\pi^2+k^2)G_1(k,q)-2G_{1\star}(k,q)-G_\star(k,q)}{4k^2-q^2}\,.
\end{align}
Note that $G_{1\star}(k,q)$ is needed only to define $G_2(k,q)$ and $G_3(k,q)$.

\allowdisplaybreaks[0]

\vskip 0.3 cm

Integrating Eq.~(\ref{eq:dd3bf_new}) for the OPE 3NF term, given in Eq.~(\ref{eq:ope}), leads to two contributions.

\vskip 0.2 cm
\noindent
{\bf OPE-1:} The first one is a tensor contribution which defines a vertex correction to a 1$\pi$ exchange nucleon-nucleon term. It is proportional to the quantity $\rho_f$, similar to what was obtained for the TPE 3NF contracted term $\widetilde V_\mathrm{TPE-1}^\mathrm{3NF}$ (see Eq.~\eqref{eq:tpe_dd_1}):
\begin{align}
\widetilde V_\mathrm{OPE-1}^\mathrm{3NF}={}&-\frac{c_D\,g_A\,\rho_f}{8\,F_\pi^4\,\Lambda_\chi}
\frac{(\boldsymbol\sigma_1\cdot{\bf q})(\boldsymbol\sigma_2\cdot{\bf q})}{q^2 + M_\pi^2}
(\boldsymbol\tau_1\cdot\boldsymbol\tau_2)\,.
\label{eq:ope_dd_1}
\end{align}
As for the $\widetilde V_\mathrm{TPE-1}^\mathrm{3NF}$ term, $\widetilde V_\mathrm{OPE-1}^\mathrm{3NF}$ is an isovector tensor term.

\vskip 0.2 cm
\noindent
{\bf OPE-2:} The second term derived from the 3NF OPE defines a vertex correction to the short-range contact nucleon-nucleon interaction. It reads:
\begin{align}
\widetilde V_\mathrm{OPE-2}^\mathrm{3NF}={}\frac{c_Dg_A}{16\pi^2F_\pi^4\Lambda_\chi}
 & \Big\{  \big( \Gamma_0(k)+2\Gamma_1(k)+\Gamma_3(k) \big)
\left[ \boldsymbol\sigma_1\cdot\boldsymbol\sigma_2\Big(2k^2-\frac{q^2}{2}\Big) 
~+~ (\boldsymbol\sigma_1\cdot{\bf q}\,\boldsymbol\sigma_2\cdot{\bf q})\Big(1-\frac{2k^2}{q^2}\Big)
\right.
\nonumber \\
&  \qquad  \left.
  ~-~  \frac{2}{q^2}\boldsymbol\sigma_1\cdot({\bf q}\times{\bf k}) \boldsymbol\sigma_2\cdot({\bf q}\times{\bf k})\frac{1}{q^2}
  \right] (\boldsymbol\tau_1\cdot\boldsymbol\tau_2)  
\nonumber \\
&\quad +~  2\Gamma_2(k)(\boldsymbol\sigma_1\cdot\boldsymbol\sigma_2) \, (\boldsymbol\tau_1\cdot\boldsymbol\tau_2)
~+~ 6{\pazocal I}(k) 
\Big\}  \,.
\label{eq:ope_dd_2}
\end{align}

\vskip .3 cm
\noindent
{\bf Exercise 11.8.} Compute Eq.~(\ref{eq:dd3bf_new}) for the contact term given in Eq.~(\ref{eq:cont}). Demonstrate that it yields a scalar central contribution to the in-medium nucleon-nucleon interaction proportional to $\rho_f$ with formal expression:
\begin{equation}
\widetilde V_\mathrm{cont}^\mathrm{3NF}=-\frac{3 c_E\rho_f}{2 F_\pi^4\Lambda_\chi}\,.
\label{eq:cont_dd}
\end{equation}
\\

{\bf Pure Neutron Matter.} In the case of pure neutron matter, the evaluation of Eq.~(\ref{eq:dd3bf_new}) is simplified. In fact,
the trace over isospin is trivial because pairs of neutrons can only be in total isospin $T=1$, thus $\boldsymbol\tau_1\cdot\boldsymbol\tau_2=1$. Consequently the exchange operators reduces only to the momentum and spin part. In operator form it reads:
\begin{equation}
P_{ij}=\frac{1+\boldsymbol\sigma_i\cdot\boldsymbol\sigma_j}{2}\,.
\label{eq:perm_op_2}
\end{equation}
Furthermore it can also be proved that for a non-local regulator, such as Eq.~\eqref{eq:regulator}, the 3NF terms proportional to $c_4$, $c_D$ and $c_E$ vanish~\cite{ch11_Tolos2008,ch11_Hebeler2010Jul}.  Therefore the only non zero density-dependent contributions in neutron matter are those containing the low-energy constants $c_1$ and $c_3$ in Eq.~(\ref{eq:tpe}).
All of their expressions seen from above  remain valid  except for the change in the trace over isospin indices. It follows that the density-dependent interacting terms obtained in neutron matter will only differ with respect to the symmetric case ones by different prefactors.

In order to obtain the correct degeneracy for neutron matter, i.e.  $\nu_d=2$, we need to replace $\rho_f \rightarrow 2\rho_f$ in the $\widetilde V_\mathrm{TPE-1}^\mathrm{3NF}$ contribution of Eq.~(\ref{eq:tpe_dd_1}) and the $\widetilde V_\mathrm{TPE-3}^\mathrm{3NF}$ contribution of Eq.~(\ref{eq:tpe_dd_3}), (see also Eq.~(\ref{eq:rho_f})). The isovector tensor terms $\widetilde V_\mathrm{TPE-1}^\mathrm{3NF}$ and  $\widetilde V_\mathrm{TPE-2}^\mathrm{3NF}$, given in Eqs.~\eqref{eq:tpe_dd_1} and~\eqref{eq:tpe_dd_2} must then change prefactor according to:
\begin{equation}
\widetilde V_\mathrm{TPE-1}^\mathrm{3NF}: \boldsymbol\tau_1\cdot\boldsymbol\tau_2 \rightarrow \frac{1}{2}\boldsymbol\tau_1\cdot\boldsymbol\tau_2\,, 
\label{eq:pnm_tpe_1}
\end{equation}
\begin{equation}
\widetilde V_\mathrm{TPE-2}^\mathrm{3NF}: \boldsymbol\tau_1\cdot\boldsymbol\tau_2 \rightarrow 
\frac{1}{4}(\boldsymbol\tau_1\cdot\boldsymbol\tau_2-2)\,.
\label{eq:pnm_tpe_2}
\end{equation}
The isoscalar part of the density-dependent potential appearing in $\widetilde V_\mathrm{TPE-3}^\mathrm{3NF}$, which contributes to both a central and spin-orbit terms, must change prefactor according to:
\begin{equation}
\widetilde V_\mathrm{TPE-3}^\mathrm{3NF}: 1 \rightarrow \frac{1}{3}\,.
\label{eq:pnm_tpe_3}
\end{equation}

\bibliographystyle{spphys}
\bibliography{chapter_11_biblio}

\end{document}